\documentclass[useAMS,usenatbib]{mn2e}

\usepackage{graphicx}
\usepackage{comment}
\usepackage{amsmath}
\usepackage{amssymb}

\newif\ifhires
\hiresfalse

\title[Photometric Proxies for Selecting Spirals]{Non-Parametric Cell-Based Photometric Proxies for Galaxy Morphology: Methodology and Application to the Morphologically-Defined Star Formation -- Stellar Mass Relation of Spiral Galaxies in the Local Universe}
\author[M. W. Grootes, et al.]
{\parbox{\textwidth}{M.~W.~Grootes$^{1}$\thanks{E-mail: \texttt{meiert.grootes@mpi-hd.mpg.de}},
R.~J.~Tuffs$^{1}$,
C.~C.~Popescu$^{2}$,
A.~S.~G.~Robotham$^{3,4}$,
M.~Seibert$^{5}$,\\
L.~S.~Kelvin$^{3,4,6}$}\vspace{0.4cm}\\
\parbox{\textwidth}{$^{1}$Max-Planck Institut f\"ur Kernphysik, Saupfercheckweg 1, 69117 Heidelberg, Germany\\
$^{2}$Jeremiah Horrocks Institute, University of Central Lancashire, Preston PR1 2HE, UK\\
$^{3}$ICRAR, The University of Western Australia, 35 Stirling Highway, Crawley, WA 6009,Australia\\
$^{4}$SUPA School of Physics \& Astronomy, University of St. Andrews, North Haugh, St. Andrews KY16 9SS, UK\\
$^{5}$Observatories of the Carnegie Institution for Science, 813 Santa Barbara Street, Pasadena, CA 91101, USA\\
$^{6}$Institut f\"ur Astro- und Teilchenphysik, Universit\"at Innsbruck, Technikerstrasse 25, 6020 Innsbruck, Austria\\
}}

\begin{document}

\date{Accepted ???? Received  ????; in original form ????}

\pagerange{\pageref{firstpage}--\pageref{lastpage}} \pubyear{2002}

\maketitle

\label{firstpage}

\begin{abstract}
We present a non-parametric cell-based method of selecting highly pure and largely complete samples of spiral galaxies using photometric and
structural parameters as provided by standard photometric pipelines and simple shape fitting algorithms. The performance of the method is quantified for different parameter combinations, using purely human-based classifications as a benchmark. The discretization of the parameter space allows a markedly superior selection than commonly used proxies relying on a fixed curve or surface of separation. Moreover, we find structural parameters derived using passbands longwards of the $g$ band and linked to older stellar populations, especially  the stellar mass surface density $\mu_*$ and the $r$ band effective radius $r_e$, to perform at least equally well as parameters more traditionally linked to the identification of spirals by means of their young stellar populations, e.g. UV/optical colours. In particular the distinct bimodality in the parameter $\mu_*$, consistent with expectations of different evolutionary paths for spirals and ellipticals, represents an often overlooked yet powerful parameter in differentiating between spiral and non-spiral/elliptical galaxies. We use the cell-based method for the optical parameter set including $r_e$ in combination with the S\'ersic index $n$ and the $i-$band magnitude to investigate the intrinsic specific star-formation rate - stellar mass relation ($\psi_* - M_*$) for a morphologically defined volume limited sample of local universe spiral galaxies. The relation is found to be well described by $\psi_* \propto M_*^{-0.5}$ over the range of $10^{9.5} M_{\odot} \le M_* \le 10^{11} M_{\odot}$ with a mean interquartile range of $0.4\,$dex. This is somewhat steeper than previous determinations based on colour-selected samples of star-forming galaxies, primarily due to the inclusion in the sample of red quiescent disks.  
\end{abstract}


\section{Introduction}\label{intro}
With the advent of large optical photometric ground and space-based surveys which are ongoing, commencing (The Sloan Digital Sky Survey 
(SDSS; \citealt{YORK2000,ABAZAJIAN2009}, The Galaxy And Mass Assembly Survey (GAMA; \citealt{DRIVER2011}), SKYMAPPER 
(\citealt{KELLER2007}), The VST Atlas, The Kilo Degree Survey (KiDS; \citealt{DEJONG2012}), The Dark Energy Survey (DES; \citealt{DES2005})) or 
scheduled to commence in the next years (e.g., EUCLID; \citealt{LAUREIJS2011}),
the number of extragalactic sources with reliable, uniform data is increasing dramatically, further opening the door to statistical studies of
the population of galaxies, both at local and intermediate redshifts.\newline
To first order, the visible matter distributions of galaxies may be classified as being best described either as an exponential disk, i.e. a largely 
rotationally-supported system, or a spheroid, i.e. a largely pressure-supported system. This dichotomy forms the basis of the standard 
morphological categorization of galaxies into late-types/spirals and early-types/ellipticals, introduced by \citet{HUBBLE1926} and in widespread 
use ever since. This basic morphological bimodality of the galaxy population appears to be mirrored in a range of physical properties, with late-
type/spiral galaxies having blue UV/optical colours and showing evidence of star formation, on average, while early-type/elliptical galaxies 
appear red on average, and mostly only display a low level of star formation, if any at all \citep[e.g.][]
{STRATEVA2001,BALDRY2004,BALOGH2004}. However, a wide variety of exceptions to this rule exist. For example, spiral galaxies may appear 
red due to the attenuation of their emission by dust in their disks, or a spiral may truly have very low star formation and red colours whilst 
maintaining its morphological identity, while, on the other hand, an elliptical galaxy may appear blue due to a localized recent burst of star-
formation.\newline
It is assumed that different modes of assembly of the stellar populations of these galaxy categories are responsible for the distinction. This in 
turn, necessitates the ability to reliably identify and distinguish between the types of galaxies when investigating the physical processes 
determining galaxy formation and evolution on the basis of large statistical samples of galaxies. Furthermore, it is clear that in any investigation 
of galaxy properties for a given morphological class, the classification itself should not introduce a bias into the property being investigated. For 
example, a pure sample of spiral galaxies used to investigate star-formation as a function of galaxy environment must include the population of 
red, passively star-forming spiral galaxies.\newline
Visual classifications of galaxy morphology by professional astronomers therefore remain the method of choice and the benchmark for robustly 
identifying the morphology of a galaxy. However, such classifications may suffer from biases arising from the individual performing the 
classification, and the uncertainty/robustness of the classification is difficult to quantify. Furthermore, in the case of marginally resolved data, 
even the ability of the human eye to identify morphological structure may be limited, so that the decreasing linear resolution as a function of 
redshift may introduce systematic biases. In such cases, quantitative photometric measures of the light-profile may be at least as reliable as 
human classifications. The overriding fact which immediately stymies the visual classification by professionals of all sources in modern imaging 
surveys such as SDSS, however, is the size of the galaxy samples provided by the surveys, and accordingly the time required for classification. 
Thus, one is forced to develop alternative schemes for obtaining morphological classifications of large samples of galaxies.\newline 
Recently, in an attempt to circumvent the limitations in sample size, reduce
the possibility of bias, and provide an objective measure of robustness, \citet{LINTOTT2008} have enlisted the help of 'citizen scientists´ in 
visually classifying a large fraction
of SDSS DR7 galaxies in the GALAXY ZOO project \citep{LINTOTT2008,LINTOTT2011}, releasing a catalogue of probability-weighted visual 
classifications into spirals and ellipticals.
Although demonstrably feasible, such an approach is nevertheless very time consuming, especially on large data-sets.\newline 
The often adopted alternative is to attempt an automatic classification of galaxies based on some proxy for a galaxy's morphology. These 
automatic classification schemes can be roughly divided into three categories: i) those relying on a detailed analysis of the full imaging products, 
ii) those using a wide variety of photometric and spectroscopic proxies, in combination with a sophisticated algorithmic decision process, and iii) 
those using one or two simple, usually photometric, parameters and a fixed or simply parameterized separator. Of course, hybrids between 
these categories also exist.\newline
Examples of the first category include the Concentration, Asymmetry, and Clumpiness (CAS, \citealp{CONSELICE2003}) parameters, derived 
directly from the data reduction and model fitting of the imaging data, as well as the Gini coefficient \citep{GINI1912,ABRAHAM2003,LOTZ2004} 
and the $\mathrm{M}_{20}$ coefficient \citep{LOTZ2004}. Forming a hybrid between this and the second category, \citet{SCARLATA2007} have 
introduced the \textit{ Zurich Estimator of Structural Types (ZEST)} based on a principle component analysis of these and other model-
independent quantities, which has been applied to various data sets. Examples of the second category are given by classification schemes based 
on neural networks \citep[e.g.][]{BANERJI2010} and making use of support vector machines \citep{HUERTAS-COMPANY2011}. Finally, the third 
category, which finds widespread use, includes, for example,  
the concentration index \citep{STRATEVA2001,STOUGHTON2002,KAUFFMANN2003}, 
the location in colour-magnitude space \citep{BALDRY2004}, the S\'ersic index \citep{BLANTON2003,BELL2004, JOGEE2004, 
RAVINDRANATH2004,BARDEN2005}, 
the location in the $NUV-r$ resp. $u-r$ vs. log($n$) plane \citep{KELVIN2012,DRIVER2012}, the location in the space defined by the SDSS 
$\mathrm{f_{dev}}$
parameter (i.e., the fraction of a galaxy's flux which is fit by the 
de Vaucouleurs profile \citep{DEVAUCOULEURS1948} in the best fit linear combination of a de Vaucouleurs and an exponential profile) and the 
axis ratio of the best fit exponential profile, $\mathrm{q_{exp}}$ \citep{TEMPEL2011}, and, in the case of high-z galaxies the location in the 
$(U-V)$ - $(V-J)$ restframe colour-colour plane \citep{PATEL2012}.\newline
Overall, the advantages and disadvantages of the automatic schemes can also be categorized in a similar manner. Schemes in the category i) 
ideally require well resolved imaging, which may be difficult to obtain for faint galaxies in wide field imaging surveys, even in the local universe. 
Furthermore they require detailed imaging products, often including intermediate data reduction products which are not archived, making an 
independent morphological classification very time consuming and/or computationally expensive, especially for large data sets. Schemes in 
category ii), on the other hand, require the implementation of a complex analysis algorithm in addition to the existence of a training set of 
objects with known morphologies, and may require assumptions about the nature of the statistical distribution of the parameters considered. 
Finally, for the third category, the simple parameterization must limit either the degree to which the selection recovers all members of a given 
morphological category, or the level at which the classification is robust against contamination, even for proxies which make use of structural 
information. Furthermore, it should be noted that the majority of the methods considered make use of parameters linked directly to ongoing 
star-formation, and as such may introduce a bias into the star-formation properties of a selected galaxy sample. For example in category i), the 
clumpiness parameter in the CAS scheme traces localized current star formation in spirals, while in category ii) both the methods of 
\citet{BANERJI2010} and \citet{HUERTAS-COMPANY2011} make use of galaxy colours, and \citet{BANERJI2010} uses texture of the imaging as 
well. Finally in category iii) a range of simple proxies make use of the colour bimodality, linked to star-formation, of the galaxy population. 
\newline
In the following we present a non-parametric method for selecting spirals based on the combinations of two and three photometric and simple 
structural parameters. The method is based on a discretization of the parameter space spanned by the parameter combination performed using 
an adaptive grid which increases the resolution in regions of high galaxy parameter space density. The division of the discretized parameter 
space into a spiral and a non-spiral subvolume is calibrated using the morphological classifications of GALAXY ZOO Data Release 1 (DR1; 
\citep{LINTOTT2011}. We quantify the performance of each parameter combination in terms of completeness and purity, identifying those
with the best performance, also investigating parameter combinations which make no use of properties directly linked to ongoing star-
formation. This approach can be considered formally analogous to the classifications of stars in discrete spectral classes as discussed in the 
review of \citet{MORGAN1973}.  \newline
We describe the data used in Sect.~\ref{data} and the method in Sect.~\ref{method}. We then investigate the
performance of the parameter combinations in Sect.~\ref{parameters} and compare the performance of our selection with other methods 
in Sect. \ref{comparison}. We discuss our results and the applicability of the method in Sect.~\ref{discussion}, and apply the selection method to 
obtain a reliable sample of spirals as a basis for investigating the intrinsic scatter in the stellar mass - specific star-formation rate relation of 
this class of galaxies in Sect.~\ref{application}. Finally, we close by summarizing our results in Sect.~\ref{outlook}.
Throughout the paper we assume an $\Omega_M = 0.3$, $\Omega_{\lambda} = 0.7$, $H_0 = 70\,
\mathrm{km}\mathrm{s}^{-1}\mathrm{Mpc}^{-1}$ cosmology.\newline

\section{Data}\label{data}
Within this work we aim to investigate the efficacy and performance as proxies of various combinations of UV/optical photometric parameters 
for the morphological selection of spiral galaxies.  To facilitate this comparison
and broaden the range of possible proxies we have endeavored to create an unbiased sample of galaxies with as much available
data as possible. We have selected all spectroscopic objects with \verb SpecClass = 2 (Galaxies) from the seventh data release (DR7) 
of SDSS \citep{ABAZAJIAN2009} which lie within the GALEX MIS depth ($1500\,$s; \citealp{MARTIN2005,MORRISSEY2007}) footprint. We have 
matched this 
sample to the catalogue of the 
MPA/JHU analysis of SDSS DR7 spectra (providing emission line fluxes) and to the catalogue of single S\'ersic fits recently 
published by \citet{SIMARD2011} using the SDSS unique identifiers, and to the preliminary NUV GALEX MIS depth unique NUV source galaxy 
catalogue GCAT MSC (Seibert et al., 2013 in prep.)
using a 4 arcsec matching radius\footnote{We note that the GCAT MSC includes a cut on $S/N > 3$}. Given the uncertainties involved with flux 
redistribution \citep[e.g.,][]{ROBOTHAM2011}, we have chosen to treat only 
one-to-one matches between SDSS and GALEX as possessing reliable UV data.\newline
Where multiple spectra are available for a single photometric object, we have 
used the spectrum corresponding to the the MPA/JHU entry. Where multiple spectra form the MPA/JHU reductions are available, we have chosen 
the spectrum
with the smallest redshift error. 
In order to obtain a reliable benchmark morphological classification, we have matched the sample to the GALAXY ZOO data release 1 (DR1)
\citep{LINTOTT2008,LINTOTT2011} catalogue of visual, red-shift debiased morphological classifications 
\citep{BAMFORD2009,LINTOTT2011} using the photometric SDSS \verb ObjId, limiting ourselves to local universe sources (redshift $z \le 
0.13$).  
This selection provides a sample of 166429 galaxies (referred to as the \textit{Opticalsample}), with a subsample of 114047 NUV detected, 
uniquely matched sources (referred to as the \textit{NUVsample}). Finally we have cross-matched these samples to the catalogue of $\sim14$k 
bright SDSS DR4 \citep{ADELMAN-MCCARTHY2006} galaxies with detailed morphological classifications of \citet{NAIR2010}. This results in a 
subsample of 6220 sources with two independent morphological classifications (which we refer to as the \textit{NAIRsample}). 4470 sources in 
the \textit{NAIRsample} have NUV detections, and we refer to this subsample as the \textit{NUVNAIRsample}.\newline   

\subsection{SDSS and GALEX photometry}\label{data_phot}
We have retrieved Petrosian magnitudes, the foreground extinction, the $f_{deV}$ and $q_{exp}$
parameters from the SDSS photometric pipeline, and the petrosian 50th ($R_{50}$) and 90th ($R_{90}$) percentile 
radii in the $u$, $g$, $r$, and $i$ pass-bands from the SDSS database using CasJobs.
To obtain total (S\'ersic) magnitudes we use the algorithms for converting SDSS petrosian magnitudes to total 
S\'ersic magnitudes 
derived by \citet{GRAHAM2005}. The obtained
magnitudes have been corrected for foreground extinction using the extinction values supplied by SDSS (derived from
 the \citet{SCHLEGEL1998} dust maps). K-corrections to $z=0$ have been performed using 
\verb kcorrect_v4.2  \citep{BLANTON2007}.\newline
GALEX sources with NUV artifact flag indicating window or dichroic reflections have been removed from the sample.
The FUV and NUV magnitudes of the matched GALEX sources have been corrected for foreground extinction using the 
\citet{SCHLEGEL1998} dust maps and $A_{\mathrm{FUV}} = 8.24\,E(B-V)$ and $A_{\mathrm{NUV}} = 8.2\,E(B-V)$ 
following \citet{WYDER2007}. \newline

Photometric stellar mass estimates have been calculated from the extinction and k-corrected magnitudes using 
the $g-i$ colour and 
the $i$-band absolute magnitude $M_i$ as
\begin{equation}
\textrm{log}(M_{*}) = -0.68 + 0.7\cdot(g - i) -0.4M_{i} + 0.4\cdot 4.58\;,
\label{Eq_SM}
\end{equation}
where the factor $4.58$ is identified as the solar $i$-band magnitude, following
the prescription provided by \citet{TAYLOR2011}.\newline

\subsection{Emission line data} \label{data_haeqw}
We make use of the emission line fluxes form the H$\alpha$, H$\beta$, [NII]6584, and [OIII]5007 emission lines, and of the underlying 
continuum flux for the H$\alpha$ emission line. Using these data we calculate the H$\alpha$ equivalent width, and the Balmer Decrement. We 
use the H$\alpha$ equivalent width as an independent observable in the investigation of possible biases in the morphological proxies for spiral 
galaxies and the Balmer Decrement in the correction of observed UV photometry for the effects of attenuation due to dust using the prescription 
of \citet{CALZETTI2000} (cf. section~\ref{application}). The ratios of H$\alpha$ to [NII]6584 and H$\beta$ to [OIII]5007 are used to identify 
galaxies hosting an AGN following the prescription of \citet{KEWLEY2006}.  
The emission line data is taken from the MPA/JHU analysis of the SDSS DR7 spectra\footnote{The data and catalogues are available
from http://www.mpa-garching.mpg.de/SDSS/} (performed by Stephane Charlot, Guineverre Kauffmann, Simon White, Tim Heckman, Christy 
Tremonti, and Jarle Brinchmann). We calculate the $H_{\alpha}$ EQWs as the ratio of emission
line to continuum flux. As the listed uncertainties are formal we multiply the uncertainties on the emission line fluxes by the factors listed
on the website, in particular by 2.473 for H$\alpha$, 2.039 for [NII]6584, 1.882 for H$\beta$, and 1.566 for [OIII]5007. These factors have 
been determined by the MPA/JHU group using comparisons of duplicate spectra of objects within the sample. 
For sources with $S/N < 3$ we use three times the uncertainty as an upper limit. For details on the data and catalogues we refer the reader to 
the MPA/JHU website.

\subsection{Single S\'ersic Profile Fits}\label{data_sers}
In constructing the parameter combinations for use as proxies,  
we have made use of the structural information supplied by the simultaneous fits in the $g$- and $r$-band of single S\'ersic profiles to SDSS 
photometry made available by \citet{SIMARD2011}, performed 
using \verb GIM2D. In particular, we have used the S\'ersic index $n$, the single S\'ersic effective radius $r_e$ (half-light semi-major axis) in 
the $r$-band, and the ellipticity $e$. \citet{SIMARD2011} find that multiple component fits are not justified for most SDSS sources given the 
resolution of the imaging, and similar issues will afflict other surveys as well. Therefore we have chosen to use the largely robust single S\'ersic 
profile fits in this work. We note, however, that \citet{BERNARDI2012} have recently argued that for the brightest sources two component fits are 
preferable over single S\'ersic fits and that for these sources the sizes derived by \citet{SIMARD2011} are systematically too small. This will not 
affect the analysis presented here, as these sources form a minority of the population considered and the effect will be accounted for in the 
calibration of the proxies.\newline

\subsection{GALAXY ZOO DR1}\label{data_gz}
The GALAXY ZOO DR1 \citep{LINTOTT2008,BAMFORD2009,LINTOTT2011} represents the largest and faintest sample of galaxies with 
morphological classifications based on visual inspection.
We have employed these morphological classifications,
specifically those of the sources with redshift debiased classifications as provided by \citet{BAMFORD2009}, as a benchmark 
morphological classification. Such a debiased estimate is only possible for sources with spectroscopic redshifts. 
Rather than a binary classification, GALAXY ZOO DR1 provides a probability for 
the source being an elliptical ($P_{\mathrm{E,DB}}$) or a spiral ($P_{\mathrm{CS,DB}}$ ($\mathrm{CS}$ denotes the combined spiral class, i.e 
summed over the sub-classes available in GALAXY ZOO DR1, i.e clockwise spiral, anti-clockwise spiral, spiral edge-on/other), based on the 
outcome of all classifications of the object\footnote{It should be noted that due to the debiasing procedure, $P_{CS,DB} + P_{E,DB}$ for a given 
galaxy is not necessarily equal to unity.}. 
It is then up to the user to decide where
to place the threshold for assuming a classification is reliable. After eyeballing a selection of galaxies 
we have chosen to treat a debiased probability of 0.7 or
greater as being a reliable classification in the context of this work. Such a choice results in three populations: i) spirals, ii) ellipticals, and iii) 
undefined. We will show that this choice leads to highly pure samples of spirals.\newline

\subsection{The Sample of \citet{NAIR2010}}
\citet{NAIR2010} have provided detailed visual morphological classifications of 14,034 galaxies in the SDSS DR4 
\citep{ADELMAN-MCCARTHY2006}, with $0.01 \le z \le 0.1$ and $g' < 16$mag. They provide T-types for each source as follows:
(c0, E0, E+): -5; (S0-): -3; (S0, S0+): -2; (S0/a): 0; (Sa): 1; (Sa/b): 2, (Sb): 3, (Sb/c): 4; (Sc): 5; (Sc/d): 6; (Sd): 7; (Sdm): 8; (Sm): 9; (Im): 10; 
(unknown/?): 99. In the context of this work we have treated the T-types 1-10 as late-types/spirals. We note, however, that this sample does 
not extend to the depth of the GALAXY ZOO sample, and that, in spite of its size, independent visual classifications are only available for $\sim 
6000$ of the galaxies in our sample. As such the population of faint and/or marginally resolved galaxies which dominate the source counts of 
current wide-field blind optical surveys, is only marginally sampled.\newline

\section{A Non-Parametric, Cell-based Classification Scheme}\label{method}
In order to obtain reliable morphological selections of galaxies based upon photometric parameters, the parameter chosen must ideally display
a distinct separation into two populations corresponding to the different morphological categories. Prominent examples of such one 
parameter separation criteria are the concentration index $C_{idx} = R_{90}/R_{50}$ \citep[e.g.,][]{STRATEVA2001} and the S\'ersic index $n$ 
\citep[e.g.,][]{BLANTON2003}.\newline
Other schemes make use of combinations of two or more parameters such as the $u-r$ colour and $r$-band absolute magnitude 
\citep{BALDRY2004}, or the 
$q_{exp}$ and $f_{deV}$ parameters, possibly in combination with $u-r$ colour information \citep{TEMPEL2011}. Recently, \citet{KELVIN2012} 
and \citet{DRIVER2012} have suggested the use of a UV/optical colour ($u-r$, resp. $NUV-r$) and the S\'ersic index $n$ in 
separating spiral and elliptical galaxies, and a variant of the $NUV-r$, $n$ selection has been used by \citet{GROOTES2013} to select spiral 
galaxies for the purpose of a radiation transfer analysis and has proven 
to be efficient.\newline
Common to all these approaches is the difficulty of selecting a curve/surface of separation between the two populations, which includes as large 
a fraction
of the desired category as possible, whilst simultaneously keeping the level of contamination as low as feasible. 
In addition, this choice may be influenced by further requirements upon the recovery fraction and purity of the sample, 
which can be envisioned to vary with application.\newline
The functional form of the curve or hypersurface providing the optimal separation of the two populations is not known a priori, and an 
appropriate choice can be non-trivial, even if the population of  spiral galaxies is easily separable from the non-spiral population by eye. 
Furthermore, the sharp division between the two is generally not exhibited by the galaxy populations which show a more gradual transition. 
Accordingly, sharp transitions in combination with simple parameterizations where the functional form may be ill-suited can give rise to large 
contaminations.\newline

\subsection{Discretizing Parameter Space}\label{cells}
\begin{figure*}
\ifhires
\includegraphics[width=0.9\textwidth]{CELLs_wCalibgals_v6b.eps}
\else 
\includegraphics[width=0.9\textwidth]{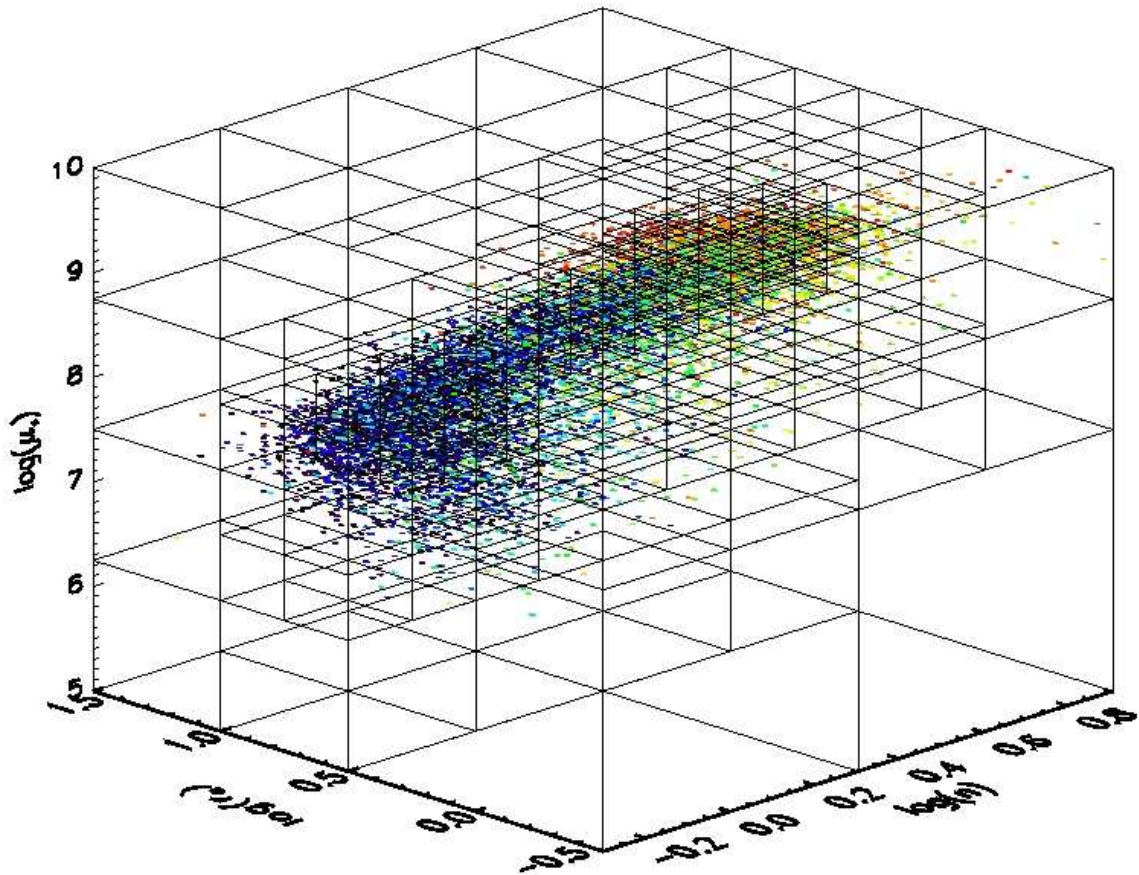}
\fi
\caption{Cell grid obtained for the parameter combination (log($n$),$\mathrm{log}(r_e)$,log($\mu_*$)) using a calibration sample of 10000 
galaxies. The 10k galaxies of the calibration sample are overplotted with colour-coding according to the probability of being a spiral (blue : 
spiral, red: non-spiral).}
\label{fig_cellplot}
\end{figure*}

Rather than making assumptions about the functional form of the separation, we discretize the space
spanned by the parameters used into individual cells. For each cell we can, using the Galaxy Zoo classifications measure the fraction of the 
galaxies residing therein which are spirals (i.e. $P_{CS,DB} \ge 0.7$), and define a subvolume of the total parameter space composed of cells 
with a fraction greater than some desired threshold fraction. This subvolume can then be associated with a population of spiral galaxies.
\newline
As further discussed in sections~\ref{sensitivity} and \ref{parameters} the discretization is performed using a random subsample of 50k 
(respectively 30k for the NUV sample) galaxies. Since the density of galaxies in parameter space is highly non-uniform,
the discretization is performed using an adaptive scheme, with the number of divisions 
along each axis increasing by a power of 2 with each level of refinement. Cells at each level are further refined to a maximum of 3 
refinement steps, i.e., to 16 subdivisions per axis, if they contain more 
than 200 galaxies. This adaptive refinement allows the resolution of the grid to adapt to the density of sources in parameter space, and
ensures that the dividing hypersurface is both well-defined and well-resolved in regions of high and low source density. The value of the 
refinement threshold has little impact on the result of the classification, provided the calibration sample is large enough that sufficient 
refinement is achievable. A high threshold in combination with a small calibration sample will lead to a low level of resolution and a potential 
increase in the level of contamination. Choosing the threshold for refinement at 200 galaxies is found to allow for sufficient resolution, whilst 
maintaining bin populations at such a level that the relative uncertainties of the spiral fraction for the most finely subdivided cells are less than 
0.3 on average. Fig.~\ref{fig_cellplot} shows the resultant grid for a possible combination of three parameters (the grids will differ for different 
parameter combinations).\newline
In each of the cells we calculate the the fraction of spirals $F_{sp}$ as
\begin{equation}
F_{sp} = \frac{N_{\mathrm{GZ,sp}}}{N_{\mathrm{cell}}}\,,
\end{equation}
where $N_{\mathrm{GZ,sp}}$ is the number of GALAXY ZOO spirals (i.e., $P_{\mathrm{CS,DB}} \ge 0.7$) in the cell and $N_{\mathrm{cell}}$ is 
the total number of galaxies in the cell. The associated relative error $\Delta F_{sp,rel}$ is calculated using Poisson statistics and error 
propagation. We then define those cells with $F_{sp} \ge \mathcal{F}_{sp}$ (where $\mathcal{F}_{sp}$ is the threshold spiral fraction) and 
$\Delta F_{sp,rel} \le 1.$ to be spiral cells, i.e., we treat every object in the cell 
as a spiral galaxy, and thus obtain a decomposition of the parameter space into a spiral and a non-spiral subvolume.
The choice of $\Delta F_{sp,rel} \le 1.$ has little effect in terms of the total population, as large values 
of $\Delta F_{sp,rel}$ correspond to scarcely populated cells. The population is obviously more sensitive to the choice of the limiting
fraction $\mathcal{F}_{sp}$, with lower values leading to larger recovery fractions but lower purity. Here we have experimented with different 
values of $\mathcal{F}_{sp}$ and find $\mathcal{F}_{sp} = 0.5$ to result in a very pure, yet nevertheless largely complete, sample of spirals. In 
this work, we continue with the choice $\mathcal{F}_{sp} = 0.5$, however, note that if a larger recovery fraction or an even greater purity is 
desired this choice can be altered.\newline 
In this work we focus on combinations of two and three parameters. While the approach is theoretically applicable to higher dimensional 
parameter spaces, 
the requirements on resolution and cell population impose an effective limit of three dimensions for the calibration sample available. We provide 
a decomposition of the parameter space for three combinations of three parameters in appendix~\ref{appendCell}, which also provide the values 
of $F_{sp}$ and $\Delta F_{sp,rel}$ for all cells. We emphasize that any reader wanting to use the discretizations provided must check 
for systematic differences between his/her data/parameters and those used in this work, and refer the reader to Sect.~\ref{appotsurv} for a 
further discussion of the application of the results presented here to other surveys.\newline

\subsection{Sensitivity to the Calibration Sample}\label{sensitivity}
\begin{figure}
\includegraphics[width=0.45\textwidth]{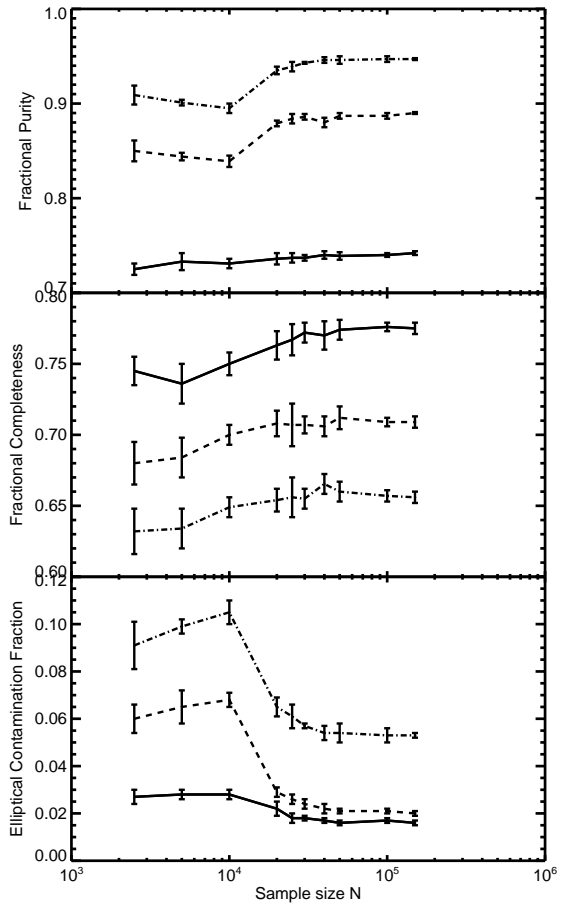}
\caption{Fractional purity (top), fractional completeness (middle), and fractional contamination by ellipticals (bottom) for a selection of spirals 
obtained using the S\'ersic index (i.e. log($n$)), the effective radius in the $r$-band (i.e. log($r_e$)), and the stellar mass surface density (i.e. 
log($\mu_*$)), as a function of the size of the calibration sample. The solid line corresponds to the results obtained when classifying the optical 
sample (i.e without the requirement of an NUV detection), while the dash-dotted line corresponds to the results obtained when classifying the 
optical sample with morphological classifications by \citet{NAIR2010} defining spirals using these detailed classifications, and the dashed line 
corresponds to the optical sample matched to the \citet{NAIR2010} catalogue but using the GALAXY ZOO visual classifications. The data points 
correspond to the mean of 5 random realizations of the calibration sample drawn form the optical galaxy sample with the error bars 
corresponding to the 1-$\sigma$ standard deviation about the mean.}
\label{fig_SAMPSIZE}
\end{figure}

In order to provide a robust and reliable decomposition of the parameter space, the calibration sample must adequately sample the parameter 
space and the galaxy population, i.e. it must contain sufficient galaxies to achieve the required level of resolution and to sufficiently populate 
the individual cells, as well as be representative of the galaxy population as a whole. On the other hand, as the calibration sample must be 
visually classified, it is desirable to understand how the performance of the method relies on the size of the calibration sample. In particular, it is 
of interest how the purity, completeness, and contamination by ellipticals of the sample depend on the size of the calibration sample.\newline
We define the purity fraction $P_{\mathrm{pure}}$ as 
\begin{equation}
P_{\mathrm{pure}} = \frac{N_{\mathrm{sel,SP}}}{N_{\mathrm{sel}}} \;,
\label{eq_Ppure}
\end{equation}
where $N_{\mathrm{sel}}$ is the number of galaxies selected as spirals by the cell-based method, and $N_{\mathrm{sel,SP}}$ is the number of 
those galaxies which are visually classified as being spiral galaxies. Analogously the contamination fraction $P_{\mathrm{cont}}$ is defined as 
the fraction of the selected galaxies which are visually classified as ellipticals, i.e.
\begin{equation}
P_{\mathrm{cont}} = \frac{N_{\mathrm{sel,E}}}{N_{\mathrm{sel}}} \;.
\label{eq_Pcont}
\end{equation}
The completeness fraction of the sample $P_{\mathrm{comp}}$ is defined as
\begin{equation}
P_{\mathrm{comp}} = \frac{N_{\mathrm{sel,SP}}}{N_{\mathrm{SP}}} \;,
\label{eq_Pcomp}
\end{equation}
where $N_{\mathrm{SP}}$ is the total number of visually classified spirals in the sample being classified by the cell-based method.\newline
Fig.~\ref{fig_SAMPSIZE} shows the fractional purity, completeness, and contamination by elliptical galaxies for samples selected using a 
combination of the parameters  S\'ersic index (log($n$)), effective radius in the $r$-band (log($r_e$)), and stellar mass surface density 
(log($\mu_*$)), as a function of the size of the calibration sample (this parameter combination is found to perform well in selecting 
simultaneously pure and complete samples of spirals; for further details on the parameters, the parameter combinations, and their performance 
we refer the reader to Sect.~\ref{parameters}). The values at each sample size correspond to the mean obtained from 5 random realizations of a 
calibration sample of that size, with the error bars corresponding to the 1-$\sigma$ standard deviation. In each case, the calibration sample is 
drawn from the whole of the GALAXY ZOO sample.\newline 
The figure shows the performance in classifying three test samples: i) the entire optical galaxy sample using the visual classifications of spirals 
provided by GALAXY ZOO (solid), ii) the optical galaxy sample with independent morphological classifications provided by \citet{NAIR2010} 
making use of these to define which galaxies really are spirals (dash-dotted), and iii) the optical galaxy sample with morphological 
classifications provided by \citet{NAIR2010}, but making use of the visual classifications provided by GALAXY ZOO (dashed). When calculating 
the contamination by ellipticals for GALAXY ZOO-based definitions we assume all sources with $P_{\mathrm{E,DB}} \ge 0.5$ to be ellipticals. For 
each of the test samples contamination decreases while the completeness and purity increase markedly with increasing size of the calibration 
sample. However, calibration sample sizes greater than $\sim50$k galaxies no longer lead to a large improvement of the performance. The 
improvement in performance with increasing size of the calibration sample is particularly striking for the optical sample matched to the bright 
galaxy sample of \citet{NAIR2010}. The increasing sample size enables a higher resolution, thus increasing purity and decreasing contamination 
by allowing regions of parameter space to be excluded, while simultaneously allowing the full extent of the parameter space occupied by spiral 
galaxies to be sufficiently sampled, increasing completeness by including other sections of the parameter space.\newline 
Even for the smallest sample sizes the performance of the method does not appear to depend strongly on the specific realization of the 
calibration sample, as shown by the errorbars. However, there is nevertheless a notable decrease in the 1-$\sigma$ uncertainty around the 
mean with increasing sample size from $\sim 1-1.5$\% to $\lesssim 0.5$\%, i.e calibration with a larger sample leads to a more robust and 
reliable discretization.\newline 
In light of these results, we have chosen a calibration sample of 50k galaxies for discretizations of the parameter space for the optical sample 
(i.e., without the requirement of an NUV detection), and a subsample of 30k of these galaxies for the discretizations of the parameter space  for 
the NUV sample (i.e., with the requirement of an NUV detection). This allows the rest of the sample to be used as a semi-independent test 
population with which to investigate the performance of a given parameter combination. As we desire the method to be applicable over the full 
redshift range considered $z\le 0.13$ we randomly select the calibration sample from this redshift range.\newline

\section{Parameter combinations}\label{parameters}
In the context of this work we focus on a suite of directly observed and derived parameters for the purpose of identifying spiral galaxies which
consists of a UV/optical colour ($u-r$, respectively $NUV-r$ for the NUV 
matched sample), the S\'ersic index $n$, the effective radius $r_e$ (half-light semi-major axis), the $i$-band absolute magnitude, 
the ellipticity $e$, the stellar mass $M_*$, and the stellar mass surface density $\mu_*$ calculated as
\begin{equation}
\mu_* = \frac{M_*}{2\pi r_{e}^2}\;.
\label{eq_mustar}
\end{equation}
 
The usefulness of the $u-r$ colour and the S\'ersic index in selecting spirals is well documented (e.g., \citealt{BALDRY2004} respectively 
\citealt{BARDEN2005}). Similarly, as spiral galaxies are often assumed to be
largely star-forming, the $NUV-r$ colour may be assumed to be of use. We have chosen to include the $i$-band
magnitude $M_i$ (a directly observable tracer of stellar mass) and the derived parameter stellar mass $M_*$, as early-type galaxies are, on 
average, more massive than 
late-types. Furthermore, at a given stellar mass, it appears likely that a rotationally-supported spiral will be more radially extended than a 
pressure-supported early-type galaxy, hence we make
use of the effective radius. This also implies that the stellar mass surface density of sources may be useful in separating spirals from
non-spirals. While for a spiral the value of $\mu_*$ derived using Eq.~\ref{eq_mustar} is readily interpretable in a physical sense \footnote{As a 
spiral galaxy can be assumed to be circular to first order, the effective radius can be used to derive a reasonable estimate of the surface area and 
consequently of the stellar mass surface density.}, the value derived in this manner for a true ellipsoid will tend to underestimate
the actual surface density of the object, as the approximation of the surface area using $r_e$ as in Eq.~\ref{eq_mustar} will tend to overestimate 
the projected surface area. Hence, any observed separation of the spiral and non-spiral populations in this parameter will represent a lower limit 
to the actual separation.
Finally we have included the observed ellipticity $e$, as the objects on the sky which appear most elliptical are likely to be spirals observed at a 
more edge-on orientation. We note, however, that the use of ellipticity as a parameter will bias any selection of spirals towards sources seen 
edge-on.\newline
Our goal is to identify (multiple) optimal sets of parameters which can be used as morphological proxies in the selection of highly pure and 
largely complete samples of spiral galaxies. As NUV data is
only available for a subset of the total sample we perform the investigations in parallel both for the \textit{OPTICALsample}, as well as the 
\textit{NUVsample}.\newline 
For the \textit{OPTICALsample} we perform the discretization of the parameter space using a sample of 50k galaxies randomly drawn from the 
\textit{OPTICALsample} (the same sample is used for all parameter combinations) and classify the performance using the 
\textit{OPTICALsample} and the \textit{NAIRsample} (i.e. the  subsample with morphological classifications from\citet{NAIR2010}). For the NUV 
preselected sample (the \textit{NUVsample}) we perform the discretizations using a sample of 30k galaxies with NUV detections (randomly 
sampled from the sample of 50k galaxies used for the \textit{OPTICALsample}), and in this case classify the performance using the entire 
\textit{NUVsample}, and the \textit{NUVNAIRsample} (i.e., the subsample of galaxies with morphological classifications from \citet{NAIR2010} 
and NUV detections.) \newline

\begin{figure*}
\begin{center}
\includegraphics[width=0.9\textwidth]{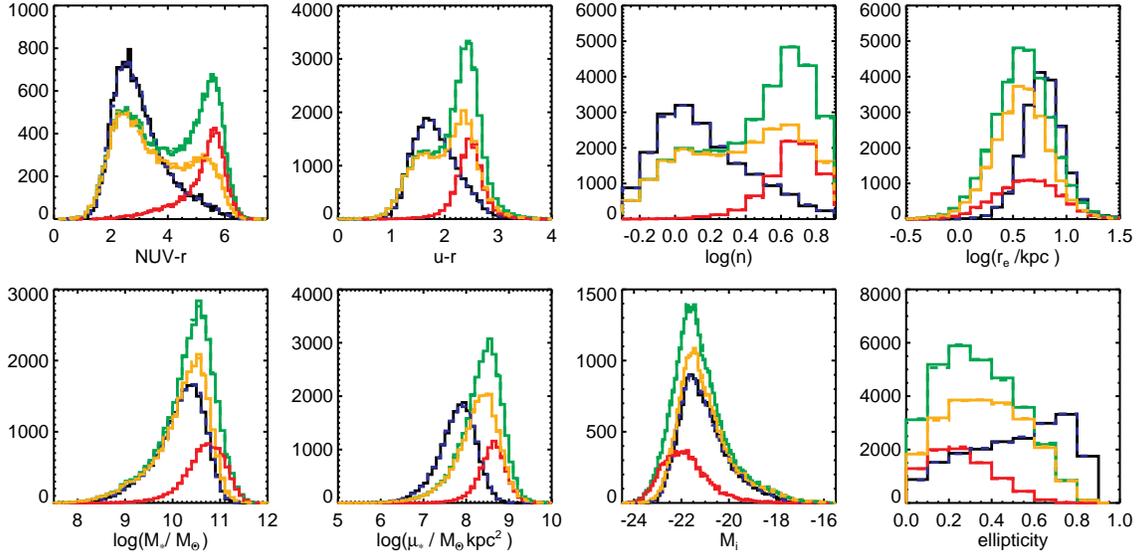}
\caption{Distribution of the parameters in the entire \textit{OPTICALsample} (dashed) and the calibration sample as defined in 
Sect.~\ref{sensitivity} for the population of spirals (blue),
ellipticals (red), non-spirals (green), and undefined (orange). The distributions of the whole sample and the calibration subsample are nearly 
indistinguishable as differences are smaller than the line width.}
\label{fig_pardistopt}
\end{center}
\end{figure*}

\begin{figure*}
\begin{center}
\includegraphics[width=0.9\textwidth]{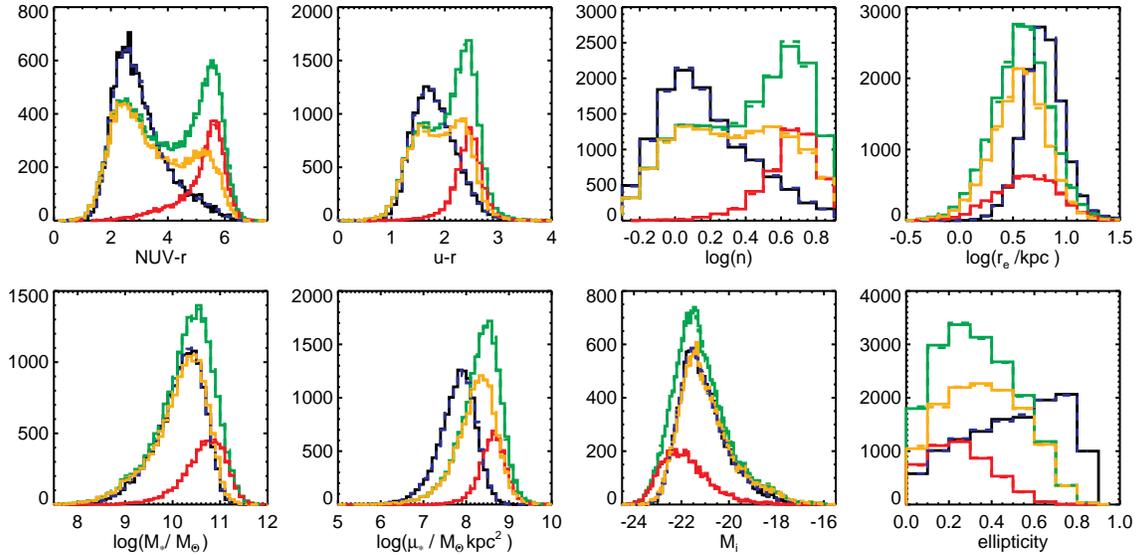}
\caption{Distribution of the parameters in \textit{NUVsample} (dashed) and NUV the calibration sample as defined in 
Sect.~\ref{sensitivity} (solid) for the population of spirals (blue),
ellipticals (red), non-spirals (green), and undefined (orange). The distributions of the whole sample and the calibration sample are nearly 
indistinguishable.}
\label{fig_pardistuv}
\end{center}
\end{figure*}

Fig.~\ref{fig_pardistopt} shows the distributions of the parameters for the entire \textit{OPTICALsample} (dashed), as well as for the 
randomly selected subset of 50k galaxies in the calibration subsample (solid). As expected, the distributions for the two samples are so similar 
as to be indistinguishable in Fig.~\ref{fig_pardistopt} with the differences being smaller than the line width\footnote{ This is quantitatively 
supported by the fact that Kolmogorov-Smirnoff tests (and two sample $\chi^2$-tests for similarity for the discrete distributions in $e$,$n$, 
and $r_e$) support the null hypothesis that the samples have the same distribution ($p\ge 0.49$).} The figure shows the distributions for the 
galaxies in the samples classified as spirals ($P_{CS,DB} \ge 0.7$, blue), 
ellipticals ($P_{E,DB} \ge 0.7$, red), non-spirals ($P_{CS,DB} < 0.7$, green), 
and undefined ( $P_{CS,DB} < 0.7$ and $P_{E,DB} < 0.7$, orange) using GALAXY ZOO.\newline
As expected, the spiral and elliptical populations are
reasonably separated in terms of UV/optical colour and S\'ersic index. However, the overlap between the spiral and undefined populations is 
nevertheless large for these parameters. Furthermore, the distribution of $\mu_*$ notably also displays a distinct separation of the two 
populations, and even shows a separation between the spiral and undefined populations. The parameters stellar mass, effective radius, and $i$-
band absolute magnitude show the expected trends in the populations as previously discussed. The distribution of ellipticities, however, is 
noteworthy. As expected, the spiral sample dominates the largest values of ellipticity and displays a separation from the undefined population at 
high ellipticity. However, at intermediate and lower values of $e$ there is considerable overlap with the other populations. Furthermore, the 
population of spirals as defined by GALAXY ZOO appears biased towards high values of ellipticity, i.e. galaxies seen edge-on\footnote{For an 
unbiased sample one would expect a flat distribution in ellipticity}. As a consequence a discretization of parameter space using this calibration 
sample and $e$ in the parameter combination will also be biased towards high values of ellipticity (even more so, than due to the intrinsic 
overlap of the spiral and non-spiral sample at low and intermediate values of $e$). However, the bias will not affect the discretization of the 
parameter space for combinations of parameters which are, to first order, independent of the orientations of the galaxies with respect to the 
observer (e.g. log($r_e$), log($M_*$), log($\mu_*$), $M_i$, log($n$)\footnote{A bias in ellipticity can potentially give rise to a slight bias 
towards redder UV/optical colours, as edge-on spirals appear redder on average. However, we have found no significant evidence of such a bias. 
Recent work by \citet{PASTRAV2013a} has also found that fully resolved dust rich galaxies seen edge-on may appear larger than when seen 
face-on, however, the strength of this effect remains to be quantified for marginally resolved sources.}. In such cases, the distribution of 
ellipticities of spiral galaxies in each of the cells may be expected to be similar to that of the entire calibration sample, hence the bias towards 
edge-on systems will have no effect.\newline
The bias of the GALAXY ZOO spiral sample must also be taken into account when quantifying the performance of different combinations of 
parameters. When using samples relying on the GALAXY ZOO classifications as test samples, the bias in $e$ can give rise to spuriously complete 
samples in combination with $e$ as a selection parameter. In spite of this bias, we nevertheless choose to use the GALAXY ZOO sample for 
calibration and testing purposes, as it represents the only large and faint sample of visually classified galaxies with a wide range of 
homogeneous ancillary data available. We check for effects arising from the ellipticity bias using the bright subsample of galaxies with 
independent visual classifications by \citet{NAIR2010}, which does not display an ellipticity bias. \newline
Fig.~\ref{fig_pardistuv} shows the same for the parameter distributions of the \textit{NUVsample} and the randomly selected subset of 
30k galaxies constituting the NUV calibration sample.\newline
Comparing the parameter distributions between the \textit{OPTICALsample} and the \textit{NUVsample} shown in 
Figs.~\ref{fig_pardistopt} \& \ref{fig_pardistuv}, the samples appear remarkably similar. Nevertheless, Kolomogorov-Smirnoff and $\chi^2$ 
tests indicate that, in spite of their similar appearance, the null hypothesis that the parameter distributions in these samples are the same has 
low probability ($p\le 0.03$).
However, if one considers only the subsamples of spirals and ellipticals, the tests find no statistically significant difference in the parameter 
distributions for the \textit{OPTICALsample} and the \textit{NUVsample} ($p\ge 0.37$), with the exception of the $u-r$ and $NUV-r$ colours 
($p\le 8\cdot 10^{-4}$), indicating that the NUV pre-selection mainly affects the undefined population and its size relative to the spiral and 
elliptical populations. Despite of these differences, overall, the use of UV-preselection only has a small effect on the parameter distributions, in 
comparison with the large shift in the distributions between the morphological categories. This qualitative impression is confirmed for the 
optical properties of spirals and ellipticals, the null hypothesis being supported with $p\ge0.37$. A might be expected, the null hypothesis is, 
however, rejected for the $NUV-r$ and $u-r$ colors ($p\le 8\cdot 10^{-4}$). The NUV pre-selection also appears to affect the undefined 
population and its size relative to the spiral and elliptical distributions, even in the optical parameters, the null hypothesis being rejected for this 
class for all parameters.\footnote{This statement is valid for the combination of UV and optical photometric depths in the dataset used in this 
work. We caution that for different datasets this may not necessarily be true.}\newline 

Our goal in this work is to identify parameter combinations which provide a pure, but also largely complete sample of spiral galaxies. As such an 
additional important figure of merit in quantifying the performance of the different parameter combinations is the bijective discrimination power 
$P_{\mathrm{bij}}$ which we define as the product of $P_{\mathrm{pure}}$ and $P_{\mathrm{comp}}$ as defined in Eqs.~\ref{eq_Ppure}, 
\ref{eq_Pcomp}, i.e.
\begin{equation}
P_{\mathrm{bij}} = P_{\mathrm{pure}}\cdot P_{\mathrm{comp}}\,.
\label{eq_Pbij}
\end{equation}
This provides a measure of the efficacy
 of the parameter combination at simultaneously selecting a pure and complete sample of spirals from the test samples. $P_{\mathrm{bij}}$ can 
 take on values between 0 and 1, with 1 corresponding to a perfectly pure and complete sample. As a reference, a
selected sample with $P_{\mathrm{pure}} = 0.75$ and $P_{\mathrm{comp}} = 0.7$ (good values of completeness and purity) would
have $P_{\mathrm{bij}} = 0.525$. Applying this metric to the GALAXY ZOO classifications as used in this work of the \citet{NAIR2010} 
sample, one finds that the GALAXY ZOO classifications attain $P_{\mathrm{pure}} = 0.984$ and $P_{\mathrm{comp}} = 0.732$, resulting in 
$P_{\mathrm{bij}} = 0.720$ for the \citet{NAIR2010} sample of bright galaxies.\newline
In the case of test samples using the visual classifications provided by GALAXY ZOO, the purity refers to the subsample of reliable spirals (i.e. 
with $P_{\mathrm{CS,DB}} \ge 0.7$). However, not all galaxies which do not fulfill this criterion will be ellipticals. Rather, a fraction may be 
spirals with a less certain classification. In order to quantify the extent to which the sample is contaminated by ellipticals we also provide the 
value of $P_{\mathrm{cont}}$ as defined in Eq.~\ref{eq_Pcont}, where we define all sources with $P_{\mathrm{E,DB}} \ge 0.5$ to be ellipticals.
\newline

\subsection{Application to optical samples}\label{WHOLESAMP}
In the following we investigate the performance of selections using parameters which can be applied to samples without the requirement of UV 
data, i.e. $u-r$ colour, log($n$), log($r_e$), log($M_*$), log($\mu_*$), $M_i$, and $e$. The figures of merit involving completeness 
$P_{\mathrm{comp}}$ and $P_{\mathrm{bij}}$ are
given in relation to the \textit{OPTICALsample} and the \textit{NAIRsample}.\newline

\subsubsection{Two parameter cells applied to optical samples}\label{WHOLESAMP_2P}
Tables~\ref{tab_2PurGZ} and \ref{tab_2PurNA} show the figures of merit achieved when testing using the \textit{OPTICALsample} and the 
\textit{NAIRsample}, respectively, for all 21 unique combinations of two parameters drawn from the suite applicable to optical samples.\newline

Testing the performance of different parameter combinations using the \textit{OPTICALsample}, we find that  
the parameters $\mathrm{log}(\mu_*)$ and $\mathrm{log}(r_e)$ are efficient at selecting complete samples, with all samples with 
$P_{\mathrm{comp}} \ge 0.7$ involving combinations including at least one of these parameters. These parameters also perform well in 
selecting pure samples, as most combinations involving them attain values of $P_{\mathrm{pure}} > 0.7$. In concert with either $\mathrm{log}
(\mu_*)$ or $\mathrm{log}(r_e)$, the parameter $\mathrm{log}(n)$ also leads to pure and complete samples of spirals (in particular 
(log($n$),log($r_e$)) attains the highest value of $P_{\mathrm{bij}} = 0.529$). Using $e$ in parameter combinations leads to selections which 
are highly pure on average ($P_{\mathrm{pure}} \gtrsim 0.71$), but have comparably low values of completeness ($P_{\mathrm{comp}} < 0.6$), 
and accordingly have low bijective discrimination power. A notable exception to this is the combination (log($\mu_*$),$e$) with 
$P_{\mathrm{pure}} = 0.710$, $P_{\mathrm{comp}} = 0.744$, and $P_{\mathrm{bij}} = 0.528$, the second highest value of $P_{\mathrm{bij}}$ 
overall. However, this may be influenced by the ellipticity bias in the test sample (see the previous discussion in Sect.~\ref{parameters}).\newline
Interestingly, use of the $u-r$ colour does not of itself lead to very pure samples, as the purity of, e.g.,
 the combinations ($u-r$,log($M_*$)) and ($u-r$,$M_i$) is only $\sim 0.6$, while similar combinations (e.g., (log($r_e$), log($M_*$)) attain 
 much greater values.  In addition, the completeness attained by using the $u-r$ colour is strongly dependent upon the second parameter used. 
 If the second parameter is more bimodal, e.g. log($\mu_*$), the combination provides good purity and completeness, while the completeness 
 drops for parameters with less separation of the populations (e.g. $M_i$). Similarly, the S\'ersic index is less efficient than expected, as the 
 bijective discrimination power of the combinations of log($n$) with
log($M_*$) and $M_i$ (but also $u-r$), is low compared to that attained in combination with log($r_e$) and log($\mu_*$). Overall, the 
combination (log($n$),log($r_e$)) has the greatest bijective discrimination power ($P_{\mathrm{bij}} = 0.529$) closely followed by the 
combination (log($\mu_*$),$e$) with ($P_{\mathrm{bij}} = 0.528$) and the combinations (log($r_e$),log($M_*$), (log($r_e$),log($\mu_*$)), 
and (log($n$),log($\mu_*$)) all with $P_{\mathrm{bij}} \approx 0.5$. Amongst these combinations (log($n$),log($r_e$)) and 
(log($n$),log($\mu_*$)) have the lowest values of contamination by ellipticals with $P_{\mathrm{cont}} \le 0.032$, i.e. the lowest values 
attained by any parameter combination.\newline

Table~\ref{tab_2PurNA} shows the values for the figures of merit obtained when testing using the \textit{NAIRsample}, using both the 
independent morphological classifications of \citet{NAIR2010} and the GALAXY ZOO visual classifications.\newline
Overall, the purity of the selections obtained when testing the parameter combinations using the \textit{NAIRsample} with GALAXY ZOO visual 
classifications is greater than for the \textit{OPTICALsample} with values of $P_{\mathrm{pure}} \sim 0.8-0.9$, indicating, that some of the 
'impurities' in the selections from the \textit{OPTICALsample} are very likely unreliably classified spirals. On the other hand, the fractional 
completeness of the selections is of order $0.05-0.1$ less than for the \textit{OPTICALsample}. An exception to this are the combinations 
including $e$, for which the fractional completeness is $\sim 0.2$ less. This stronger decrease in completeness reflects the bias towards large 
values of $e$ in the \textit{OPTICALsample} which is not present in the \textit{NAIRsample}.  As for the \textit{OPTICALsample}, the parameter 
combination with the greatest bijective discrimination power is (log($n$,log($r_e$)). Unlike for the \textit{OPTICALsample}, however, the 
combination with the second largest value of $P_{\mathrm{bij}}$ is (log($n$),log($\mu_*$)), which also attains the lowest value of 
contamination by ellipticals, rather than (log($\mu_*$),$e$) (likely due to the removal of the ellipticity bias as previously discussed). As for the 
\textit{OPTICALsample} the 5 combinations with the highest values of $P_{\mathrm{bij}}$ ((log($n$),log($r_e$)), (log($n$),log($\mu_*$), ($u-
r$,log($\mu_*$)), (log($r_e$),log($M_*$)), (log($\mu_*$),$M_i$)) all include either log($r_e$) or log($\mu$). Furthermore, log($n$) again leads 
to very pure and complete selections in combination with log($r_e$) or log($\mu_*$). In addition, its efficiency in combination with other 
parameters is also increased (e.g., (log($n$),$M_i$)).\newline 
Testing using the \textit{NAIRsample} with the independent classifications of \citet{NAIR2010} leads to very similar results. However, the 
fractional purity of the selections is even larger, further underscoring the conclusion that a large contribution to the 'impurity' of the selections is 
due to unreliably classified spirals. 
which also has amongst the lowest contamination by ellipticals. 
The combinations with the highest bijective discrimination power again include either log($r_e$), log($\mu_*$), and/or log($n$), supporting the 
previous findings.\newline 
Overall, the parameters log($\mu_*$), log($r_e$), and log($n$) appear to be most efficient at selecting pure and complete samples of spirals.
\newline

\begin{table}
 \centering
 \caption{$N_{\mathrm{sel}}$,$P_{\mathrm{pure}}$, $P_{\mathrm{comp}}$, $P_{\mathrm{bij}}$, $P_{\mathrm{cont}}$ for combinations of two 
 parameters applied to the \textit{OPTICALsample}. }
  \begin{tabular}{@{}lccccc@{}}
  \hline
 Parameter combination & $N_{sel}$& $P_{\mathrm{pure}}$ & $P_{\mathrm{comp}}$ & $P_{bij}$ & $P_{\mathrm{cont}}$ \\ 
 \hline
 ($u-r$,log($n$)) & 67436 & 0.617 & 0.655 & 0.404 & 0.060\\ 
 ($u-r$,log($r_e$)) &57168 & 0.710 & 0.639 & 0.453 & 0.054  \\
($u-r$,log($M_*$)) &  63194 & 0.580 & 0.577 & 0.334 & 0.084\\
($u-r$,log($\mu_*$)) & 65254 & 0.690 & 0.709 & 0.489 & 0.054\\ 
 ($u-r$,$M_i$) &61275 & 0.584 & 0.563 & 0.329 & 0.079  \\
($u-r$,$e$) & 47567 & 0.719 & 0.538 & 0.387 & 0.042 \\
(log($n$),log($r_e$)) & 64179 & 0.724 & 0.731 & 0.529 & 0.032 \\ 
 (log($n$),log($M_*$) &67304 & 0.623 & 0.660 & 0.412 & 0.055  \\
(log($n$),log($\mu_*$) & 67026 & 0.688 & 0.726 & 0.499 & 0.027 \\ 
(log($n$),$M_i$ & 71707 & 0.615 & 0.694 & 0.427 & 0.055 \\
 (log($n$),$e$) &55547 & 0.685 & 0.599 & 0.410 & 0.038  \\
(log($r_e$),log($M_*$)) & 63985 & 0.711 & 0.716 & 0.509 & 0.048  \\
(log($r_e$),log($\mu_*$)) & 61678 & 0.721 & 0.700 & 0.504 & 0.048 \\ 
(log($r_e$),$M_i$) & 61263 & 0.699 & 0.674 & 0.471 & 0.071 \\
 (log($r_e$),$e$) &44938 & 0.760 & 0.538 & 0.409 & 0.051  \\
 (log($M_*$),log($\mu_*$)) &60231 & 0.724 & 0.686 & 0.496 & 0.040 \\ 
(log($M_*$),$M_i$) &  45243 & 0.578 & 0.412 & 0.238 & 0.069  \\
(log($M_*$),$e$) & 34862 & 0.737 & 0.405 & 0.298 & 0.062  \\
 (log($\mu_*$),$M_i$) &65086 & 0.697 & 0.714 & 0.497 & 0.049  \\
(log($\mu_*$),$e$) & 66627 & 0.710 & 0.744 & 0.528 & 0.035  \\
($M_i$,$e$) & 35006 & 0.730 & 0.402& 0.293 & 0.072  \\
\hline                      
\end{tabular}
  \label{tab_2PurGZ}
\end{table}

\begin{table*}
 \centering
 \caption{$N_{\mathrm{sel}}$,$P_{\mathrm{pure}}$, $P_{\mathrm{comp}}$, $P_{\mathrm{cont}}$, and $P_{\mathrm{bij}}$ for combinations of 
 two parameters applied to \textit{NAIRsample} using the GALAXY ZOO visual classifications (columns 3-6) and the independent classifications 
 of \citet[][columns 7-9]{NAIR2010}. In the case of the independent classifications the contamination fraction is taken to be the complement of 
 the purity (i.e. this includes sources with T-type = 99).  }
  \begin{tabular}{@{}lcccccccc@{}}
  \hline
  \multicolumn{2}{c}{} &\multicolumn{4}{c}{GALAXY ZOO}    &     \multicolumn{3}{c}{Nair \& Abraham (2010)} \\
 Parameter combination & $N_{sel}$& $P_{\mathrm{pure}}$ & $P_{\mathrm{comp}}$ & $P_{bij}$ & $P_{\mathrm{cont}}$ & $P_{\mathrm{pure}}$ 
 & $P_{\mathrm{comp}}$ & $P_{bij}$  \\ 
 \hline
($u-r$, log($n$)) & 2104 & 0.839 & 0.601 & 0.505 & 0.048 & 0.923 & 0.575 & 0.530 \\
 ($u-r$, log($r_e$)) & 1828 & 0.882 & 0.549 & 0.485 & 0.040&  0.9234 & 0.496 & 0.458 \\
($u-r$, log($M_*$)) & 1856 & 0.799 & 0.505 & 0.403 & 0.075&  0.883 & 0.481 & 0.425 \\
($u-r$, log($\mu_*$)) & 2053 & 0.884 & 0.618 & 0.546 & 0.030 & 0.950 & 0.572 & 0.544 \\
 ($u-r$, $M_i$)  &1815 & 0.803 & 0.496 & 0.398 & 0.068&  0.888 & 0.473 & 0.420 \\
($u-r$, $e$) &  1111 & 0.832 & 0.315 & 0.262 & 0.038&  0.926 & 0.302 & 0.280 \\
(log($n$), log($r_e$)) & 2479 & 0.821 & 0.693 & 0.569 & 0.086 &  0.874 & 0.641 & 0.560 \\
 (log($n$), log($M_*$) & 2173 & 0.824 & 0.609 & 0.502 & 0.055&  0.904 & 0.581 & 0.525 \\ 
(log($n$), log($\mu_*$) & 2124 & 0.873 & 0.631 & 0.551 & 0.023& 0.950 & 0.597 & 0.567 \\
(log($n$), $M_i$ &  2382 & 0.811& 0.657 & 0.533 & 0.063& 0.894 & 0.630 & 0.563 \\ 
 (log($n$), $e$) &1435 & 0.833 & 0.407 & 0.339 & 0.033&  0.929 & 0.394 & 0.366 \\
(log($r_e$), log($M_*$)) &  2006 & 0.893 & 0.610 & 0.545 & 0.026 &  0.947 & 0.558 & 0.528 \\
(log($r_e$), log($\mu_*$)) &  1948 & 0.901 & 0.598 & 0.538 & 0.024 &  0.956 & 0.546 & 0.523 \\
(log($r_e$), $M_i$) & 1868 & 0.866 & 0.551 & 0.477 & 0.050&  0.926 & 0.507 & 0.469 \\
 (log($r_e$), $e$) & 1354 & 0.792 & 0.365 & 0.289 & 0.091&  0.854 & 0.339 & 0.290\\
(log($M_*$), log($\mu_*$))& 1858 & 0.906 & 0.573 & 0.519 & 0.021&   0.959 & 0.523 & 0.502 \\
(log($M_*$), $M_i$) &   1351 & 0.827 & 0.380 & 0.314 & 0.057&   0.899 & 0.356 & 0.320 \\
(log($M_*$), $e$) &  798 & 0.786 & 0.213 & 0.168 & 0.056&  0.905 & 0.212 & 0.192 \\
 (log($\mu_*$), $M_i$) &2012 & 0.891 & 0.610 & 0.543 & 0.027&  0.953 & 0.562 & 0.535 \\
(log($\mu_*$), $e$) & 1880 & 0.874 & 0.559 & 0.489 & 0.023 & 0.950 & 0.522 & 0.497 \\
($M_i$, $e$) & 793 & 0.784 & 0.212 & 0.166 & 0.067& 0.898 & 0.209 & 0.187 \\
\hline                      
\end{tabular}
  \label{tab_2PurNA}
\end{table*}

\subsubsection{Three parameter cells applied to optical samples}\label{WHOLESAMP_3P}
While the performance of selections using only two parameters is already encouraging, it seems likely that the purity and completeness, and 
hence the bijective discrimination power, as well as the fractional contamination, can be improved by using more information in the selection, 
i.e. by using a third parameter.\newline
Tables~\ref{tab_3PurGZ} and \ref{tab_3PurNA} show the figures of merit achieved when testing using the \textit{OPTICALsample} and the 
\textit{NAIRsample}, respectively, for all 35 unique combinations of three parameters drawn from the suite applicable to optical samples.
\newline

Testing the performance of different combinations of three parameters using the \textit{OPTICALsample}, we find that both the purity and 
completeness attained are greater, on average, than for combinations of two parameters, as shown in Table~\ref{tab_3PurGZ}. In most cases, 
the use of additional information in the form of a third parameter leads to a simultaneous increase in purity and completeness. In some cases, 
however, the deprojection along the additional third axis can lead to the inclusion of more parameter space, causing an increase of 
completeness at the cost of a decrease in purity or, vice versa, to the exclusion of parameter space, increasing purity at the expense of 
completeness (e.g., (log($r_e$),log($M_*$)) with $P_{\mathrm{pure}} = 0.711$ \&  $P_{\mathrm{comp}} = 0.716$ and  (log($r_e$),log($M_*$),
$M_i$) with $P_{\mathrm{pure}} = 0.707$ \&  $P_{\mathrm{comp}} = 0.739$, respectively (log($n$),$M_i$) with $P_{\mathrm{pure}} = 0.615$ 
\&  $P_{\mathrm{comp}} = 0.694$ and (log($n$),$M_i$,$e$) with $P_{\mathrm{pure}} = 0.708$ \&  $P_{\mathrm{comp}} = 0.641$).\newline
As for the combinations of two parameters, combinations of three parameters including $e$ attain high values of purity (13/15 with 
$P_{\mathrm{pure}} \ge 0.7$, and  6/15 with $P_{\mathrm{pure}} \ge 0.75$). Of these combinations those which include two other parameters 
which efficiently select pure and complete samples of spirals (e.g., log($r_e$) and log($\mu_*$)) also attain very high values of completeness 
($\gtrsim 0.7$), leading to high values of $P_{\mathrm{bij}}$ (of the 10 combinations with the highest values of $P_{\mathrm{bij}}$, the first 6 
include $e$). However, as for the combinations of two parameters, these high values of completeness are partially due to the ellipticity bias of 
the \textit{OPTICALsample}. We will discuss the performance of these combinations on the basis of tests using the \textit{NAIRsample} below. 
However, we note that all six combinations with the highest values of $P_{bij}$ include log($r_e$) and/or log($\mu_*$) . The remaining four 
parameter combinations of the 10 with the highest values of $P_{\mathrm{bij}}$ are (in descending order) (log($n$),log($r_e$),$M_i$) with 
$P_{\mathrm{bij}} = 0.576$, (log($n$),log($r_e$),log($\mu_*$)) with $P_{\mathrm{bij}} = 0.572$, (log($n$),log($M_*$),log($\mu_*$)) with 
$P_{\mathrm{bij}} = 0.565$, and (log($n$),log($r_e$),log($M_*$)) with $P_{\mathrm{bij}} = 0.564$, all of which also include the parameters 
log($r_e$) and/or log($\mu_*$) in addition to log($n$), indicating the potential of these parameters to select pure and complete samples of 
spirals. In addition these four combinations exhibit the lowest contamination by ellipticals with $P_{\mathrm{cont}} \lesssim 0.02$.  As for 
combinations of two parameters, however, log($n$) is only efficient in combination with another efficient parameter. The same is true for the 
parameter $u-r$ colour. Finally, the parameters  $M_i$, and log($M_*$), are efficient in combination with combinations of log($r_e$), 
log($\mu_*$), and log($n$).\newline

Testing the performance of three-parameter combinations using the \textit{NAIRsample} with GALAXY ZOO visual classifications 
(Table~\ref{tab_3PurNA}), we again find again find that the values of $P_{\mathrm{pure}}$ and $P_{\mathrm{comp}}$ are greater than for 
combinations of two parameters. Comparison of the values of purity with those obtained for the \textit{OPTICALsample} also again indicate that 
a fraction of the 'impurity' arises from the unreliable classification of spirals.\newline
Of the 10 combinations with the highest values of $P_{\mathrm{bij}}$ none include $e$, indicating that the high values attained for the 
\textit{OPTICALsample} are, at least partially, due to the ellipticity bias. In descending order, the combinations with the greatest bijective 
discrimination power are (log($n$),log($r_e$),log($\mu_*$)), (log($n$),log($M_*$),log($\mu_*$)) , (log($n$),log($\mu_*$),$M_i$), 
(log($n$),log($r_e$),$M_i$), and (log($n$),log($r_e$),log($M_*$)), supporting the results obtained using the \textit{OPTICALsample}.\newline
Testing using the \textit{NAIRsample} with the independent classifications of \citet{NAIR2010} again leads to very similar results. In terms of 
choice of the most effective parameters, the 5 parameter combinations with the greatest values of  $P_{\mathrm{bij}}$ are the same as found 
when using the GALAXY ZOO visual classifications, although the combination with the overall greatest bijective discrimination power is 
(log($n$),log($\mu_*$),$M_i$) rather than (log($n$),log($r_e$),log($\mu_*$)).\newline

Overall we find that the optimum results in terms of purity and simultaneous completeness for optical samples are obtained by combinations of 
three parameters including log($r_e$), log($\mu_*$), log($n$), and log($M_*$) or $M_i$, notably (log($n$),log($r_e$),log($\mu_*$)), 
(log($n$),log($r_e$),$M_i$), and (log($n$),log($\mu_*$),$M_i$).\newline
  
\begin{table*}
 \centering
 \caption{$N_{\mathrm{sel}}$,$P_{\mathrm{pure}}$, $P_{\mathrm{comp}}$, $P_{\mathrm{bij}}$, and $P_{\mathrm{cont}}$  for combinations of 
 three parameters applied to the \textit{OPTICALsample}.}
  \begin{tabular}{@{}lrrrrrr@{}}
  \hline
Parameter combination & $N_{\mathrm{sel}}$ & $P_{\mathrm{pure}}$ & $P_{\mathrm{comp}}$ & $P_{\mathrm{bij}}$ & $P_{\mathrm{cont}}$\\   
\hline
($u-r$, log($n$), log($r_e$)) & 65154 & 0.724 & 0.743 & 0.539 & 0.024 \\
($u-r$, log($n$), log($M_*$)) & 69906 & 0.625 & 0.688 & 0.430 & 0.058 \\
($u-r$, log($n$), log($\mu_*$)) & 66453 & 0.709 & 0.741 & 0.526 & 0.033\\
($u-r$, log($n$), $M_i$) & 70880 & 0.623 & 0.695 & 0.433 & 0.058  \\
($u-r$, log($n$), $e$) & 60259 & 0.682 & 0.647 & 0.442 & 0.042 \\
($u-r$, log($r_e$), log($M_*$)) & 65727 & 0.713 & 0.737 & 0.525 & 0.038 \\
($u-r$, log($r_e$), log($\mu_*$)) & 63633 & 0.720 & 0.721 & 0.520 & 0.042 \\
($u-r$, log($r_e$), $M_i$) & 67015 & 0.710 & 0.749 & 0.532 & 0.047 \\
($u-r$, log($r_e$), $e$) & 63993 & 0.764 & 0.770 & 0.588 & 0.022\\ 
($u-r$, log($M_*$), log($\mu_*$)) & 62888 & 0.719 & 0.712 & 0.512 & 0.039 \\ 
($u-r$, log($M_*$), $M_i$) & 64714 & 0.582 & 0.593 & 0.345 & 0.082 \\
($u-r$, log($M_*$), $e$) & 56811 & 0.701 & 0.626 & 0.439 & 0.045 \\
($u-r$, log($\mu_*$), $M_i$) & 62289 & 0.720 & 0.706 & 0.508 & 0.037\\ 
($u-r$, log($\mu_*$), $e$) & 66140 & 0.735 & 0.766 & 0.563 & 0.023 \\
($u-r$, $M_i$, $e$) & 56083 & 0.713 & 0.629 & 0.449 & 0.045 \\
(log($n$), log($r_e$), log($M_*$)) & 65708 & 0.738 & 0.764 & 0.564 & 0.018\\  
(log($n$), log($r_e$), log($\mu_*$)) & 66581 & 0.739 & 0.774 & 0.572 & 0.017\\ 
(log($n$), log($r_e$), $M_i$) &  66937 & 0.740 & 0.779 & 0.576 & 0.021 \\
(log($n$), log($r_e$), $e$) & 60988 & 0.776 & 0.745 & 0.577 & 0.019 \\
(log($n$), log($M_*$), log($\mu_*$)) & 67149 & 0.731 & 0.773 & 0.565 & 0.019 \\
(log($n$), log($M_*$), $M_i$) & 68977 & 0.624 & 0.678 & 0.423 & 0.052  \\
(log($n$), log($M_*$), $e$) & 58955 & 0.692 & 0.643 & 0.445 & 0.042 \\
(log($n$), log($\mu_*$), $M_i$) & 68151 & 0.716 & 0.768 & 0.549 & 0.018\\ 
(log($n$), log($\mu_*$), $e$) & 67837 & 0.715 & 0.763 & 0.546 & 0.020 \\
(log($n$), $M_i$, $e$) & 57541 & 0.708 & 0.641 & 0.454 & 0.036 \\
(log($r_e$), log($M_*$), log($\mu_*$)) & 63189 & 0.717& 0.713 & 0.511 & 0.044\\  
(log($r_e$), log($M_*$), $M_i$) & 66491 & 0.706 & 0.739 & 0.521 & 0.052 \\
(log($r_e$), log($M_*$), $e$) &  64608 & 0.754 & 0.767 & 0.579 & 0.027\\
(log($r_e$), log($\mu_*$), $M_i$) & 66374 & 0.707 & 0.739 & 0.523 & 0.055\\ 
(log($r_e$), log($\mu_*$), $e$) & 65079 & 0.759 & 0.777 & 0.590 & 0.026\\ 
(log($r_e$), $M_i$, $e$) & 58887 & 0.753 & 0.698 & 0.525 & 0.038 \\
(log($M_*$), log($\mu_*$), $M_i$) & 63574 & 0.713 & 0.713 & 0.509 & 0.045\\ 
(log($M_*$), log($\mu_*$), $e$) & 65408 & 0.754 & 0.776 & 0.585 & 0.027 \\
(log($M_*$), $M_i$, $e$) &  49084 & 0.686 & 0.530 & 0.363 & 0.061 \\
(log($\mu_*$), $M_i$, $e$) &  66104 & 0.745 & 0.775 & 0.577 & 0.033 \\
\hline
                  
\end{tabular}
  \label{tab_3PurGZ}
\end{table*}

\begin{table*}
 \centering
	\caption{$N_{\mathrm{sel}}$, $P_{\mathrm{pure}}$, $P_{\mathrm{comp}}$, $P_{\mathrm{cont}}$, and $P_{\mathrm{bij}}$ for combinations 
	of three parameters applied to \textit{NAIRsample} using the GALAXY ZOO visual classifications (columns 3-6) and the independent 
	classifications of \citet[][columns 7-9]{NAIR2010}. In the case of the independent classifications the contamination fraction is taken to be the 
	complement of the purity (i.e. this includes sources with T-type = 99).  }
  \begin{tabular}{@{}lcccccccc@{}}
  \hline
  \multicolumn{2}{c}{} &\multicolumn{4}{c}{GALAXY ZOO}    &     \multicolumn{3}{c}{Nair \& Abraham (2010)} \\
 Parameter combination & $N_{sel}$& $P_{\mathrm{pure}}$ & $P_{\mathrm{comp}}$ & $P_{bij}$ & $P_{\mathrm{cont}}$ & $P_{\mathrm{pure}}$ 
 & $P_{\mathrm{comp}}$ & $P_{bij}$  \\ 
 \hline
($u-r$, log($n$), log($r_e$)) &  2339 & 0.867 & 0.690 & 0.598 & 0.041 & 0.925 & 0.640 & 0.592 \\
($u-r$, log($n$), log($M_*$)) & 2280 & 0.829 & 0.643 & 0.533 & 0.053 & 0.910 & 0.614 & 0.559 \\
($u-r$, log($n$), log($\mu_*$)) & 2270 & 0.872 & 0.674 & 0.588 &  0.033 & 0.941 & 0.632 & 0.595 \\
($u-r$, log($n$), $M_i$) & 2353 & 0.826 & 0.662 & 0.546 & 0.052 & 0.909 & 0.633 & 0.576 \\
($u-r$, log($n$), $e$) & 1627 & 0.846 & 0.469 & 0.396 & 0.030 & 0.930 & 0.448 & 0.416 \\
($u-r$, log($r_e$), log($M_*$)) & 2100 & 0.897 & 0.641 & 0.575 & 0.020 & 0.951 & 0.587 & 0.558 \\
($u-r$, log($r_e$), log($\mu_*$)) & 2068 & 0.894 & 0.630 & 0.563 & 0.024 & 0.951 & 0.577 & 0.549 \\
($u-r$, log($r_e$), $M_i$) & 2059 & 0.888 & 0.622 & 0.553 & 0.030 & 0.944 & 0.571 & 0.538 \\
($u-r$, log($r_e$), $e$) & 1872 & 0.888 & 0.566 & 0.502 & 0.017 & 0.947 & 0.521 & 0.493 \\
($u-r$, log($M_*$), log($\mu_*$)) & 1995 & 0.896 & 0.609 & 0.546 & 0.022 & 0.956 & 0.560 & 0.535 \\
($u-r$, log($M_*$), $M_i$) & 2066 & 0.809 & 0.569 & 0.460 & 0.071 & 0.886 & 0.537 & 0.476 \\
($u-r$, log($M_*$), $e$) & 1375 & 0.834 & 0.391 & 0.326 & 0.038 & 0.919 & 0.371 & 0.341 \\
($u-r$, log($\mu_*$), $M_i$) &  1992 & 0.896 & 0.608 & 0.545 & 0.020 & 0.958 & 0.560 & 0.536 \\
($u-r$, log($\mu_*$), $e$) & 1932 & 0.893 & 0.587 & 0.524 & 0.019 & 0.962 & 0.546 & 0.525 \\
($u-r$, $M_i$, $e$) & 1452 & 0.842 & 0.416 & 0.351 & 0.035 & 0.915 & 0.390 & 0.356 \\
(log($n$), log($r_e$), log($M_*$)) & 2319 & 0.881 & 0.696 & 0.613 & 0.024 & 0.941 & 0.646 & 0.608 \\
(log($n$), log($r_e$), log($\mu_*$)) & 2364 & 0.884 & 0.712 & 0.629 & 0.024 & 0.945 & 0.660 & 0.624 \\
(log($n$), log($r_e$), $M_i$) &  2360 & 0.879 & 0.706 & 0.621 & 0.032 & 0.935 & 0.652 & 0.610 \\
(log($n$), log($r_e$), $e$) & 2142 & 0.867 & 0.632 & 0.548 & 0.045 & 0.920 & 0.582 & 0.536 \\
(log($n$), log($M_*$), log($\mu_*$)) & 2347 & 0.885 & 0.707 & 0.626 & 0.024 & 0.946 & 0.657 & 0.621 \\
(log($n$), log($M_*$), $M_i$) & 2283 & 0.833 & 0.647 & 0.539 & 0.049 & 0.908 & 0.613 & 0.557 \\
(log($n$), log($M_*$), $e$) & 1703 & 0.847 & 0.491 & 0.416 & 0.039 & 0.926 & 0.466 & 0.432 \\
(log($n$), log($\mu_*$), $M_i$) & 2363 & 0.881 & 0.709 & 0.625 & 0.020 & 0.950 & 0.664 & 0.631 \\
(log($n$), log($\mu_*$), $e$) & 1989 & 0.873 & 0.591 & 0.516 & 0.019 & 0.953 & 0.560 & 0.534 \\
(log($n$), $M_i$, $e$) & 1686 & 0.856 & 0.492 & 0.421 & 0.035 & 0.921 & 0.459 & 0.422 \\
(log($r_e$), log($M_*$), log($\mu_*$)) &  1983 & 0.901 & 0.608 & 0.548 & 0.023 & 0.955 & 0.556 & 0.531 \\
(log($r_e$), log($M_*$), $M_i$) & 2098 & 0.884 & 0.631 & 0.558 & 0.032 & 0.939 & 0.578 & 0.543 \\
(log($r_e$), log($M_*$), $e$) &  1888 & 0.895 & 0.575 & 0.514 & 0.019 & 0.953 & 0.528 & 0.504 \\
(log($r_e$), log($\mu_*$), $M_i$) & 2091 & 0.885 & 0.630 &  0.557 & 0.035 & 0.940 & 0.577 & 0.542 \\
(log($r_e$), log($\mu_*$), $e$) & 1908 & 0.899 & 0.584 & 0.525 & 0.018 & 0.958 & 0.536 & 0.514 \\
(log($r_e$), $M_i$, $e$) &  1731 & 0.870 & 0.513 & 0.446 & 0.034 & 0.932 & 0.473 & 0.441 \\
(log($M_*$), log($\mu_*$), $M_i$) & 1980 & 0.893 & 0.602 & 0.538 & 0.028 & 0.952 & 0.552 & 0.526 \\
(log($M_*$), log($\mu_*$), $e$) &  1926 & 0.899 & 0.590 & 0.530 &0.017 & 0.958 & 0.541 & 0.518 \\
(log($M_*$), $M_i$, $e$) &   1447 & 0.838 & 0.413 & 0.346 & 0.048 & 0.909 & 0.430 & 0.391 \\
(log($\mu_*$), $M_i$, $e$) &  1922 & 0.900 & 0.589 & 0.530 & 0.017 & 0.957 & 0.539 & 0.516 \\
\hline                      
\end{tabular} 
  \label{tab_3PurNA}
\end{table*}

\subsection{Application to NUV-preselected samples}\label{NUVSAMP}
Spirals are very often found to be systems with on-going star formation, consequently possessed of a younger stellar population emitting in the 
UV (FUV and NUV) and 
displaying blue UV/optical colours. Early-type galaxies on the other hand are generally found to be more quiescent and redder. Where available,
the use of UV properties of sources may thus prove efficient in the selection of spiral galaxies. Similarly, a pre-selection on UV emission will 
enhance the purity of a sample of star-forming spiral galaxies, at the expense of removing UV-faint, quiescent spirals.
In the following we investigate the performance of selections using parameters which can be applied to samples preselected on the availability of 
NUV data (the \textit{NUVsample} and \textit{NUVNAIRsample} in this case), i.e. $NUV-r$ colour, log($n$), log($r_e$), log($M_*$), 
log($\mu_*$), $M_i$, and $e$. 
The figures of merit involving completeness $P_{\mathrm{comp}}$ and $P_{\mathrm{bij}}$ are
given in relation to the NUV preselected samples ($P_{\mathrm{comp,n}}$ and $P_{\mathrm{bij,n}}$) and to the optical samples for comparison 
($P_{\mathrm{comp,o}}$ and $P_{\mathrm{bij,o}}$).

\subsubsection{Two parameter cells applied to the NUV samples}\label{NUVSAMP_2P}
Tables~\ref{tab_2PNUVrGZ} and \ref{tab_2PNUVrNA} show the figures of merit for all 21 unique combinations of two parameters applied to the 
NUV preselected samples.\newline

Testing using the \textit{NUVsample}, the combinations with the greatest values of $P_{\mathrm{bij,n}}$ are (log($\mu_*$),$e$) with 
$P_{\mathrm{bij,n}} = 0.542$ (although the completeness may be influenced by the ellipticity bias), (log($r_e$),log($M_*$)) with 
$P_{\mathrm{bij,n}} = 0.532$, (log($n$),log($r_e$)) with $P_{\mathrm{bij,n}}= 0.529$, (log($r_e$),log($\mu_*$)) with $P_{\mathrm{bij,n}} = 
0.525$, and (log($r_e$),$M_i$) with $P_{\mathrm{bij,n}} = 0.523$. The parameters log($r_e$) and log($\mu_*$) again result in the most 
simultaneously  pure and complete samples, particularly in combination with log($M_*$), $M_i$, or log($n$). In particular log($\mu_*$) leads to 
selections with high purity (4/5 with $P_{\mathrm{pure}} \ge 0.7$ and 2/5  with $P_{\mathrm{pure}} \ge 0.74$. While the $NUV-r$ colour and 
S\'ersic index are less efficient at selecting pure and complete samples than expected, only attaining values of  $P_{\mathrm{pure}} \gtrsim 0.6$ 
in combination with another strongly bimodal parameter, the use of the $NUV-r$ colour does, however, predominantly lead to samples with 
high completeness ($\gtrsim 0.68$),even in combination with log($M_*$) and $M_i$.\newline
  
Making use of the \textit{NUVNAIRsample} with GALAXY ZOO visual classifications we find that the combinations with the greatest bijective 
discrimination power are ($NUV-r$,log($r_e$)) with $P_{\mathrm{bij,n}} = 0.624$, ($NUV-r$,log($M_*$)) with $P_{\mathrm{bij,n}} = 0.612$ 
and ($NUV-r$,$M_i$) with $P_{\mathrm{bij,n}} = 0.608$, followed by (log($n$,log($r_e$))  with $P_{\mathrm{bij,n}} = 0.568$ and 
(log($n$,log($\mu_*$))  with $P_{\mathrm{bij,n}} = 0.567$. The use of $NUV-r$ and a marginally efficient parameter applied to the NUV 
preselected sample leads to highly complete samples ($P_{\mathrm{comp,n}} \sim 0.74$), while $NUV-r$ in combination with efficient 
parameters leads to pure samples ( e.g. ($NUV-r$,log($\mu_*$)) with $P_{\mathrm{pure}} = 0.888$). Combinations with log($\mu_*$) all result 
in very pure samples with $P_{\mathrm{pure}} > 0.87$, usually, however, at the cost of completeness.\newline
Using the independent morphological classifications of \citet{NAIR2010} we obtain very similar results, with the most bijectively powerful 
combinations including $NUV-r$ with $M_i$, log($M_*$), or log($r_e$) followed by those combining log($n$), log($r_e$), and log($\mu_*$).
\newline
For the bright subsample of \citet{NAIR2010} $NUV-r$ efficiently selects pure and complete samples of spirals, however, the efficiency of the 
parameters log($M_*$) and log($r_e$) also remains high.\newline
Overall, the parameters log($n$), log($r_e$), and log($\mu_*$) appear efficient in selecting pure and complete samples of spirals as for optical 
samples. In addition, the $NUV-r$ colour in combination with NUV preselection is also efficient in this respect.\newline

A comparison of the figures of merit of the selections applied to the NUV pre-selected samples with those of comparable parameter 
combinations applied to the optical samples indicates that the use of such a preselection enhances the ability of the method to select pure and 
complete samples of spirals, with $P_{\mathrm{bij,n}}$ being, on average, greater than $P_{\mathrm{bij}}$ for comparable parameter 
combinations applied to the optical samples. This is due to the NUV preselection removing non-spiral contaminants, thus enlarging the spiral 
subvolume by making spirals more dominant and increasing the purity of spiral cells. In many cases both the completeness and the purity of the 
selections increase (e.g., (log($r_e$), log($M_*$))). However, in some cases the increase in completeness is accompanied by a (slight) decrease in 
the purity, indicating that the enlargement of parameter space is the dominant effect.\newline
Nevertheless, it must be born in mind that these samples are complete with respect to the preselected sample and may be biased against 
intrinsically UV faint spiral galaxies as well as strongly attenuated spirals seen edge-on if these sources lie below the NUV detection threshold. 
\newline

\begin{table*}
 \centering
 	\caption{Purity, completeness, bijective discrimination power, and contamination  for combinations of two parameters applied to 
 	\textit{NUVsample}. Completeness and bijective discrimination power are listed w.r.t. the \textit{OPTICALsample} ($P_{\mathrm{comp,o}}$ 
 	and $P_{\mathrm{bij,o}}$) and the \textit{NUVsample} ($P_{\mathrm{comp,n}}$ and $P_{\mathrm{bij,n}}$). }
  \begin{tabular}{@{}lccccccc@{}}
  \hline
  Parameter combination & $N_{sel}$ & $P_{\mathrm{pure}}$ & $P_{\mathrm{comp,n}}$& $P_{\mathrm{bij,n}}$ & $P_{\mathrm{cont}}$ & 
  $P_{\mathrm{comp,o}}$ & $P_{\mathrm{bij,o}}$  \\   
 \hline
 ($NUV-r$, log($n$)) & 53285 & 0.603 & 0.678 & 0.408 & 0.069 & 0.506 & 0.305 \\
($NUV-r$, log($r_e$)) &46791 & 0.722 & 0.713 & 0.514 & 0.042 & 0.532 & 0.384\\
($NUV-r$, log($M_*$)) & 56682 & 0.581 & 0.695 & 0.404 & 0.082 & 0.518 & 0.301\\
($NUV-r$, log($\mu_*$)) & 47516 & 0.717 & 0.719 & 0.516 & 0.031 & 0.536 & 0.385\\
($NUV-r$, $M_i$) & 55825 & 0.582 & 0.685 & 0.399 & 0.081 & 0.511 & 0.298\\
($NUV-r$, $e$) & 40000 & 0.714 & 0.603 & 0.431 & 0.041 & 0.450 & 0.321\\
(log($n$), log($r_e$)) & 46867 & 0.731 & 0.723 & 0.529 & 0.033 & 0.540 & 0.395\\
(log($n$), log($M_*$) & 53124 & 0.608 & 0.681 & 0.414 & 0.063 & 0.508 & 0.309\\
(log($n$), log($\mu_*$) & 51284 & 0.688 & 0.744 & 0.512 & 0.032 & 0.555 & 0.382\\
(log($n$), $M_i$ & 54617 & 0.606 & 0.698 & 0.423 & 0.064 & 0.521 & 0.315\\
(log($n$), $e$) & 37343 & 0.705 & 0.556 & 0.392 & 0.044 & 0.415 & 0.293\\
(log($r_e$), log($M_*$)) & 47184 & 0.731 & 0.727 & 0.532 & 0.039 & 0.543 & 0.397\\
(log($r_e$), log($\mu_*$)) & 45305 & 0.741 & 0.708 & 0.525 & 0.036 & 0.529 & 0.392\\
(log($r_e$), $M_i$) & 49531 & 0.707 & 0.739 & 0.523 & 0.070 & 0.552 & 0.390\\
(log($r_e$), $e$) & 40215 & 0.734 & 0.623 & 0.457 & 0.083 & 0.465 & 0.341\\
(log($M_*$), log($\mu_*$)) & 44472 & 0.742 & 0.696 & 0.517 & 0.032 & 0.520 & 0.386\\
(log($M_*$), $M_i$) & 38529 & 0.567 & 0.461 & 0.262 & 0.097 & 0.344 & 0.195\\
(log($M_*$), $e$) & 28449 & 0.731 & 0.439 & 0.321 & 0.075 & 0.327 & 0.239\\
(log($\mu_*$), $M_i$) & 47342 & 0.718 & 0.717 & 0.515 & 0.037 & 0.535 & 0.384\\
(log($\mu_*$), $e$) & 49323 & 0.721 & 0.751 & 0.542 & 0.030 & 0.560 & 0.404\\
($M_i$, $e$) & 24399 & 0.767 & 0.395 & 0.302 & 0.061 & 0.294 & 0.226\\
\hline                      
\end{tabular}
 \label{tab_2PNUVrGZ}
\end{table*}

\begin{table*}
 \centering
	\caption{Purity, completeness, bijective discrimination power, and contamination  for combinations of two parameters applied to 
	\textit{NUVNAIRsample} using the GALAXY ZOO visual classifications (columns 3-8) and the independent classifications of \citet[][columns 
	9-13]{NAIR2010}. Completeness and bijective discrimination power are listed w.r.t. the \textit{OPTICALsample} ($P_{\mathrm{comp,o}}$ and 
	$P_{\mathrm{bij,o}}$) and the \textit{NUVsample} ($P_{\mathrm{comp,n}}$ and $P_{\mathrm{bij,n}}$). In the case of the independent 
	classifications the contamination fraction is taken to be the complement of the purity (i.e. this includes sources with T-type = 99). }
  \begin{tabular}{@{}lcccccccccccc@{}}
  \hline
  \multicolumn{2}{c}{} &\multicolumn{6}{c}{GALAXY ZOO}    &     \multicolumn{5}{c}{Nair \& Abraham (2010)} \\
 Parameter combination & $N_{sel}$& $P_{\mathrm{pure}}$ & $P_{\mathrm{comp,n}}$ & $P_{bij,n}$ & $P_{\mathrm{cont}}$ & 
 $P_{\mathrm{comp,o}}$ & $P_{bij,o}$  & $P_{\mathrm{pure}}$ & $P_{\mathrm{comp,n}}$ & $P_{bij,n}$ &  $P_{\mathrm{comp,o}}$ & $P_{bij,o}$ 
 \\ 
 \hline
($NUV-r$, log($n$)) &1551 & 0.853 & 0.607 & 0.518 & 0.053 & 0.450 & 0.384 & 0.919 & 0.565 & 0.519& 0.418 & 0.384\\
($NUV-r$, log($r_e$)) &1801 & 0.869 & 0.719 & 0.624 & 0.044 & 0.533 & 0.463 & 0.914 & 0.650 & 0.594& 0.483 & 0.441\\
($NUV-r$, log($M_*$)) & 1970 & 0.822 & 0.744 & 0.612 & 0.064 & 0.552 & 0.454 & 0.895 & 0.695 & 0.622 & 0.517 & 0.463\\
($NUV-r$, log($\mu_*$)) & 1497 & 0.888 & 0.611 & 0.543 & 0.030 & 0.453 & 0.402 & 0.948 & 0.560 & 0.531& 0.416 & 0.394\\
($NUV-r$, $M_i$) & 1950 & 0.824 & 0.738 & 0.608 & 0.064 & 0.547 & 0.451 & 0.896 & 0.689 & 0.617 & 0.512 & 0.459\\
($NUV-r$, $e$) & 1127 & 0.859 & 0.444 & 0.382 & 0.031 & 0.330 & 0.283 & 0.933 & 0.415 & 0.387& 0.308 & 0.287\\
(log($n$), log($r_e$)) & 1790 & 0.831 & 0.683 & 0.568 & 0.084 & 0.507 & 0.421 & 0.879 & 0.623 & 0.548 & 0.461 & 0.405\\
(log($n$), log($M_*$) & 1591 & 0.813 & 0.594 & 0.482 & 0.069 & 0.440 & 0.358 & 0.894 & 0.564 & 0.504& 0.417 & 0.373\\
(log($n$), log($\mu_*$)) & 1616 & 0.873 & 0.648 & 0.566 & 0.032 & 0.480 & 0.419 & 0.942 & 0.603 & 0.568 & 0.446 & 0.421\\
(log($n$), $M_i$ & 1706 & 0.813 & 0.637 & 0.518 & 0.070 & 0.472 & 0.384 & 0.896 & 0.606 & 0.543& 0.448 & 0.402\\
(log($n$), $e$) & 944 & 0.815 & 0.353 & 0.288 & 0.049 & 0.262 & 0.213 & 0.915 & 0.342 & 0.313& 0.253 & 0.232\\
(log($r_e$), log($M_*$)) & 1512 & 0.900 & 0.625 & 0.562 & 0.026 & 0.463 & 0.417 & 0.950 & 0.567 & 0.539& 0.421 & 0.400\\
(log($r_e$), log($\mu_*$)) & 1447 & 0.902 & 0.599 & 0.540 & 0.025 & 0.444 & 0.401 & 0.956 & 0.546 & 0.522& 0.405 & 0.388\\
(log($r_e$), $M_i$) & 1630 & 0.842 & 0.630 & 0.531 & 0.075 & 0.467 & 0.394 & 0.890 & 0.572 & 0.509& 0.425 & 0.378\\
(log($r_e$), $e$) & 1488 & 0.728 & 0.498 & 0.363 & 0.160 & 0.369 & 0.269 & 0.776 & 0.456 & 0.354& 0.339 & 0.263\\
(log($M_*$), log($\mu_*$)) & 1387 & 0.906 & 0.577 & 0.523 & 0.021 & 0.428 & 0.388 & 0.960 & 0.525 & 0.504& 0.390 & 0.374\\
(log($M_*$), $M_i$) & 1263 & 0.792 & 0.459 & 0.364 & 0.097 & 0.340 & 0.270 & 0.859 & 0.428 & 0.368& 0.318 & 0.273\\
(log($M_*$), $e$) &  728 & 0.731 & 0.244 & 0.178 & 0.092 & 0.181 & 0.132 & 0.865 & 0.249 & 0.215& 0.185 & 0.160\\
(log($\mu_*$), $M_i$) & 1488 & 0.898 & 0.613 & 0.551 & 0.026 & 0.455 & 0.408 & 0.953 & 0.559 & 0.533& 0.416 & 0.396\\
(log($\mu_*$), $e$) & 1397 & 0.886 & 0.568 & 0.504 & 0.022 & 0.422 & 0.374 & 0.953 & 0.525 & 0.500& 0.390 & 0.372\\
($M_i$, $e$) &  631 & 0.751 & 0.218 & 0.163 & 0.094 & 0.161 & 0.121 & 0.876 & 0.218 & 0.191& 0.162 & 0.142\\
\hline                      
\end{tabular}
  \label{tab_2PNUVrNA}
\end{table*}

\subsubsection{Three parameter cells applied to the NUV samples}\label{NUVSAMP_3P}
Application of combinations of three parameters to the NUV preselected samples has much the same effect as for the optical samples, i.e. the 
purity and completeness, and consequently the bijective discrimination power, increase with respect to selections based on two parameters. The 
same processes as discussed in Sect.~\ref{WHOLESAMP_3P} apply. Tables~\ref{tab_3PNUVrGZ} and \ref{tab_3PNUVrNA} show the figures of 
merit for combinations of three parameters applied to the \textit{NUVsample} and \textit{NUVNAIRsample}.\newline

The combination of three parameters with the highest value of $P_{\mathrm{bij}}$ when applied to the \textit{NUVsample} is ($NUV-
r$,log($r_e$),$e$) with 
$P_{\mathrm{bij,n}} = 0.617$ ($P_{\mathrm{pure}} = 0.777$, $P_{\mathrm{comp,n}} = 0.794$). Of the 10 combinations with the greatest 
bijective discrimination power, the first 7 again include $e$( and are likely affected by the ellipticity bias). However, all 10 combinations include 
log($r_e$), log($\mu_*$) and/or log($n$). The three most efficient parameter combinations not including $e$ are 
(log($n$),log($r_e$),log($\mu_*$)) ($P_{\mathrm{pure}} = 0.744$, $P_{\mathrm{comp,n}} = 0.780$), (log($n$),log($r_e$),$M_i$) 
($P_{\mathrm{pure}} = 0.749$, $P_{\mathrm{comp,n}} = 0.775$), and ($NUV-r$,log($r_e$),$M_i$) ($P_{\mathrm{pure}} = 0.731$, 
$P_{\mathrm{comp,n}} = 0.789$). Overall, the use of three parameter combinations applied to the NUV preselected\textit{NUVsample} leads to 
very complete selections. Of the combinations not including $e$ 18/20 have $P_{\mathrm{comp,n}} > 0.7$, 6 of which have 
$P_{\mathrm{comp,n}} > 0.77$. In particular, $NUV-r$ in combination with at least one efficient parameter leads to very complete selections 
with $P_{\mathrm{comp,n}} \gtrsim 0.73$. \newline

Testing the performance of combinations of three parameters using the \textit{NUVNAIRsample} with GALAXY ZOO visual classifications the 
most bijectively powerful combination is ($NUV-r$,log($r_e$),$e$) with $P_{\mathrm{bij,n}} = 0.645$ ($P_{\mathrm{pure}} = 0.908$, 
$P_{\mathrm{comp,n}} = 0.711$; this result in not influenced by a bias in the test sample towards large values of $e$). However, of the ten 
most efficient combinations, this is the only one including $e$. The following 5 combinations with the highest values of $P_{\mathrm{bij,n}}$ 
are (in descending order): ($NUV-r$,log($n$),log($r_e$)), ($NUV-r$),log($r_e$),$M_i$), (log($n$),log($r_e$),log($M_*$)), ($NUV-
r$,log($n$),log($M_*$)), and ($NUV-r$,log($n$),log($\mu_*$)). Clearly $NUV-r$ applied in combination with another efficient parameter and 
NUV preselection leads to very pure and complete selections recovered from the bright subsample. Similar purity, but at the cost of 
completeness is also achieved by the parameter log($\mu_*$), even without the parameter $NUV-r$ (e.g. (log($r_e$),log($M_*$),log($\mu_*$)).
\newline
Testing using the \textit{NUVNAIRsample} with the independent morphological classifications of \citet{NAIR2010} supports the importance of 
$NUV-r$ as a parameter for selecting pure and complete samples of spirals under NUV preselection. The combinations with the largest bijective 
discrimination power are ($NUV-r$,log($n$),log($M_*$)), ($NUV-r$,log($n$),log($r_e$)), and ($NUV-r$,log($r_e$),$e$), with the use of $NUV-
r$ leading to very complete samples, as visible in the comparison of ($NUV-r$,log($n$),log($r_e$)) with (log($n$),log($r_e$),log($\mu_*$)), or 
(log($n$),log($r_e$),$M_i$).\newline

To summarize, we find that for NUV preselected samples the use of $NUV-r$ as a parameter leads to very complete, and in the case of the 
bright subsample of \citet{NAIR2010} also pure, selections of spiral galaxies. This is particularly the case in combination with log($r_e$) and 
log($n$), while combinations with log($\mu_*$) are also efficient, but mostly improve the purity of selections at the expense of completeness. A 
comparison of the figures of merit for comparable parameter combinations applied to the optical and NUV samples shows, as for the 
combinations of two parameters, that the use of NUV preselection increases both purity and completeness on average. We again note, however, 
that the values of completeness are with respect to the NUV samples, and will be biased against UV-faint sources (these may be intrinsically UV 
faint or UV faint due to being seen edge-on and experiencing severe attenuation due to dust).\newline

Overall, the parameters log($r_e$), log($\mu_*$), and log($n$) appear efficient at selecting pure and complete samples of spirals, as for the 
optical samples. Under NUV preselection however, the $NUV-r$ colour becomes efficient at selecting complete and pure spiral samples, much 
more so that the $u-r$ colour for the optical samples. The most efficient combinations include ($NUV-r$,log($r_e)$,$e$), ($NUV-
r$,log($n$),log($r_e$)), and (log($n$),log($r_e$),log($\mu_*$)).\newline

\begin{table*}
 \centering
 \caption{Purity, completeness, bijective discrimination power, and contamination  for combinations of three parameters applied to 
 \textit{NUVsample}. Completeness and bijective discrimination power are listed w.r.t. the \textit{OPTICALsample} ($P_{\mathrm{comp,o}}$ and 
 $P_{\mathrm{bij,o}}$) and the \textit{NUVsample} ($P_{\mathrm{comp,n}}$ and $P_{\mathrm{bij,n}}$). }
  \begin{tabular}{@{}lccccccc@{}}
  \hline
  Parameter combination & $N_{sel}$ & $P_{\mathrm{pure}}$ & $P_{\mathrm{comp,n}}$& $P_{\mathrm{bij,n}}$ & $P_{\mathrm{cont}}$ & 
  $P_{\mathrm{comp,o}}$ & $P_{\mathrm{bij,o}}$  \\   
 \hline
($NUV-r$, log($n$), log($r_e$)) & 50514 & 0.726 & 0.774 & 0.562 & 0.028 & 0.577 & 0.419 \\
($NUV-r$, log($n$), log($M_*$)) &56380 & 0.617 & 0.733 & 0.452 & 0.064 & 0.547 & 0.337\\
($NUV-r$, log($n$), log($\mu_*$)) &48707 & 0.716 & 0.736 & 0.527 & 0.032 & 0.549 & 0.39\\
($NUV-r$, log($n$), $M_i$) & 56496 & 0.616 & 0.734 & 0.452 & 0.064 & 0.548 & 0.337\\
($NUV-r$, log($n$), $e$) & 43708 & 0.695 & 0.641 & 0.445 & 0.044 & 0.478 & 0.332\\
($NUV-r$, log($r_e$), log($M_*$)) &48885 & 0.736 & 0.759 & 0.559 & 0.029 & 0.567 & 0.417\\
($NUV-r$, log($r_e$), log($\mu_*$)) & 49163 & 0.737 & 0.765 & 0.564 & 0.029 & 0.571 & 0.421\\
($NUV-r$, log($r_e$), $M_i$)  & 51151 & 0.731 & 0.789 & 0.577 & 0.033 & 0.589 & 0.430\\
($NUV-r$, log($r_e$), $e$) &48396 & 0.777 & 0.794 & 0.617 & 0.014 & 0.592 & 0.460 \\
($NUV-r$, log($M_*$), log($\mu_*$)) & 46269 & 0.746 & 0.728 & 0.543 & 0.029 & 0.543 & 0.405\\
($NUV-r$, log($M_*$), $M_i$) &56066 & 0.582 & 0.689 & 0.401 & 0.085 & 0.514 & 0.299 \\
($NUV-r$, log($M_*$), $e$) & 43874 & 0.730 & 0.676 & 0.493 & 0.035 & 0.504 & 0.368\\
($NUV-r$, log($mu_*$), $M_i$) & 48991 & 0.730 & 0.755 & 0.551 & 0.030 & 0.563 & 0.411\\
($NUV-r$, log($mu_*$), $e$) &49430 & 0.748 & 0.780 & 0.583 & 0.015 & 0.582 & 0.435\\
($NUV-r$, $M_i$, $e$) &44092 & 0.734 & 0.683 & 0.501 & 0.033 & 0.509 & 0.374\\
(log($n$), log($r_e$), log($M_*$)) &49304 & 0.744 & 0.773 & 0.575 & 0.020 & 0.577 & 0.429\\
(log($n$), log($r_e$), log($\mu_*$)) & 49665 & 0.744 & 0.780 & 0.580 & 0.022 & 0.582 & 0.433\\
(log($n$), log($r_e$), $M_i$) & 49054 & 0.749 & 0.775 & 0.580 & 0.023 & 0.578 & 0.433\\
(log($n$), log($r_e$), $e$) &47441 & 0.765 & 0.766 & 0.586 & 0.029 & 0.571 & 0.437\\
(log($n$), log($M_*$), log($\mu_*$)) &49945 & 0.736 & 0.775 & 0.571 & 0.020 & 0.579 & 0.426\\
(log($n$), log($M_*$), $M_i$) & 53302 & 0.611 & 0.687 & 0.420 & 0.062 & 0.513 & 0.313\\
(log($n$), log($M_*$), $e$) & 41242 & 0.702 & 0.611 & 0.429 & 0.044 & 0.456 & 0.320\\
(log($n$), log($\mu_*$), $M_i$) & 50378 & 0.719 & 0.764 & 0.550 & 0.019 & 0.570 & 0.410\\
(log($n$), log($\mu_*$), $e$) & 51054 & 0.715 & 0.770 & 0.551 & 0.026 & 0.575 & 0.411\\
(log($n$), $M_i$, $e$)  & 42160 & 0.705 & 0.627 & 0.443 & 0.046 & 0.468 & 0.330\\
(log($r_e$), log($M_*$), log($\mu_*$)) & 46264 & 0.738 & 0.721 & 0.532 & 0.033 & 0.538 & 0.397\\
(log($r_e$), log($M_*$), $M_i$) & 48838 & 0.727 & 0.749 & 0.545 & 0.042 & 0.559 & 0.407\\
(log($r_e$), log($M_*$), $e$) & 48793 & 0.764 & 0.786 & 0.600 & 0.028 & 0.586 & 0.448\\
(log($r_e$), log($\mu_*$), $M_i$) & 48671 & 0.729 & 0.749 & 0.546 & 0.045 & 0.559 & 0.407\\
(log($r_e$), log($\mu_*$), $e$) & 49571 & 0.762 & 0.797 & 0.607 & 0.027 & 0.595 & 0.453\\
(log($r_e$), $M_i$, $e$) & 46084 & 0.757 & 0.736 & 0.556 & 0.043 & 0.549 & 0.415\\
(log($M_*$), log($\mu_*$), $M_i$) &47355 & 0.729 & 0.729 & 0.531 & 0.039 & 0.544 & 0.397\\
(log($M_*$), log($\mu_*$), $e$) & 49250 & 0.762 & 0.791 & 0.603 & 0.028 & 0.590 & 0.450 \\
(log($M_*$), $M_i$, $e$) & 40952 & 0.698 & 0.603 & 0.421 & 0.065 & 0.450 & 0.314\\
(log($\mu_*$), $M_i$, $e$) &49331 & 0.757 & 0.787 & 0.596 & 0.031 & 0.588 & 0.445\\
\hline                      
\end{tabular}
 \label{tab_3PNUVrGZ}
\end{table*}

\begin{table*}
 \centering
	\caption{Purity, completeness, bijective discrimination power, and contamination  for combinations of three parameters applied to 
	\textit{NAIRsample} using the GALAXY ZOO visual classifications (columns 3-6) and the independent classifications of \citet[][columns 7-9]
	{NAIR2010}. Completeness and bijective discrimination power are listed w.r.t. the \textit{NAIRsample} ($P_{\mathrm{comp,o}}$ and 
	$P_{\mathrm{bij,o}}$) and the \textit{NUVNAIRsample} ($P_{\mathrm{comp,n}}$ and $P_{\mathrm{bij,n}}$). In the case of the independent 
	classifications the contamination fraction is taken to be the complement of the purity (i.e. this includes sources with T-type = 99).  }
  \begin{tabular}{@{}lcccccccccccc@{}}
  \hline
  \multicolumn{2}{c}{} &\multicolumn{6}{c}{GALAXY ZOO}    &     \multicolumn{5}{c}{Nair \& Abraham (2010)} \\
 Parameter combination & $N_{sel}$& $P_{\mathrm{pure}}$ & $P_{\mathrm{comp,n}}$ & $P_{bij,n}$ & $P_{\mathrm{cont}}$ & 
 $P_{\mathrm{comp,o}}$ & $P_{bij,o}$  & $P_{\mathrm{pure}}$ & $P_{\mathrm{comp,n}}$ & $P_{bij,n}$ &  $P_{\mathrm{comp,o}}$ & $P_{bij,o}$ 
 \\ 
 \hline
($NUV-r$, log($n$), log($r_e$)) & 1879 & 0.864 & 0.745 & 0.644 & 0.047 & 0.553 & 0.477 & 0.915 & 0.681 & 0.623 & 0.504 & 0.461\\
($NUV-r$, log($n$), log($M_*$)) & 1934 & 0.841 & 0.747 & 0.628 & 0.055 & 0.554 & 0.466 & 0.906 & 0.694 & 0.629 & 0.514 & 0.465\\
($NUV-r$, log($n$), log($\mu_*$)) &1564 & 0.878 & 0.630 & 0.553 & 0.033 & 0.467 & 0.410 & 0.943 & 0.584 & 0.551 & 0.432 & 0.408\\
($NUV-r$, log($n$), $M_i$) & 1906 & 0.839 & 0.735 & 0.617 & 0.055 & 0.545 & 0.457 & 0.902 & 0.681 & 0.615 & 0.504 & 0.455\\
($NUV-r$, log($n$), $e$) & 1299 & 0.856 & 0.511 & 0.437 & 0.038 & 0.379 & 0.324 & 0.928 & 0.478 & 0.443 & 0.354 & 0.328\\
($NUV-r$, log($r_e$), log($M_*$)) &1687 & 0.893 & 0.691 & 0.617 & 0.027 & 0.513 & 0.458 & 0.942 & 0.627 & 0.591 & 0.466 & 0.439\\
($NUV-r$, log($r_e$), log($\mu_*$)) & 1713 & 0.891 & 0.701 & 0.624 & 0.025 & 0.520 & 0.463 & 0.941 & 0.636 & 0.599 & 0.473 & 0.445\\
($NUV-r$, log($r_e$), $M_i$) & 1770 & 0.884 & 0.718 & 0.635 & 0.034 & 0.533 & 0.471 & 0.928 & 0.648 & 0.602 & 0.482 & 0.447\\
($NUV-r$, log($r_e$), $e$) &1705 & 0.908 & 0.711 & 0.645 & 0.014 & 0.527 & 0.479 & 0.956 & 0.643 & 0.615 & 0.478 & 0.457\\
($NUV-r$, log($M_*$), log($\mu_*$)) & 1594 & 0.897 & 0.657 & 0.589 & 0.025 & 0.487 & 0.437 & 0.946 & 0.595 & 0.563 & 0.442 & 0.418\\
($NUV-r$, log($M_*$), $M_i$) &1970 & 0.815 & 0.737 & 0.601 & 0.069 & 0.547 & 0.446 & 0.887 & 0.690 & 0.612 & 0.512 & 0.455\\
($NUV-r$, log($M_*$), $e$) & 1478 & 0.884 & 0.600 & 0.531 & 0.020 & 0.445 & 0.394 & 0.941 & 0.549 & 0.516 & 0.408 & 0.384\\
($NUV-r$, log($mu_*$), $M_i$) &1647 & 0.888 & 0.672 & 0.597 & 0.029 & 0.498 & 0.442 & 0.943 & 0.613 & 0.578 & 0.455 & 0.429\\
($NUV-r$, log($mu_*$), $e$) &1494 & 0.908 & 0.623 & 0.566 & 0.017 & 0.462 & 0.420 & 0.967 & 0.570 & 0.551 & 0.424 & 0.410\\
($NUV-r$, $M_i$, $e$) &1467 & 0.883 & 0.595 & 0.526 & 0.022 & 0.441 & 0.390 & 0.938 & 0.543 & 0.509 & 0.403 & 0.378\\
(log($n$), log($r_e$), log($M_*$)) & 1745 & 0.886 & 0.710 & 0.629 & 0.028 & 0.526 & 0.466 & 0.940 & 0.650 & 0.611 & 0.481 & 0.452\\
(log($n$), log($r_e$), log($\mu_*$)) & 1736 & 0.885 & 0.705 & 0.624 & 0.028 & 0.523 & 0.463 & 0.940 & 0.646 & 0.607 & 0.478 & 0.449\\
(log($n$), log($r_e$), $M_i$) & 1757 & 0.874 & 0.705 & 0.617 & 0.042 & 0.523 & 0.457 & 0.923 & 0.642 & 0.593 & 0.475 & 0.438\\
(log($n$), log($r_e$), $e$) &1754 & 0.831 & 0.669 & 0.556 & 0.078 & 0.496 & 0.412 & 0.884 & 0.615 & 0.543 & 0.455 & 0.402\\
(log($n$), log($M_*$), log($\mu_*$)) &1698 & 0.894 & 0.697 & 0.623 & 0.025 & 0.517 & 0.462 & 0.948 & 0.638 & 0.605 & 0.472 & 0.448\\
(log($n$), log($M_*$), $M_i$) &1695 & 0.820 & 0.638 & 0.523 & 0.069 & 0.473 & 0.388 & 0.895 & 0.601 & 0.538 & 0.445 & 0.398\\
(log($n$), log($M_*$), $e$) & 1189 & 0.834 & 0.455 & 0.380 & 0.049 & 0.338 & 0.282 & 0.918 & 0.432 & 0.396 & 0.320 & 0.293\\
(log($n$), log($\mu_*$), $M_i$) & 1694 & 0.888 & 0.691 & 0.614 & 0.021 & 0.512 & 0.455 & 0.950 & 0.638 & 0.606 & 0.472 & 0.449\\
(log($n$), log($\mu_*$), $e$) & 1545 & 0.869 & 0.617 & 0.536 & 0.029 & 0.457 & 0.397 & 0.939 & 0.575 & 0.540 & 0.425 & 0.400\\
(log($n$), $M_i$, $e$) & 1307 & 0.828 & 0.497 & 0.411 & 0.060 & 0.368 & 0.305 & 0.896 & 0.464 & 0.416 & 0.343 & 0.308\\
(log($r_e$), log($M_*$), log($\mu_*$)) & 1465 & 0.903 & 0.607 & 0.549 & 0.024 & 0.450 & 0.407 & 0.954 & 0.552 & 0.526 & 0.410 & 0.391\\
(log($r_e$), log($M_*$), $M_i$) &1567 & 0.886 & 0.637 & 0.564 & 0.036 & 0.473 & 0.419 & 0.936 & 0.579 & 0.542 & 0.430 & 0.403\\
(log($r_e$), log($M_*$), $e$) & 1528 & 0.889 & 0.624 & 0.554 & 0.026 & 0.462 & 0.411 & 0.944 & 0.569 & 0.537 & 0.423 & 0.399\\
(log($r_e$), log($\mu_*$), $M_i$) & 1567 & 0.880 & 0.633 & 0.557 & 0.041 & 0.470 & 0.413 & 0.934 & 0.577 & 0.539 & 0.429 & 0.400\\
(log($r_e$), log($\mu_*$), $e$) & 1536 & 0.896 & 0.632 & 0.566 & 0.022 & 0.469 & 0.420 & 0.951 & 0.577 & 0.548 & 0.428 & 0.407\\
(log($r_e$), $M_i$, $e$) & 1450 & 0.870 & 0.579 & 0.504 & 0.044 & 0.430 & 0.374 & 0.916 & 0.524 & 0.480 & 0.389 & 0.357\\
(log($M_*$), log($\mu_*$), $M_i$) &1516 & 0.888 & 0.618 & 0.549 & 0.032 & 0.458 & 0.407 & 0.942 & 0.563 & 0.531 & 0.419 & 0.394\\
(log($M_*$), log($\mu_*$), $e$) & 1556 & 0.894 & 0.639 & 0.571 & 0.021 & 0.474 & 0.423 & 0.951 & 0.584 & 0.555 & 0.434 & 0.413\\
(log($M_*$), $M_i$, $e$) & 1154 & 0.792 & 0.420 & 0.332 & 0.074 & 0.311 & 0.246 & 0.885 & 0.403 & 0.356 & 0.299 & 0.265\\
(log($\mu_*$), $M_i$, $e$) &1548 & 0.897 & 0.637 & 0.571 & 0.023 & 0.473 & 0.424 & 0.946 & 0.578 & 0.547 & 0.429 & 0.406\\
\hline                      
\end{tabular}
  \label{tab_3PNUVrNA}
\end{table*}

\subsubsection{Effects of NUV selection}
As shown in Sect.~\ref{NUVSAMP_3P},
the use of NUV preselection results, on average, in samples with greater completeness and often also greater purity for comparable 
combinations of selection parameters. Under NUV preselection the parameter $NUV-r$ leads to efficient selections of complete samples of 
spirals, while attaining high values of purity for the bright subsample. As spiral galaxies are often star forming systems, this result is 
unsurprising. However, as discussed, NUV preselection will bias samples of spirals against intrinsically UV-faint systems, as well as against 
systems which are UV-faint due to severe attenuation (e.g. on account of being seen edge-on).\newline
Overall, the efficiency of the considered parameter combinations in selecting pure and complete (under the aforementioned caveat) samples is 
enhanced by NUV preselection, with larger volumes of the parameter space being included in the spiral volume than for the whole sample, as 
indicated by increases in completeness accompanied by slight reductions in purity when using comparable parameter combinations with and 
without preselection. In addition, especially for combinations of three parameters, NUV preselection can also lead to an increase in purity 
accompanied by a decrease in completeness, as regions marginally dominated by spirals in the whole sample are excluded. On average, 
however, in both cases the value of $P_{\mathrm{bij,n}}$ is larger than $P_{\mathrm{bij}}$ for a comparable parameter combination applied to 
the \textit{OPTICALsample}. Thus,  depending upon the science goal of the selection, UV information could be a valuable asset in selecting 
samples of spirals. However, we caution that, in addition to the biases previously discussed, if the depth of the UV coverage is not such that it 
matches the depth of the optical data and encompasses the entire (realistic) colour range, UV preselection will strongly suppress the 
completeness attainable and introduce biases into any selections.\newline  
In light of these effects, the greater completeness of using only optical parameters applied to optical samples, as evidenced by the values of 
$P_{comp,o}$ in, for example, Table~\ref{tab_3PNUVrGZ} and the robustness against bias will likely outweigh the gain in purity achievable by 
NUV preselection for most applications.\newline

\subsection{Investigation of possible biases}
Based on the figures of purity, completeness, and bijective discrimination power it is readily apparent that the use of combinations of three 
parameters generally leads to purer and simultaneously more complete samples of spirals than using only two parameters.  Furthermore, the 
most important parameters appear to be log($r_e$) \& log($\mu$), which provide the most efficient selection when complemented by log($n$) 
and/or $M_i$. Applying an NUV preselection appears to further improve the attainable purity, and makes $NUV-r$ a further important selection 
parameter. However, although the purity, completeness, and bijective discrimination power are good indicators of a selection's performance, 
they provide little information about possible biases in the selections. While the cell-based method allows for a flexible surface of separation, 
any boundary in parameter space used in classifying objects entails that reliable spirals with strongly outlying values in the selection parameters 
may be missed, and that the selection may not be fully representative of the actual population of spirals.\newline
In the following we will investigate the potential biases caused by the selection on the basis of four different representative  combinations of 
three parameters (($u-r$,log($r_e$),$e$) resp. ($NUV-r$,log($r_e$),$e$), (log($n$),log($r_e$),log($\mu_*$)), (log($n$),log($r_e$),$M_i$), and 
(log($n$,log($M_*$),log($\mu_*$))), chosen to be amongst the most bijectively powerful. We will consider the distributions of the suite of 
parameters investigated for these selections, as well as consider the the distributions of the H$\alpha$ equivalent width as an independent 
observable and the T-type classification given by \citet{NAIR2010} to investigate possible biases in the selections of spiral galaxies. Finally, we 
will investigate the redshift dependence of the selections of spiral galaxies.\newline

\subsubsection{Distributions of the parameter suite}
\begin{figure*}
\begin{center}
\includegraphics[width=0.9\textwidth]{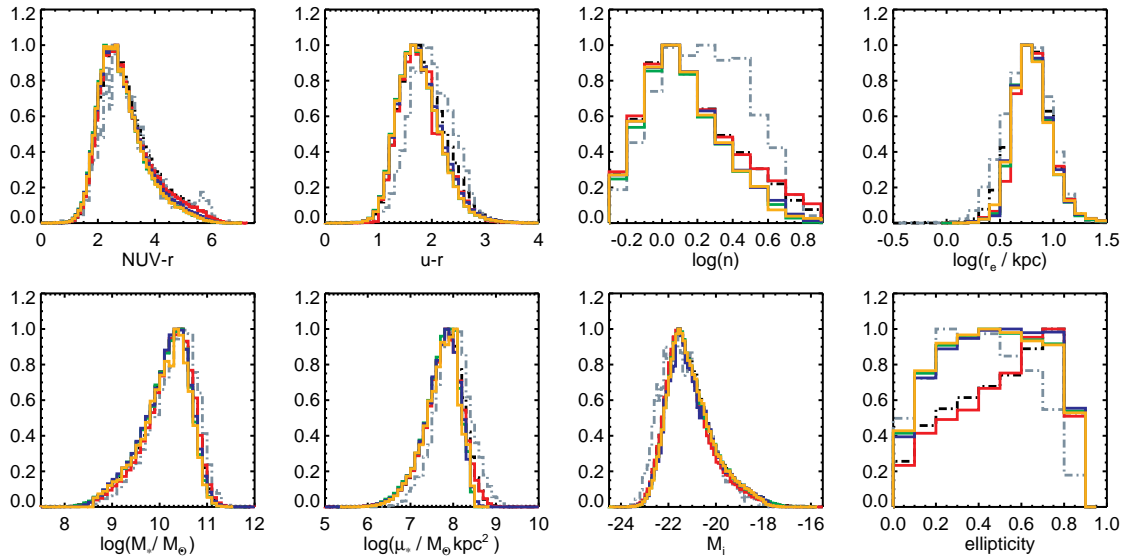}
\caption{Normalized distribution of the suite of 8 parameters as recovered for all GALAXY ZOO reliable spirals in the \textit{OPTICALsample} 
(black dashed) and the selections defined using ($u-r$,log($r_e$),$e$) (red) ,(log($n$),log($r_e$),log($\mu_*$))  (green), (log($n$),log($r_e$),
$M_i$) (blue), and (log($n$,log($M_*$),log($\mu_*$)) (orange), applied to the \textit{OPTICALsample}. The parameter distribution of spirals as 
defined by the classifications of \citet{NAIR2010} in the \textit{NAIRsample} is shown as a grey dash-dotted line.}
\label{fig_pardistU}
\end{center}
\end{figure*}

\begin{figure*}
\begin{center}
\includegraphics[width=0.9\textwidth]{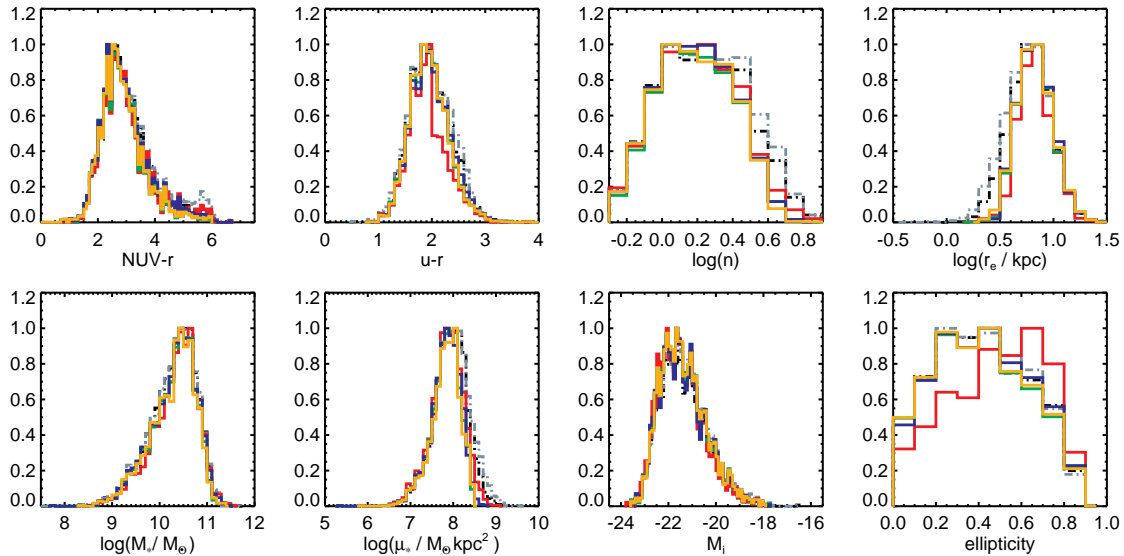}
\caption{As Fig.~\ref{fig_pardistU} but for the \textit{NAIRsample}.}
\label{fig_pardistUNA}
\end{center}
\end{figure*}

Figs.~\ref{fig_pardistU} \& \ref{fig_pardistUNA} show the normalized distributions of all eight parameters in the suite investigated, after 
selection by four different representative  combinations of three parameters (($u-r$,log($r_e$),$e$) resp. ($NUV-r$,log($r_e$),$e$) in red, 
(log($n$),log($r_e$),log($\mu_*$)) in green, (log($n$),log($r_e$),$M_i$) in blue, and (log($n$,log($M_*$),log($\mu_*$)) in orange), chosen to 
be amongst the most bijectively powerful, applied to both the \textit{OPTICALsample} (Fig.~\ref{fig_pardistU}) and to the \textit{NAIRsample} 
(Fig.~\ref{fig_pardistUNA}). For comparison the parameter's distribution for reliable spirals in the respective sample as defined by GALAXY ZOO 
is shown as a dash-dotted black line. Finally, the parameter's distribution for reliable spirals as defined by the independent morphological 
classifications of \citet{NAIR2010}, i.e. in the \textit{NAIRsample}, is shown as a grey dash-dotted line.\newline   
Overall, the distributions of the parameters derived from the selections applied to the \textit{OPTICALsample} (Fig.~\ref{fig_pardistU}) coincide 
well with that of the GALAXY ZOO defined sample, indicating that the non-parametric method using three parameters is neither heavily 
influencing the parameter ranges available to the sample, nor is itself introducing large biases. Similarly, the parameter combinations for the 
selections applied to the \textit{NAIRsample} also agree well with the parameter's distributions as defined by the GALAXY ZOO and 
\citet{NAIR2010} visual classifications. Nevertheless, the effect of the individual choice of parameter combinations is visible in the distributions, 
with this being more pronounced for the application to the \textit{NAIRsample}. For example, all combinations involving log($n$) are biased 
towards lower values of this parameter than the visually defined samples, while the combination ($u-r$,log($r_e$),$e$) traces them with higher 
fidelity. The discontinuous steep fall-off towards redder $u-r$ colours of the selection determined by ($u-r$,log($r_e$),$e$) (most pronounced 
in the \textit{NAIRsample}), is also an example of the effects of the discretization.\newline  
The largest differences, both between the selections and the visually-defined samples, as well as between the selections themselves, are visible, 
however, in the distributions of ellipticity. While the distribution of $e$ is more or less flat in the \textit{NAIRsample}, as is to be expected for an 
unbiased sample, the GALAXY ZOO-defined spiral subsample of the \textit{OPTICALsample} displays a bias towards high values of $e$. Using 
$e$ as selection parameter, as in the combination ($u-r$,log($r_e$),$e$), gives rise to a bias in the distribution of $e$ for the selected sample 
as visible in Fig.~\ref{fig_pardistUNA}, causing the selection provided by ($u-r$,log($r_e$),$e$) to largely coincide with the GALAXY ZOO 
defined spiral sample for the \textit{OPTICALsample}. This bias may also give rise to the agreement between the $NUV-r$ colour distributions 
of the GALAXY ZOO defined sample and the ($u-r$,log($r_e$),$e$) selection in Fig.~\ref{fig_pardistU} (i.e. for the \textit{OPTICALsample}), 
which extend to redder colours than the other selections, as NUV emission from highly inclined galaxies will be strongly attenuated, more so 
than in optical bands \citep[e.g.,][]{TUFFS2004}. In contrast to the selection using ($u-r$,log($r_e$),$e$), the other investigated parameter 
combinations show distributions which are more or less flat in $e$, also justifying the use of the GALAXY ZOO sample as a calibration sample.
\newline  
Comparison of the distribution of the parameters in the selections applied to the \textit{OPTICALsample} with those of the galaxies classified as 
spirals in the \textit{NAIRsample} using the classifications of \citet{NAIR2010}, shows a systematic difference in the distributions of the 
parameters between these samples. Overall, the spiral galaxies in the \textit{NAIRsample} are more weighted towards redder $NUV-r$ and $u-
r$ colours, as well as towards larger values of log($M_*$) and log($\mu_*$), and brighter $i$-band absolute magnitudes. Furthermore, the 
distributions of log($n$) and log($r_e$) are weighted towards larger values of $n$ and lower values of $r_e$, respectively. The observable 
differences are largely consistent with the bright \textit{NAIRsample} ($g'$-band mag $\le 16$) being more weighted towards large spirals 
which, on average, are more massive and redder than lower mass spiral galaxies. Furthermore, they often also have more dominant bulges, 
increasing the values of $n$ and decreasing those of $r_e$, while simultaneously decreasing the value of $e$, in agreement with the observed 
distributions. However, the differences may also be due, in part, to the fact that the cell-based selection misses regions of parameter space 
which are sparsely populated by spirals and in which they do not represent the dominant galaxy population. Nevertheless, 
Fig.~\ref{fig_pardistUNA} shows that the selections using combinations of three parameters trained on the GALAXY ZOO visual classifications of 
the \textit{OPTICALsample} perform well at recovering the \textit{NAIRsample}.\newline  

\begin{figure*}
\begin{center}
\includegraphics[width=0.9\textwidth]{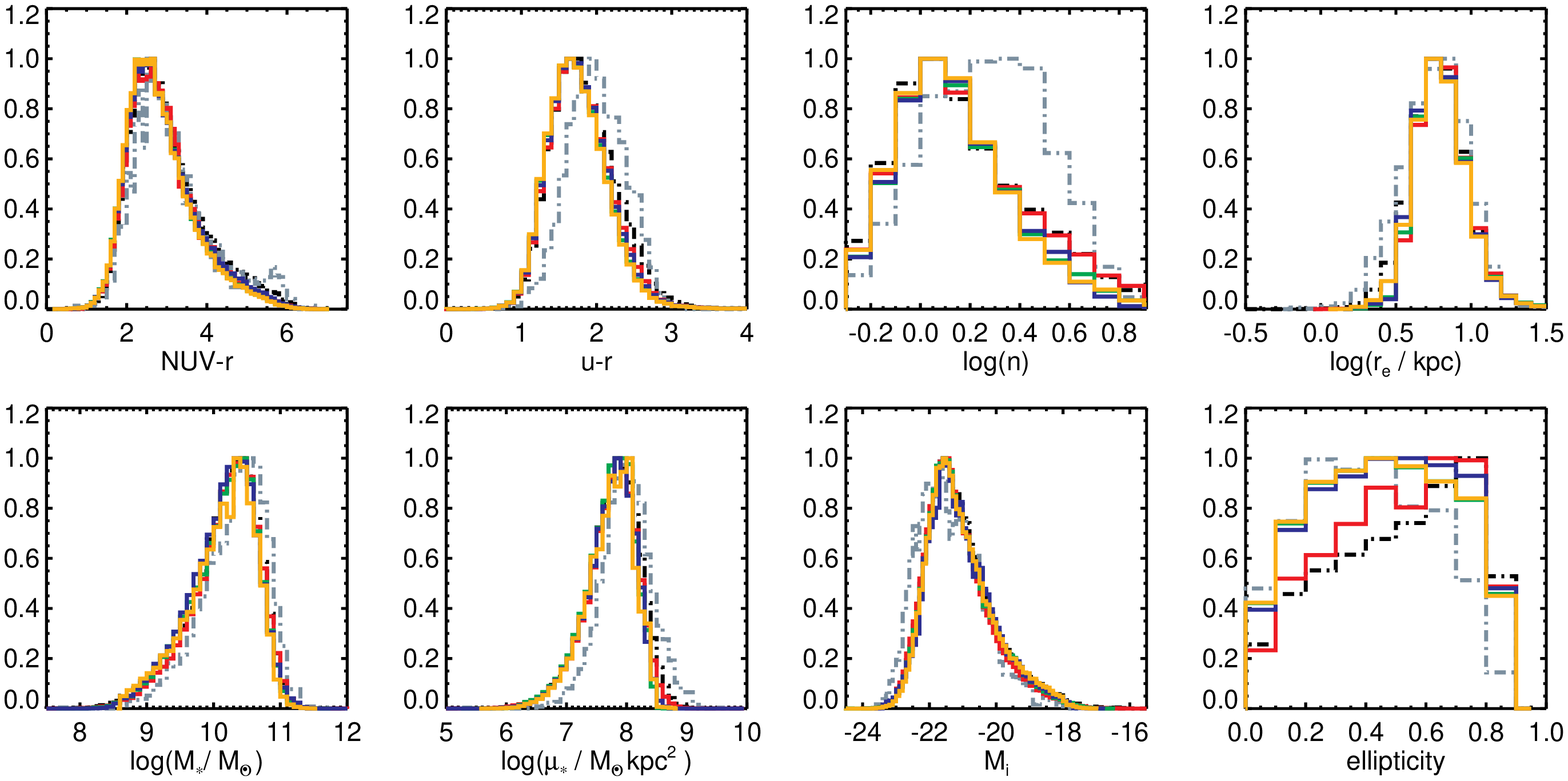}
\caption{Normalized distribution of the suite of 8 parameters as recovered for all GALAXY ZOO reliable spirals in the \textit{NUVsample} (black 
dashed) and the selections defined using ($NUV-r$,log($r_e$),$e$) (red) ,(log($n$),log($r_e$),log($\mu_*$))  (green), (log($n$),log($r_e$),
$M_i$) (blue), and (log($n$,log($M_*$),log($\mu_*$)) (orange), applied to the \textit{NUVsample}. The parameter distribution of spirals as 
defined by the classifications of \citet{NAIR2010} in the \textit{NUVNAIRsample} is shown as a grey dash-dotted line.}
\label{fig_pardistNUV}
\end{center}
\end{figure*}

Fig.~\ref{fig_pardistNUV} shows the parameter distributions for the combinations applied to the \textit{NUVsample} (we make use of ($NUV-
r$,log($r_e$),$e$) instead of ($u-r$,log($r_e$),$e$)). The results of applying the combinations to the \textit{NUVsample} are nearly identical to 
those obtained for the \textit{OPTICALsample}. However, the use of NUV preselection does bias the selected galaxy populations towards bluer 
objects as can be seen in the shift of the distributions of the $u-r$ and to lesser extent the $NUV-r$ colour, between Figs.~\ref{fig_pardistU} 
\& \ref{fig_pardistNUV}. The use of NUV preselection and $NUV-r$ colour also slightly lessens the bias against sources with low values of $e$ 
selected using the combination ($NUV-r$,log($r_e$),$e$), rendering the distribution in $e$ of this selection flatter than that of the GALAXY 
ZOO defined sample. The overall similarity to the results obtained for the optical samples show that the requirement of an NUV detection itself is 
only mildly influencing the selections.\newline

\subsubsection{T-type and H$\alpha$ equivalent width as independent observables}
Although the agreement between the parameter distributions of the visually defined samples and the selections is very good, the fact that a bias 
towards bluer $u-r$ and $NUV-r$ colours is discernible, and that the selections slightly favour lower values of log($n$) and log($\mu_*$) and 
higher values of log($r_e$), raises the possibility that the selections may nevertheless be biased against a subclass of spirals. \newline

\begin{figure}
\includegraphics[width=0.5\textwidth]{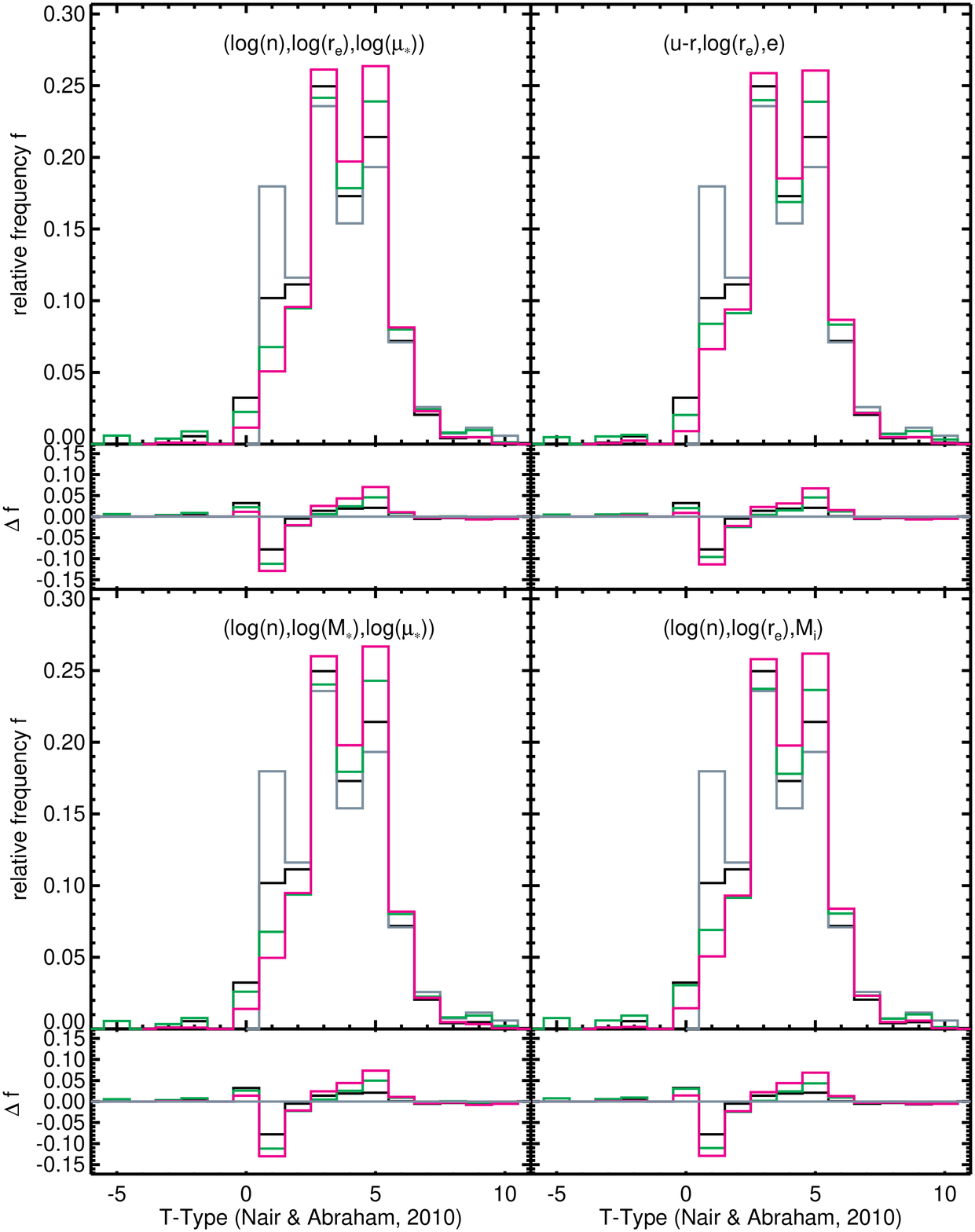}
\caption{Distribution of T-types for galaxies in the \textit{NAIRsample} classified as spirals based on the classifications of \citet{NAIR2010} 
(gray), GALAXY ZOO (black), and the parameter combination listed top left (green). The T-type distribution of galaxies with $P_{\mathrm{CS,DB 
\ge 0.7}}$ located in cells associated with spiral galaxies is shown in magenta. The inset panel below each distribution shows the 
distribution of the difference in relative frequency for this galaxy type relative to those of the \citet{NAIR2010} classifications.}
\label{fig_ttypeU}
\end{figure}

\paragraph*{T-type distributions of the \textit{NAIRsample}}$\,$\newline
In order to investigate to what extent such a bias may be present, we first make use of the distributions of the T-type classifications of 
\citet{NAIR2010}. Fig.~\ref{fig_ttypeU} shows the normalized distributions of the T-type values for the four selections, compared with the 
distributions of the visually classified spiral samples (GALAXY ZOO: black, \citet{NAIR2010}:grey). The distribution of the T-types of galaxies 
classified as spirals by the selection is shown in green, while the magenta line shows the T-type distributions of the GALAXY ZOO defined 
reliable spirals located in spiral cells following the selection.  For the \textit{NAIRsample} the GALAXY ZOO classifications (black solid line) 
appear moderately biased against early type spirals (mainly against Sa, and less against Sa/b). The selections based on the 
combinations of three parameters (green line) display a similar, but more pronounced bias, favoring spiral galaxies of type Sa/b, Sb and later, 
underscored by the stronger bias against early type spirals of GALAXY ZOO spirals in spiral cells (magenta line). Overall, the parameter based 
selections recover relatively more earlier type spirals than the GALAXY ZOO classifications, in line with the findings that a large fraction of the 
'impurity' arises from spiral galaxies which fail to meet the $P_{CS,DB} \ge 0.7$ requirement. All combinations considered display very 
similar performance in terms of the relative fractions of galaxy types recovered, although the bias against Sa/b galaxies of the selections using 
the parameters colour, effective radius and ellipticity is slightly less pronounced than for the other parameter combinations which involve more 
structural information ( the use of structural information may be more sensitive to the presence of a prominent bulge in early-type spirals). 
\newline

\begin{figure}
\includegraphics[width=0.5\textwidth]{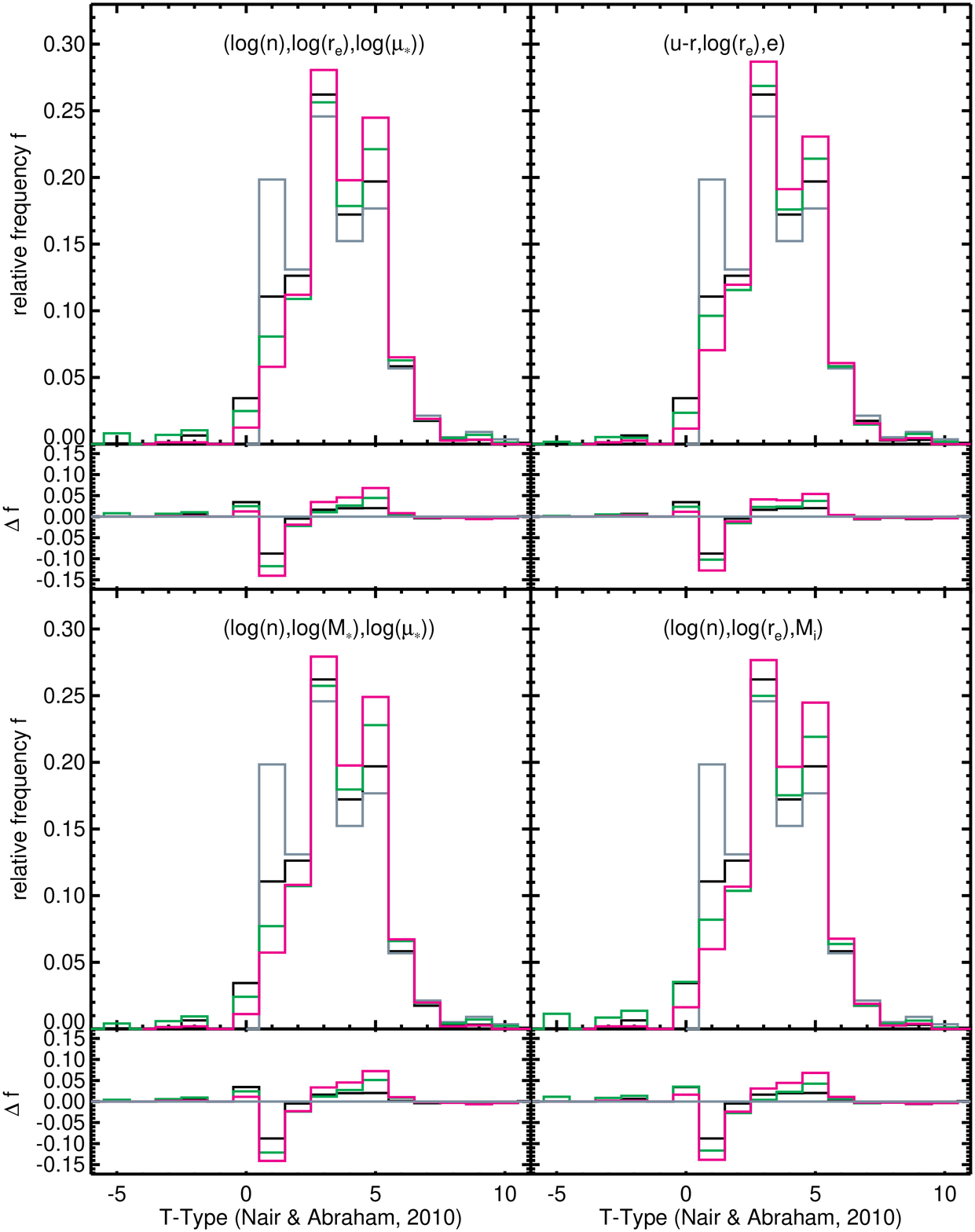}
\caption{As Fig.~\ref{fig_ttypeU} but for galaxies in the \textit{NUVNAIRsample}.}
\label{fig_ttypeNUV}
\end{figure}

\paragraph*{T-type distributions of the \textit{NUVNAIRsample}}$\,$\newline
Fig.~\ref{fig_ttypeNUV} shows the resultant distributions of T-types for the selections applied to the \textit{NUVNAIRsample} (using $NUV-r$ 
rather than $u-r$). Overall, the results are very similar, with both the GALAXY ZOO classified spirals and the spirals selected by the parameter 
combinations being more weighted towards later type galaxies than the classifications of \citet{NAIR2010}. We note the fact that the 
\textit{NUVNAIRsample} is more weighted towards earlier type spirals than the \textit{NAIRsample}.\newline

\begin{figure}
\includegraphics[width=0.5\textwidth]{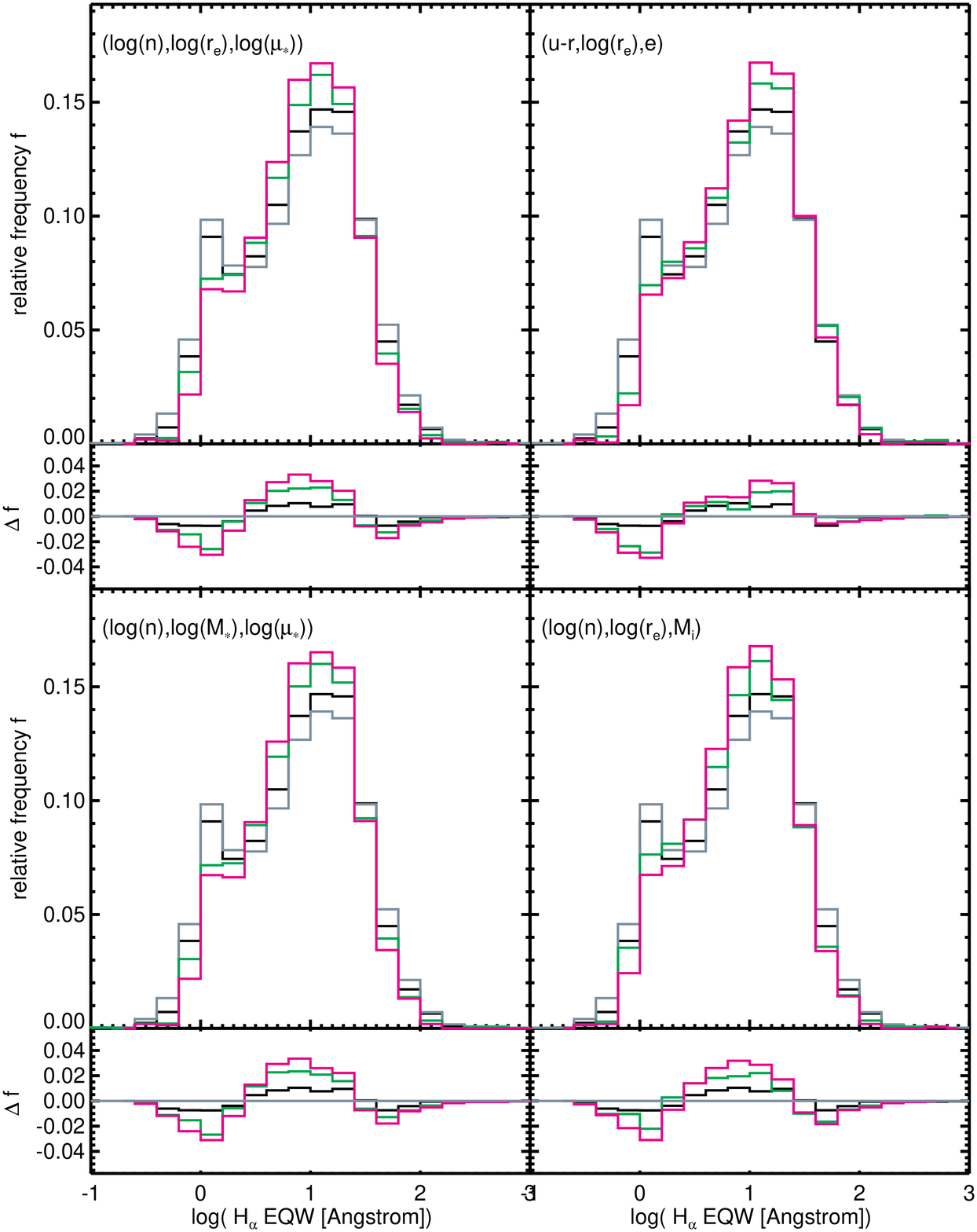}
\caption{Distribution of H$\alpha$ EQW for galaxies in the \textit{NAIRsample} classified as spirals based on the classifications of 
\citet{NAIR2010} (gray), GALAXY ZOO (black), and the parameter combination listed top left (green). The H$\alpha$ EQW distribution of galaxies 
with $P_{\mathrm{CS,DB \ge 0.7}}$ located in cells associated with spiral galaxies is shown in magenta. The inset panel below each 
distribution shows the distribution of the difference in relative frequency for each bin in H$\alpha$ EQW relative to that of the \citet{NAIR2010} 
classifications.}
\label{fig_haeqwUNA}
\end{figure}

\begin{figure}
\includegraphics[width=0.5\textwidth]{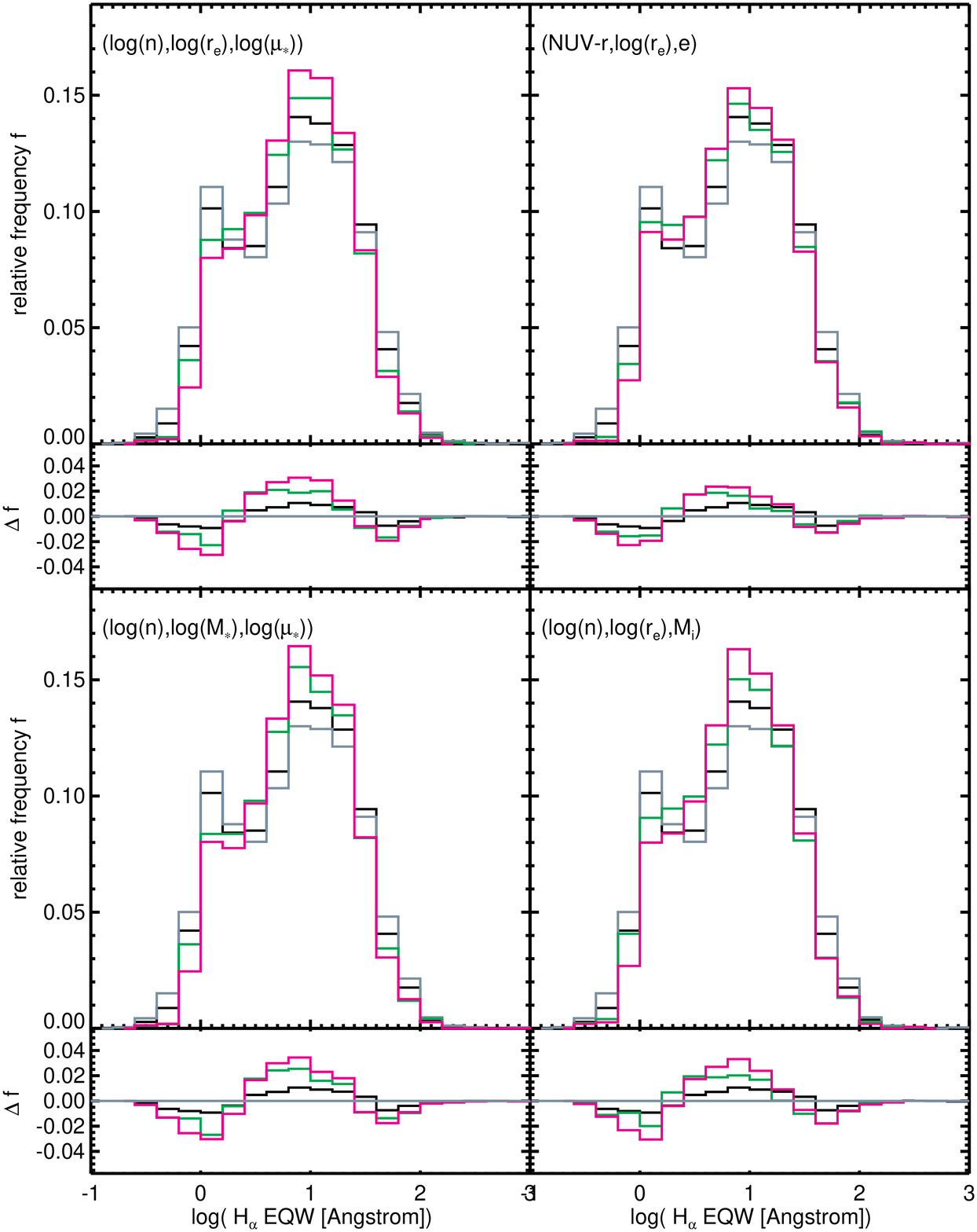}
\caption{As Fig.~\ref{fig_haeqwUNA} but for galaxies in the \textit{NUVNAIRsample}.}
\label{fig_haeqwNUVNA}
\end{figure}

\paragraph*{H$\alpha$ Equivalent Width Distribution of the \textit{NAIRsample} \& \textit{NUVNAIRsample}}$\,$\newline
A similar investigation of the possible bias against subclasses of spiral galaxies for the \textit{OPTICALsample}, respectively for the 
\textit{NUVsample}, is not possible, as these lack independent visual classifications and T-Types. However, to at least gain a qualitative insight 
into the possible biases for these larger samples, we make use of the distributions of H$\alpha$ equivalent width (EQW), an observable used 
neither in our classification nor in that supplied by GALAXY ZOO.\newline
Based on H$\alpha$ EQW, galaxies are often divided into two main populations, 'line-emitting' galaxies (i.e. galaxies with non-negligible Balmer 
line emission, usually actively star forming) and passive galaxies (very little/no line emission, usually quiescent). In general, spirals tend to 
exhibit H$\alpha$ line emission (although a non-negligible fraction has very small H$\alpha$ EQWs indicative of passive systems), while early-
types are predominantly passive. Similarly, earlier type spirals often have smaller values of H$\alpha$ EQW than later types (see e.g., 
\citealt{ROBOTHAM2013}. for a detailed discussion).\newline
Figs.~\ref{fig_haeqwUNA} \& \ref{fig_haeqwNUVNA} show the distributions of H$\alpha$ EQW for the \textit{NAIRsample} and 
\textit{NUVNAIRsample}, respectively. The distribution of the samples defined using the classifications of \citet{NAIR2010} is again shown in 
gray, with that of the sample defined by GALAXY ZOO in black. In both cases the GALAXY ZOO defined sample is weighted more towards 
intermediate values of H$\alpha$ EQW with respect to the classifications of \citet{NAIR2010}, showing evidence of a bias against low values of 
H$\alpha$ EQW as well as, to a lesser extent, against the highest values. The distributions of H$\alpha$ EQW of the samples defined by the 
selections (green) all display a similar, yet more pronounced bias against low values of  H$\alpha$ EQW. The selections, with the exception of 
($u-r$,log($r_e$,$e$), all also appear weighted against the highest values of H$\alpha$ EQW.     
These biases against low values of H$\alpha$ EQW may be considered to be consistent with the distributions of the T-types in the samples, with 
the selections favoring later type spirals. 
\newline 
In summary, we find that the GALAXY ZOO classifications display a simultaneous mild bias against early type spirals and systems with low values 
of H$\alpha$ EQW for the \textit{NAIRsample} and \textit{NUVNAIRsample}, and that this bias is slightly more pronounced for the parameter 
combination based selections.\newline 

\begin{figure}
\includegraphics[width=0.5\textwidth]{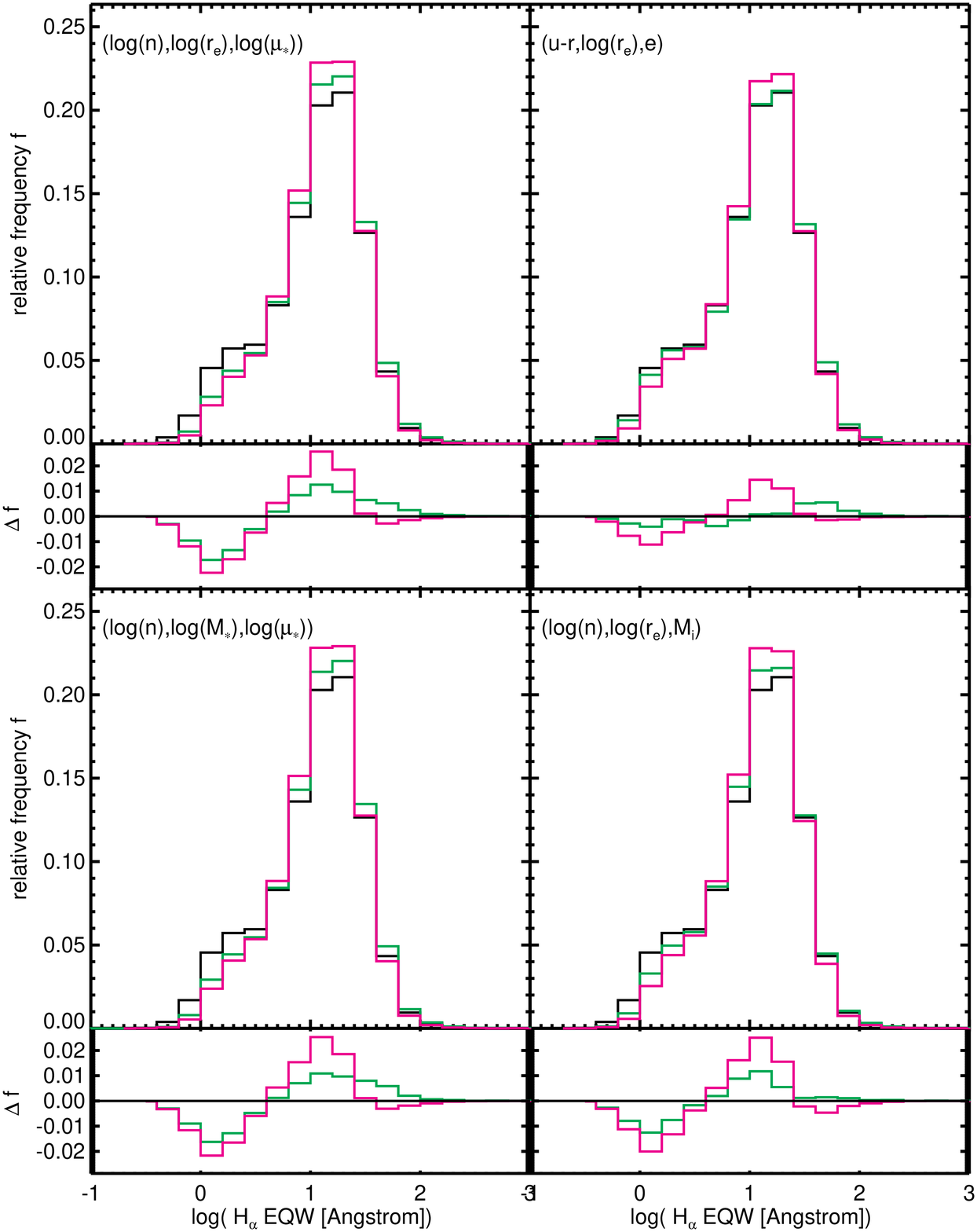}
\caption{Distribution of H$\alpha$ EQW for galaxies in the \textit{OPTICALsample} classified as spirals by GALAXY ZOO (black), and the 
parameter combination listed top left (green). The  H$\alpha$ EQW distribution of galaxies with $P_{\mathrm{CS,DB \ge 0.7}}$ located in cells 
associated with spiral galaxies is shown in magenta. The inset panel below each distribution shows the distribution of the difference in 
relative frequency for each bin in H$\alpha$ EQW relative to that of the GALAXY ZOO classifications.}
\label{fig_haeqwOPT}
\end{figure}

\begin{figure}
\includegraphics[width=0.5\textwidth]{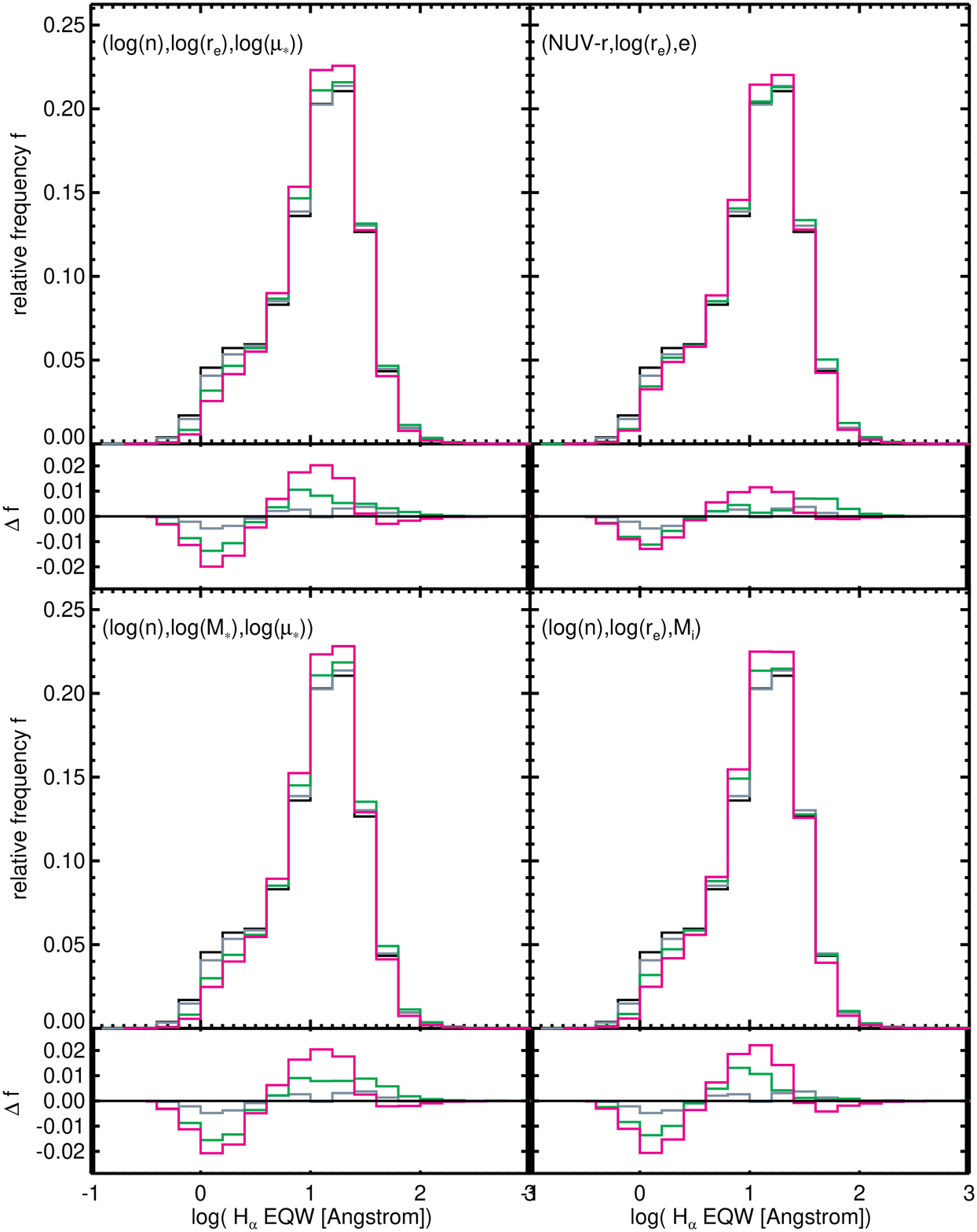}
\caption{As Fig.~\ref{fig_haeqwOPT} but for galaxies in the \textit{NUVsample}.}
\label{fig_haeqwNUV}
\end{figure}

\paragraph*{H$\alpha$ Equivalent Width Distribution of the \textit{Opticalsample} \& \textit{NUVsample}}$\,$\newline   
Bearing this mild simultaneous bias in mind, we consider the distributions of H$\alpha$ EQW for parameter combinations as applied to the 
\textit{OPTICALsample} and the\textit{NUVsample}, shown in Figs.~\ref{fig_haeqwOPT} \& \ref{fig_haeqwNUV}, respectively.\newline
The samples selected by the same parameter combinations as previously applied to the \textit{NAIRsample} display a bias against low values of 
H$\alpha$ EQW when applied to the \textit{OPTICALsample}, similar to that observed for their application to the \textit{NAIRsample}. Overall, all 
the considered parameter combinations recover the peak in the H$\alpha$ EQW corresponding to star-forming galaxies well, with high values of 
H$\alpha$ EQW being only minimally favored with respect to the GALAXY ZOO defined sample. However, all selections display a bias against 
very low values of H$\alpha$ EQW, least so for the combination ($u-r$,log($r_e$,$e$). The general trends in the distributions of H$\alpha$ 
EQW appear very similar to those identified for the selections applied to the \textit{NAIRsample}, hence we expect that the selections applied to 
the \textit{OPTICALsample} will also exhibit a similar bias towards later type spirals.\newline
It is important to note the very good agreement between the H$\alpha$ EQW distributions of all reliable spirals in the \textit{OPTICALsample} 
(black) and \textit{NUVsample} (gray) shown in the panels of Fig.~\ref{fig_haeqwNUV}. This indicates that the NUV preselection itself is not 
introducing a strong bias. Nevertheless, NUV preselection does appear to lead to a slight bias against systems with low H$\alpha$ EQW, favoring 
high H$\alpha$ EQW systems.\newline
As for the \textit{OPTICALsample} the selections applied to the \textit{NUVsample} display a bias against low values of H$\alpha$ EQW, 
although the bias is reduced under NUV preselection. However, the parameter combinations are slightly more weighted towards high values of 
H$\alpha$ EQW than for the \textit{OPTICALsample}. Overall, the trends in the H$\alpha$ EQW distributions are similar to those observed in the 
selections drawn from the \textit{OPTICALsample}, the \textit{NAIRsample}, and the  \textit{NUVNAIRsample}. Accordingly, we expect that the 
parameter based selections will be, to some extent, biased against early type spirals. 

\subsubsection{Redshift dependence of the spiral fraction}\label{zdep}
\begin{figure}
\includegraphics[width=0.5\textwidth]{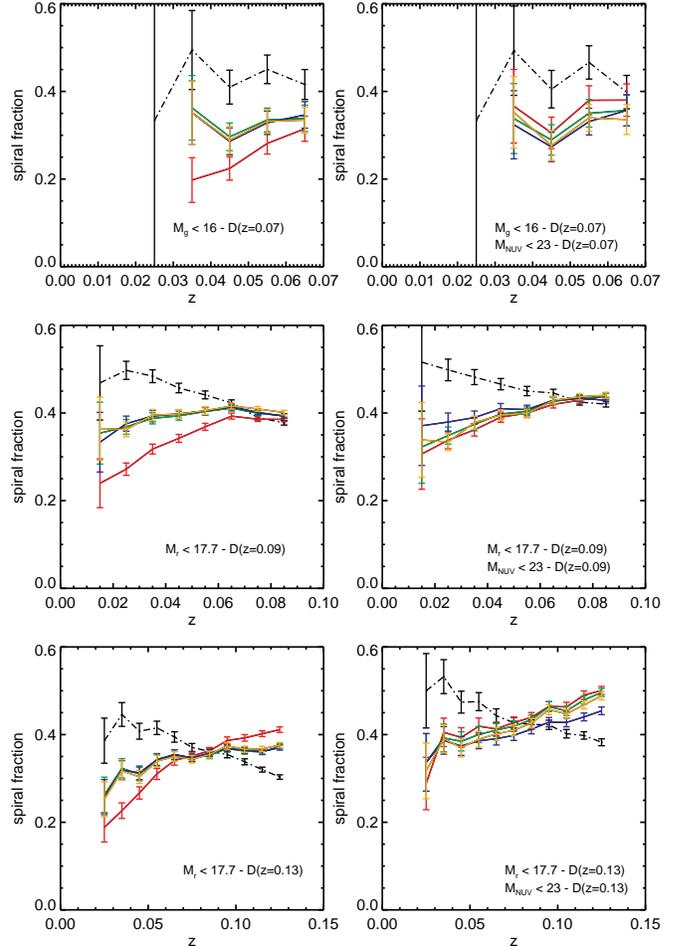}
\caption{Spiral fraction as a function of redshift $z$ in bins of width 0.01 for selections defined using ($u-r$,log($r_e$),$e$) resp. ($NUV-
r$,log($r_e$),$e$) (red) ,(log($n$),log($r_e$),log($\mu_*$))  (green), (log($n$),log($r_e$),$M_i$) (blue), and (log($n$,log($M_*$),log($\mu_*$)) 
(orange), respectively. The top left panel shows the results for the combinations applied to a volume-limited subsample of the 
\textit{NAIRsample} (the selection criteria are indicated in each panel). The redshift dependence of the spiral fraction defined by the 
classifications of \citet{NAIR2010} in the considered subsample is shown black as a dash-dotted line. Error bars indicate Poisson 1-$\sigma$ 
uncertainties. The top right panel shows the same, but applied to a subsample of the \textit{NUVNAIRsample} as defined in the panel. The 
middle and bottom left panels show the redshift dependence of the spiral fraction for the selection applied to two volume-limited subsamples of 
the \textit{OPTICALsample} with the GALAXY ZOO defined reliable spiral fraction shown as a black dash-dotted line. The middle and bottom 
right panels show the same for the \textit{NUVsample}.}
\label{fig_zfracs} 
\end{figure}

A final avenue of possible bias we address here, is the dependence of the performance of the selection on the distance/redshift of the sources. 
This is of particular interest, as the parameters with the best performance are largely structural or related parameters, e.g. log($n$), log($r_e$), 
log($\mu_*$), and as
such may depend on the resolution of the images in terms of physical sizes. \newline
Over the time span corresponding to the redshift range of $z=0-0.13$ we do not expect the distribution of galaxy morphologies to evolve in a 
significant manner \citep[e.g.][]{BAMFORD2009}, hence the fraction of spirals should be approximately constant. However, as massive bright 
galaxies are less likely to be spirals than less massive, fainter galaxies, this will only be the case for volume-limited samples. In 
Fig.~\ref{fig_zfracs} we show the fraction of galaxies classified as spirals by the parameter combinations ($u-r$,log($r_e$),$e$), resp. ($NUV-
r$,log($r_e$),$e$) in the case of NUV-preselection, (log($n$),log($r_e$),log($\mu_*$)),  (log($n$),log($r_e$),$M_i$), and  
(log($n$),log($M_*$),log($\mu_*$))  for different volume-limited samples of galaxies. At top left we show the spiral fractions as a function of 
$z$ for a volume-limited subsample of the \textit{NAIRsample} extending to $z=0.07$ (i.e. $M_g < 16 - D(z=0.07)$, where $D(z)$ is the 
distance module and $M_g$ is the absolute magnitude in the $g$ band). We find that the spiral selections recovered by the parameter 
combinations (with the exception of ($u-r$,log($r_e$),$e$)) are flat in $z$, and are in good agreement with the $z$ dependence of the spiral 
selection for this sample defined by the visual classifications of \citet{NAIR2010} (black dash-dotted line). The middle left panel shows that the 
distribution of spirals selected from a volume-limited subsample of the \textit{OPTICALsample} extending to $z=0.09$ (i.e. $M_r < 17.7 - 
D(z=0.09)$, thus extending to fainter galaxies) is also largely flat in $z$ for the selections not using colour as a parameter, while the bottom 
left panel shows a similar result for a volume-limited subsample of the \textit{OPTICALsample} extending to $z=0.13$ (i.e. $M_r < 17.7 - 
D(z=0.13)$, covering the full considered range in $z$). In the latter two panels, the dash-dotted black line indicates the $z$ dependence of the 
spiral fraction as defined by the GALAXY ZOO visual classifications. The decline in the spiral fraction is largely due to the certainty of the 
classifications decreasing with increasing $z$. If the assumption of a constant spiral faction as a function of $z$ is valid, these results may be 
seen to imply that for marginally resolved sources, the automatic cell-based non-parametric classification schemes may be superior to the 
GALAXY ZOO DR1 classifications.\newline

The right hand panels of Fig.~\ref{fig_zfracs} show the results of applying the parameter combinations to NUV preselected samples, taking into 
account the UV sensitivity limits (i.e. with the additional requirement on the samples that $M_{NUV} < 23 - D(z_{sel})$, where $z_{sel}$ is the 
limiting redshift of the sample). For volume-limited subsample of the \textit{NUVNAIRsample} we find, as for the \textit{NAIRsample}, that the 
spiral fraction is flat in $z$. For the other volume-limited samples, although the selections are largely flat in $z$, there is nevertheless an 
increase with increasing redshift.
Notably, the spiral fraction of selections which only depend on parameters determined at long wavelengths (e.g. (log($n$),log($r_e$),$M_i$)), 
and which have spiral distributions which are flat in $z$ without the requirement of an NUV detection, also display an increase of the spiral 
fraction with $z$ under NUV preselection. This can most readily be understood in the context of an evolution in the UV properties of the 
volume-limited samples of spirals considered, with an increasing fraction of spiral galaxies with NUV emission as a function of increasing 
redshift $z$. Such a scenario is consistent with the observed decline in star-formation rate density from $z~1-0$ \citep[e.g.][]{HOPKINS2008} 
and the increase in the population of quiescent galaxies in the mass range $M_* \gtrsim 10^{10} M_{\odot}$ over this redshift range \citep[]
[and references therein]{MOUSTAKAS2013}. The volume-limited samples considered will be dominated by galaxies in this mass range and be 
accordingly sensitive to such evolutionary effects.\newline

We note that as the redshift range spans over a Gyr in lookback time, some evolution in the spiral fraction may be expected linked to a slight 
decline in the fraction of spirals with decreasing $z$, i.e. we do not expect a perfectly constant fraction of spirals. Nevertheless, the lack of any 
major dependence on the spiral fraction as a function of redshift, implies that no major redshift dependent biases are introduced into the 
selection when using combinations of three parameters with the non-parametric cell-based method, and that the method may even prove to be 
more reliable than visual classifications.\newline

\section{Comparison with other proxies}\label{comparison}
Using the cell-based method presented in Sect.~\ref{method} we have identified combinations of parameters including log($r_e$), 
log($\mu_*$), log($n$), log($M_*$), and $M_i$, in particular (log($n$),log($r_e$),log($\mu_*$)), (log($n$),log($r_e$),$M_i$), and 
(log($n$),log($M_*$),log($\mu_*$)), to result in simultaneously pure and complete samples of spirals. These selections appear to be robust 
against redshift dependent biases, and to be largely unbiased in their parameter distributions, only displaying a slight bias against early type 
spirals. Accordingly, the cell-based method using these combinations appears well suited to selecting samples of spiral galaxies. In the 
following we investigate the contribution of the cell-based method to the demonstrable success, and compare its performance to a selection of 
widely used morphological proxies, as well as to a novel algorithmic approach based on support vector machines 
\citep{HUERTAS-COMPANY2011}.\newline 

\subsection{The Importance of the Cell-based Method}
While the use of the parameter combinations in concert with the cell-based method presented in sect.~\ref{method} can lead to simultaneously 
pure and complete samples of spiral galaxies, the use of the cell-based method requires a training sample, ideally of $\gtrsim30$k galaxies (cf. 
Fig.~\ref{fig_SAMPSIZE}) In contrast to this, the advantage of simple hard cuts on parameters is that they require no (or much smaller) such 
calibration samples. In our investigations we have made use of a suite of parameters including ones traditionally used in the morphological 
classification of spirals (e.g. $n$), as well as novel parameters such as $\mu_*$. In order to investigate to what extent the demonstrable success 
is due to the parameters used, and what the effect of the cell-based algorithm is, we have applied the combinations ($u-r$,log($r_e$),$e$), 
(log($n$),log($r_e$),log($\mu_*$)), (log($n$),log($r_e$),$M_i$), and (log($n$),log($M_*$),log($\mu_*$)) to the \textit{OPTICALsample} and the 
\textit{NAIRsample} using fixed boundaries derived by eye from the parameter distributions shown in Fig.~\ref{fig_pardistopt}. In this context 
we have chosen to treat galaxies with $u-r \le 2.1$, $\mathrm{log}(r_e)\le 0.65$, $e \ge 0.3$, $\mathrm{log}(n) \le 0.4$, $\mathrm{log}
(\mu_*) \le 8.3$, $\mathrm{log}(M_*) \le 10.7$, and $M_i \ge -22$ as spirals. The results tabulated in  Table~\ref{tab_fixedbounds} show that 
the bijective discrimination power of the selections using fixed boundaries is much lower than when the same parameter combinations are used 
with the cell-based method. It is clear that the use of fixed boundaries entails a strong trade-off between purity and completeness. Although the 
parameter combinations ($u-r$,log($r_e$),$e$), (log($n$),log($r_e$),log($\mu_*$)), and (log($n$),log($r_e$),$M_i$) all attain high values of 
purity (even $\sim 0.05$ greater than with the cell based method), they, however, are highly incomplete, with completeness values 
$\sim0.2-0.3$ less than attained with the cell-based method. The parameter combination (log($n$),log($M_*$),log($\mu_*$)), on the other 
hand, attains a completeness only $\sim0.07$ less than the cell-based method, but with the purity of the selection reduced by $\sim0.1$. The 
high values of completeness, attained simultaneously to the high values of purity when making use of the parameter combinations together with 
the cell-based method, thus appear largely due to the flexibility of the boundaries given by the cell-based method. \newline

\begin{table*}
 \centering
	\caption{Purity, completeness, bijective discrimination power, and contamination  for the combinations ($u-r$,log($r_e$),$e$), 
	(log($n$),log($r_e$),log($\mu_*$)), (log($n$),log($r_e$),$M_i$), and (log($n$),log($M_*$),log($\mu_*$)) using fixed boundaries, applied to 
	the \textit{OPTICALsample} (columns 2-5) and the \textit{NAIRsample} using the GALAXY ZOO visual classifications (columns 6-9) as well as 
	the independent classifications of \citet[][columns 10-12]{NAIR2010}.  }
  \begin{tabular}{@{}lccccccccccc@{}}
  \hline
  \multicolumn{1}{c}{} &\multicolumn{4}{c}{\textit{OPTICALsample}}    &     \multicolumn{7}{c}{\textit{NAIRsample}} \\
 \multicolumn{1}{c}{} &\multicolumn{4}{c}{}    &  \multicolumn{4}{c}{GALAXY ZOO} &   \multicolumn{3}{c}{NAIR \& Abraham 2010} \\  
 Parameter combination & $P_{\mathrm{pure}}$ & $P_{\mathrm{comp}}$ & $P_{\mathrm{bij}}$ & $P_{\mathrm{cont}}$ & $P_{\mathrm{pure}}$ & 
 $P_{\mathrm{comp}}$ & $P_{\mathrm{bij}}$ & $P_{\mathrm{cont}}$ & $P_{\mathrm{pure}}$ & $P_{\mathrm{comp}}$ & $P_{\mathrm{bij}}$\\
 \hline
($u-r$,log($r_e$),$e$) & 0.793 & 0.398 & 0.316 & 0.015 & 0.911 & 0.257 & 0.234 & 0.006 & 0.961 & 0.236 & 0.227 \\
(log($n$),log($r_e$),log($\mu_*$)) & 0.794 & 0.567 & 0.450 & 0.006 & 0.934 & 0.487 & 0.455 & 0.007 & 0.976 & 0.442 & 0.431 \\
(log($n$),log($r_e$),$M_i$)) & 0.782 & 0.507 & 0.396 & 0.007 & 0.922 & 0.372 & 0.343 & 0.013 & 0.965 & 0.339 & 0.327 \\
(log($n$),log($M_*$),log($\mu_*$)) & 0.654 & 0.700 & 0.458 & 0.028 & 0.861 & 0.573 & 0.493 & 0.023 & 0.946 & 0.547 & 0.517 \\
\hline                      
\end{tabular}
  \label{tab_fixedbounds}
\end{table*}

\subsection{Comparison with widely used proxies}
Having identified the cell-based method used with combinations of three parameters including log($r_e$), log($\mu_*$), log($n$), log($M_*$), 
and $M_i$, in particular (log($n$),log($r_e$),log($\mu_*$)), (log($n$),log($r_e$),$M_i$), and (log($n$),log($M_*$),log($\mu_*$)), as a method 
to select simultaneously pure and complete samples of spirals we compare its performance to that of a selection of widely used morphological 
proxies, as well as to that of a novel algorithmic approach based on support vector machines \citep{HUERTAS-COMPANY2011}.\newline

Two well-known proxies for the general morphological type of a galaxy are the concentration index in the $r$ band, defined as $C_r = 
\frac{R_{90,r}}{R_{50,r}}$ where $R_{90,r}$ and
$R_{50,r}$ are the radii within which 90 resp. 50 per cent of the galaxy's (petrosian) flux are contained, and the S\'ersic index $n$,
i.e., the index obtained for the best fit of a S\'ersic profile \citep{SERSIC1968} to the galaxy's light distribution. \citet{STRATEVA2001} suggest 
the use of the
concentration index as a proxy for morphological classification with galaxies with $C_r < 2.6$ considered to be late-types/spirals, while 
\citet{BARDEN2005} suggest that galaxies with $n < 2.5$ can be considered to be late-types/spirals.\newline 
Alternatively, \citet{BALDRY2004} have suggested a separation into blue and red galaxies which they equate to late- and early-types, based on a 
galaxies position in the $u-r$ colour vs. absolute $r$ magnitude diagram, with the separator parameterized by a combination of a constant and 
a tanh function dependent on the absolute $r$ band magnitude (their Eq.~11). \newline
A different approach, also making use of two parameters, has been adopted by \citet{TEMPEL2011}. They define a subvolume in the two
dimensional space spanned by the SDSS parameters $f_{deV}$(i.e., the fraction of a galaxy's flux which is fit by the 
de Vaucouleurs profile \citep{DEVAUCOULEURS1948} in the best fit linear combination of a de Vaucouleurs and an exponential profile) and 
$q_{exp}$
(the axis ratio of the SDSS best fit exponential profile) associated with spiral galaxies and calibrated on visual classifications of SDSS galaxies in
the Sloan Great Wall region \citep{EINASTO2010} and GALAXY ZOO.\newline
Recently \citet{HUERTAS-COMPANY2011} have published a catalogue of morphological classifications of SDSS DR7 spectroscopic galaxies
based on support vector machines, which compare well with GALAXY ZOO classifications of the same sample. Similarly to GALAXY ZOO
\citet{HUERTAS-COMPANY2011} assign probabilities to the possible galaxy classes, so that for the purposes of our comparison we have
chosen to treat objects with a probability greater than 70 per cent of being a spiral as a spiral, analogously to our treatment the 
GALAXY ZOO sample\footnote{\citet{HUERTAS-COMPANY2011} provide probabilistic morphological classifications for all but 311 of the sources 
in our sample}.\newline

\begin{figure}
\includegraphics[width=0.5\textwidth]{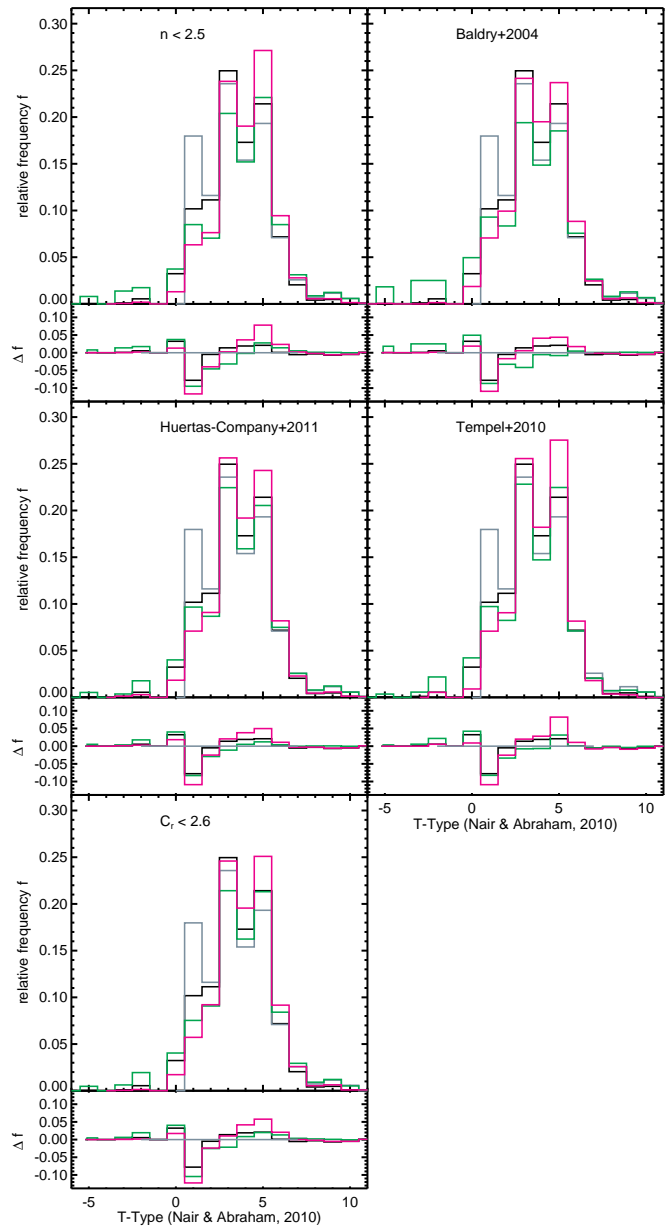}
\caption{T-type distributions of the discussed selection methods applied to the \textit{NAIRsample} indicated top left in each panel. The 
distribution of GALAXY ZOO spirals with $P_{CS,DB} > 0.7$ is shown in black. The distribution of sources selected by the method indicated is 
shown in green, while the the distribution of sources selected by the cell-based method with $P_{CS,DB} > 0.7$ is shown in magenta. The inset panel below each distribution shows the distribution of the difference in relative frequency for galaxy type relative to that of the 
\citet{NAIR2010} classifications.}  
\label{fig_Ttproxycomp}
\end{figure}

Table~\ref{tab_Pproxies} shows the purity, completeness, and bijective discrimination power for the five morphological proxies discussed above 
as well as the three parameter combinations applied to the \textit{OPTICALsample} and the \textit{NAIRsample}. All morphological proxies, with 
the exception of that proposed by \citet{TEMPEL2011}, attain values of completeness similar to, or larger than, that of the cell based method 
when applied to the \textit{OPTICALsample}, although only the classification of \citet{HUERTAS-COMPANY2011} achieves a completeness 
notably exceeding that of the cell-based method ($P_{\mathrm{comp}} = 0.903$). However, these proxies fail to attain samples with a purity 
greater than 60 per cent when applied to the \textit{OPTICALsample}, much lower than the value of $\approx 75$ per cent achieved by the cell-
based method, the exception again being the method of \citet{TEMPEL2011}. As a result, the bijective discrimination power of these 
selections is lower than that achieved by the optimal combinations of three parameters, using the cell-based method, with only the method of 
\citet{HUERTAS-COMPANY2011} attaining a comparable value of $P_{\mathrm{bij}}$. However, the contamination by ellipticals introduced by the  
proxies considered is at least a factor 3 greater than that resulting from the cell based method.\newline
Applied to the brighter \textit{NAIRsample} the purity of the considered proxies increases notably, while the completeness slightly decreases. 
The purity of the selections resulting from the use of the considered proxies remains significantly lower than that achieved by the parameter 
combinations, both when using the GALAXY ZOO visual classifications as well as those of \citet{NAIR2010}, as can also be seen in the 
distributions of the T-types in the samples selected by the considered proxies (Fig.~\ref{fig_Ttproxycomp}). The completeness, on the other 
hand, is greater than for the parameter based selections, so that the bijective discrimination power of the considered proxies is comparable to 
that of the parameter based selections when applied to the \textit{NAIRsample}.\newline

\begin{figure}
\includegraphics[width=0.5\textwidth]{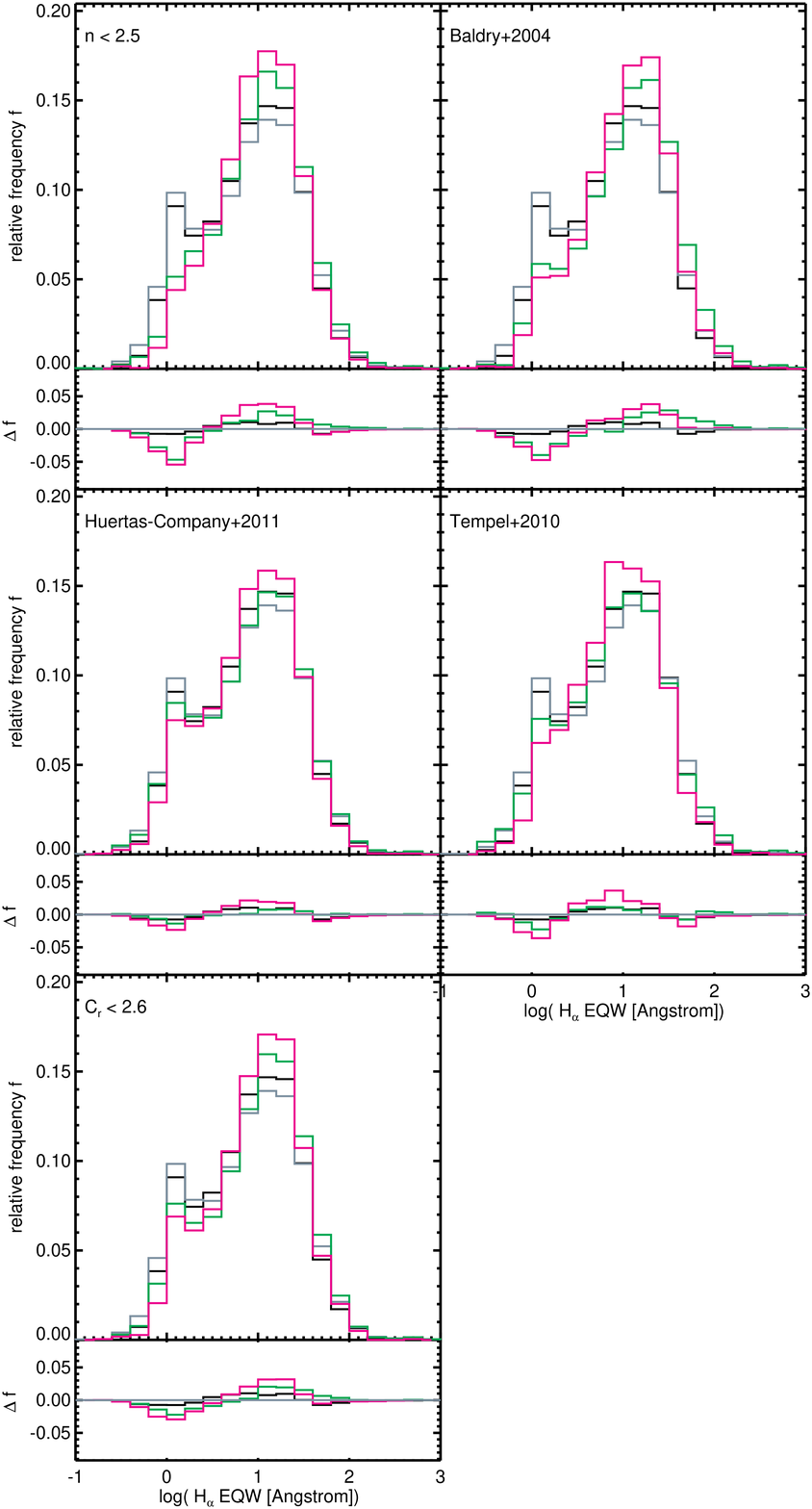}
\caption{ $H_{\alpha}$ EQW distributions of the discussed selection methods indicated top left in each panel applied to the \textit{NAIRsample}. 
The distribution of GALAXY ZOO spirals with $P_{CS,DB} > 0.7$ is shown in black. The distribution of sources selected by the method indicated 
is shown in green, while the the distribution of sources selected by the method with $P_{CS,DB} > 0.7$ is shown in magenta. The inset 
panel below each distribution shows the distribution of the difference in relative frequency for each bin in H$\alpha$ EQW relative to that of the 
\citet{NAIR2010} classifications.}  
\label{fig_HAUNAproxycomp}
\end{figure}

\begin{figure}
\includegraphics[width=0.5\textwidth]{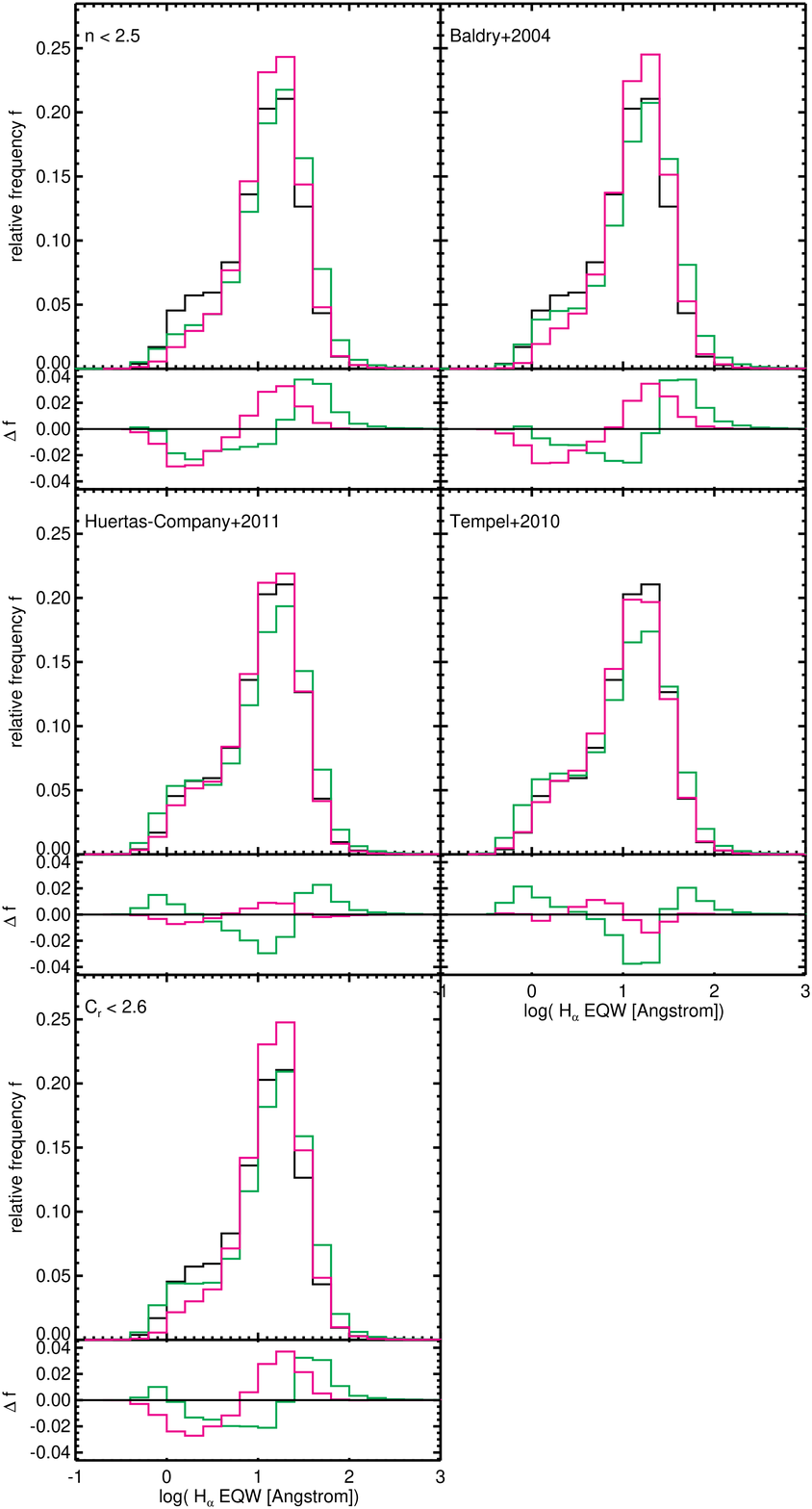}
\caption{ $H_{\alpha}$ EQW distributions of the discussed selection methods indicated top left in each panel applied to the 
\textit{OPTICALsample}. The distribution of GALAXY ZOO spirals with $P_{CS,DB} > 0.7$ is shown in black. The distribution of sources selected 
by the method indicated is shown in green, while the the distribution of sources selected by the method with $P_{CS,DB} > 0.7$ is shown in 
magenta. The inset panel below each distribution shows the distribution of the difference in relative frequency for each bin in 
H$\alpha$ EQW relative to that of the GALAXY ZOO classifications.}  
\label{fig_HAproxycomp}
\end{figure}

As can be seen in Fig.~\ref{fig_Ttproxycomp}, the T-type distributions of the considered proxies display a bias towards later type spirals 
very similar to that of the cell-based selections. However, the bias against Sa and Sa/b galaxies appears to be slightly less pronounced, 
with the relative frequency of early type galaxies being marginally higher for the samples recovered by the proxies than by the cell-based 
selections. On the other hand, the T-type distributions in Fig.~\ref{fig_Ttproxycomp} also show the considerably larger contamination by 
ellipticals not present in the cell based selections.\newline
Considering the distributions of H$\alpha$ EQW for the samples obtained by these proxies applied to the \textit{NAIRsample} as shown 
in Fig.~\ref{fig_HAUNAproxycomp} one finds that the samples recovered by the proxies (with the exception of the methods of 
\citealt{HUERTAS-COMPANY2011} and \citealt{TEMPEL2011}) display a bias towards sources with large values of  H$\alpha$ EQW, considerably 
more so than the 
cell-based selections, with $\sim 10$\% more of the sample consisting of high H$\alpha$ EQW sources than in the samples recovered by the 
cell-based method. This result is most pronounced for the samples selected by the concentration index, the S\'ersic index and the method of 
\citet{BALDRY2004}. Similar but more pronounced results are obtained if one considers the distributions of H$\alpha$ EQW for the samples 
obtained by these proxies applied to the \textit{OPTICALsample}, as shown in Fig.~\ref{fig_HAproxycomp}. In contrast, the selections based on 
the method of \citet{TEMPEL2011} and \citet{HUERTAS-COMPANY2011} appear to be weighted more towards high \textit{and} low values of 
H$\alpha$ EQW than the GALAXY ZOO reference and the selections based on the parameter combinations used in concert with the cell-based 
method.\newline     
Overall, we find that the selections resulting from the proxies are similar to, or more biased than, the selections based on the cell-based 
method, and are clearly more contaminated.\newline

In conclusion, we thus find that for the purpose of selecting a pure, yet nevertheless largely complete, sample of spiral galaxies, not limited to 
the brightest galaxies, the use of the cell-based method presented in combination with one of the optimal parameter combinations is preferable 
over the investigated well-established proxies, and at least comparable to the sophisticated approach of \citet{HUERTAS-COMPANY2011}.
\newline

\begin{table*}
 \centering
	\caption{Purity, completeness, bijective discrimination power, and contamination  for other widely used morphological proxies, applied to the 
	\textit{OPTICALsample} (columns 2-5) and the \textit{NAIRsample} using the GALAXY ZOO visual classifications (columns 6-9) as well as the 
	independent classifications of \citet[][columns 10-12]{NAIR2010}. The values attained by the combinations 
	(log($n$),log($r_e$),log($\mu_*$)), (log($n$),log($r_e$),$M_i$), and (log($n$),log($M_*$),log($\mu_*$)) are shown for comparison. }
  \begin{tabular}{@{}lccccccccccc@{}}
  \hline
  \multicolumn{1}{c}{} &\multicolumn{4}{c}{\textit{OPTICALsample}}    &     \multicolumn{7}{c}{\textit{NAIRsample}} \\
 \multicolumn{1}{c}{} &\multicolumn{4}{c}{}    &  \multicolumn{4}{c}{GALAXY ZOO} &   \multicolumn{3}{c}{NAIR \& Abraham 2010} \\  
 Method & $P_{\mathrm{pure}}$ & $P_{\mathrm{comp}}$ & $P_{\mathrm{bij}}$ & $P_{\mathrm{cont}}$ & $P_{\mathrm{pure}}$ & 
 $P_{\mathrm{comp}}$ & $P_{\mathrm{bij}}$ & $P_{\mathrm{cont}}$ & $P_{\mathrm{pure}}$ & $P_{\mathrm{comp}}$ & $P_{\mathrm{bij}}$\\
 \hline
 (log($n$), log($r_e$), log($\mu_*$))  & 0.739 & 0.774 & 0.572 & 0.017  & 0.884 & 0.712 &  0.629 & 0.024 & 0.945 & 0.660 & 0.624\\ 
(log($n$), log($r_e$), $M_i$)  & 0.740 & 0.779 & 0.576 & 0.021 & 0.879 & 0.706 & 0.621 & 0.032 & 0.935 & 0.652 & 0.610\\
(log($n$), log($M_*$), log($\mu_*$))  & 0.731 & 0.773 & 0.565 & 0.019  & 0.885 & 0.707 & 0.626 & 0.024 & 0.946 & 0.657 & 0.621\\
Huertas-Company et al., 2011 & 0.588 & 0.903 & 0.531 & 0.077 & 0.806 & 0.836 & 0.673 & 0.054 & 0.898 & 0.802 & 0.720 \\
Baldry et al., 2004 & 0.522 & 0.802 & 0.419 & 0.081 & 0.745 & 0.747 & 0.557 & 0.115 & 0.834 & 0.721 & 0.601 \\
Tempel et al., 2011 & 0.648 & 0.411 & 0.266 & 0.078 & 0.786 & 0.387 & 0.304 & 0.064 & 0.896 & 0.380 & 0.340\\
$n < 2.5$ & 0.575 & 0.805 & 0.463  & 0.105 & 0.780 & 0.732 & 0.571 & 0.079 & 0.875 & 0.707 & 0.619\\
$C_r < 2.6$ & 0.547 & 0.762 & 0.417 & 0.105 & 0.810 & 0.750 & 0.608 & 0.066 & 0.896 & 0.715 & 0.641 \\ 
\hline                      
\end{tabular}
  \label{tab_Pproxies}
\end{table*}

\section{Discussion}\label{discussion}

\subsection{Choosing a parameter combination}\label{choice}
Using the non-parametric cell-based method presented, we have successfully identified several combinations of three parameters which allow 
for an efficient and rapid selection of pure and simultaneously complete, largely unbiased samples of spiral galaxies. When applied to parent 
samples not limited to the brightest galaxies, these are superior in performance, in terms of bijective discrimination power and bias (e.g. in 
H$\alpha$ EQW), to the widely established simple morphological proxies investigated, such as the concentration index $C_r$,  the S\'ersic 
index $n$, and the division into red and blue galaxies. Furthermore, they are at least comparable in performance to the algorithmic approach 
using SVMs of \citet{HUERTAS-COMPANY2011}. \newline
However, depending upon the effort required to obtain a given parameter, either in terms of data processing or acquisition, the `cost' of 
parameters, and hence of parameter combinations, will vary. For example, a parameter combination including only quantities such as $r_e$, 
$M_i$, $u-r$, and $e$ which can, at least for reasonably resolved sources, often be measured directly by \verb SExtractor  \citep{BERTIN1996} 
is `cheaper'  than a combination involving parameters which require additional data reduction such as fitting S\'ersic profiles using., e.g. 
\texttt{GIM2D} \citep{SIMARD2002} or \texttt{GALFIT}  \citep{PENG2002} \footnote{Where high resolution imaging is available these codes 
themselves 
present a different method of automatic morphological classification, as they can perform multiple component fits which can be used to 
determine the morphological type of a galaxy. However, the requirements on resolution are severe and fitting multiple components is often not 
justified \citep{SIMARD2011}.}. Similarly the relative `cost' of additional NUV data is much higher than that of relying solely on optical pass-
bands, as it involves the use of additional observational facilities.\newline
Encouragingly, we find that various parameter combinations perform similarly well, allowing for a choice of parameter combination informed by 
both the envisioned science application as well as the relative `expense' of the parameters used.\newline
Overall, the most important parameters in selecting a sample of spiral galaxies are the effective radius log($r_e$), the stellar mass surface 
density log($\mu_*$), and the S\'ersic index log($n$). These parameters perform especially well in combination with the stellar mass or a tracer 
thereof (e.g $M_i$).
We find the combinations (log($n$),log($r_e$),log($\mu_*$)), (log($n$),log($r_e$), $M_i$), and (log($n$),log($M_*$),log($\mu_*$)) to be those 
with the greatest bijective discrimination power when applied to the \textit{OPTICALsample}. These are also amongst the most powerful under 
NUV preselection, although the combination ($NUV-r$,log($r_e$),$M_i$) is comparably powerful. In the latter case, however, the selection 
appears to be driven by the parameters $M_i$ and, in particular, log($r_e$). In terms of relative `expense' the combinations requiring NUV pre-
selection are more `expensive' than those applicable to the whole sample. Although the best-performing combinations all require S\'ersic 
profiles to be fit, the cost is strongly ameliorated by the fact that only single S\'ersic profiles are required.\newline
Unsurprisingly, the ellipticity $e$ proves to be an effective parameter, as only spirals seen edge-on appear strongly elliptical. In this sense, it 
even counters the bias against edge-on spirals, which can be introduced by using UV-optical colours as selection parameters, as dusty edge-on 
spirals may drop out of a colour selection due to attenuation of their UV emission. However, selections using $e$ as a parameter are strongly 
biased against any spirals seen approximately face-on, respectively \textit{not} edge-on. Thus, while the observed ellipticity represents a 
powerful criterion for selecting a pure sample of spirals and has a low relative cost, it leads to generally less complete samples, which are 
strongly biased towards edge-on systems.\newline
Although our results indicate that simple structural parameters derived at longer wavelengths are efficient at selecting spirals, the combinations 
($NUV-r$,log($r_e$),$M_i$), and to a lesser extent ($u-r$,log($n$),log($r_e$)), indicate that UV/optical colours linked to younger stellar 
populations do provide valuable information for selecting spiral galaxies. As mentioned above, however, use of UV-optical colour as a 
parameter can lead to biases in the selection. Dust in spirals will cause galaxies seen edge-on to appear very red, hence, the use of a 
UV-optical colour can bias the selection against these systems. Furthermore UV-optical colour selection can introduce a bias 
against any spirals which appear intrinsically red due to lack of star formation. This is the case both for the $u-r$ and $NUV-r$ colours. Finally, 
when using a colour as a parameter (in particular a UV colour) the possibility of different depths of photometry must be accounted for, i.e., the 
photometry in both bands must be deep enough to ensure that the entire range of colour normally attributed to the galaxy population is covered 
over the entire redshift range of the sample. Failure to do so will give rise to both additional incompleteness, as well as a colour bias in the 
resulting sample.\newline
Depending on the science application for which the sample is intended, and on the availability of data different combinations may be optimal in 
selecting spiral galaxies. For example, using the combination (log($n$), log($r_e$), $M_i$) would be appropriate to obtain a selection of spiral 
galaxies for a project aiming at investigating the total star formation rates of a large sample of spiral galaxies as derived from the UV. Such a 
selection would avoid a bias against quiescent systems, as would be introduced by using a NUV preselection or a UV-optical colour,
 while also guarding against any orientation biases which could arise if $e$ was used as a selection parameter. Accordingly such a sample would 
 be largely unbiased with respect to star formation characteristics. Another suitable combination for such an application would be 
 (log($n$), log($r_e$), log($\mu_*$)), which is also largely independent of UV-optical colours\footnote{The stellar mass estimate used in 
 deriving $\mu_*$ does depend on an optical color, i.e the $g-i$ colour, however, this colour is linked mainly to intermediate age and old 
 stellar populations. Given photometry of sufficient depth, the $g-i$ colour does not present a direct selection criterion but is only used in 
 calculating the stellar mass, such that $M_*$ and $\mu_*$ can be considered unbiased in terms of star-formation properties. Further more the 
 stellar mass $M_*$ derived in this manner is largely independent of dust attenuation \citep{BELL2001,NICOL2011,TAYLOR2011}.}\newline   
Conversely, however, a sample which required the greatest achievable purity should include both NUV preselection and $e$ as a parameter.
Thus, the selection can and should be adapted to the science case at hand, although the lack of requirement of UV data allows the method to be 
easily applied to very large samples with minimum requirements on wavelength coverage.\newline

\subsection{The physical basis for optical proxies }
As discussed in Sect.~\ref{choice}, we find that the most important parameters in selecting spirals are the effective radius log($r_e$), the stellar 
mass surface density log($\mu_*$), and the S\'ersic index log($n$) in combination with the stellar mass or a tracer thereof (e.g. $M_i$). In 
addition $e$ leads to very pure if incomplete selections. All these properties are derived in pass-bands normally associated with older stellar 
populations ($g$, $r$, and $i$), rather than with recent star formation. The success achieved by using parameters not obviously directly related 
to the young stellar population is remarkable and implies that the spiral and non-spiral population are more or less distinct in these parameters. 
While the success of $e$ bases on the appearance in projection of spiral galaxies, that of log($r_e$) and log($\mu_*$), on the other hand, 
entails that the radial extent and in particular the ratio of mass to size of the old stellar population is distinctly different in spirals and ellipticals. 
Rotationally supported systems (i.e. spirals)  appear to be significantly more extended than pressure supported systems (i.e 
spheroidals/ellipticals) at a given stellar mass\footnote{This size dichotomy can be boosted further by the presence of dust in the disks, which 
can increase the apparent size of disks relative to the intrinsic size \citep{MOELLENHOFF2006,PASTRAV2013a}}.\newline
This is consistent with the notion that
the stellar populations evolve via distinct evolutionary tracks for disks and spheroids, with the evolution of present day spirals thought to involve 
a smooth infall of gas and inside-out star formation, with merger activity restricted to minor mergers. \newline
In contrast, ellipticals are thought to be the products of major mergers in which angular momentum is redistributed making the central system 
more compact \citep[e.g.,][ and references therein]{BOURNAUD2007}.\newline
In light of our results we emphasize that parameters linked to the old stellar population of galaxies, normally not employed in the classification 
of spirals, may provide valuable information on the morphology of a galaxy. In particular the stellar mass surface density and/or the radial 
extent (together with another parameter, e.g. $M_i$) may be powerful due to the physically motivated characterization parameters.\newline

\subsection{Applicability of the method to other surveys}\label{appotsurv}
We have shown the cell-based method to work well for SDSS galaxies, in particular a subset of the SDSS spectroscopic sample. Hence we expect 
the method to be applicable to samples of similar depth and similar angular resolution, and thus be applicable to upcoming surveys similar to 
SDSS, e.g. SKYMAPPER (The Skymapper Southern Sky Survey; \citealt{KELLER2007}). Many upcoming surveys (DES, VST ATLAS, KiDS, and GAMA 
(Galaxy And Mass Assembly; \citealt{DRIVER2011}), as well as SDSS itself, however, extend to greater photometric depths than the sample used 
here.\newline
To answer the question of how applicable the method is to other, deeper surveys we have used a sample consisting of the 50k $r$-band 
brightest galaxies in the \textit{OPTICALsample} (i.e $m_r < 16.48$) as a calibration sample and have subsequently classified the faintest 50k 
galaxies ($m_r > 17.24$) using the parameter combinations (log($n$),log($r_e$),log($\mu_*$)), (log($n$),log($r_e$),$M_i$), and 
(log($n$),log($M_*$),log($\mu_*$)) . The results are shown in Table~\ref{tab_faint}, where we have included the results obtained using the 
calibration sample employed in sect.~\ref{parameters}, as well as the results obtained using the widely used proxies discussed in 
sect.~\ref{comparison} for comparison. Using the bright subsample to classify the faint subsample we find that the selections are very complete, 
yet appear to be less pure than when classifying the entire \textit{OPTICALsample}. However, this is largely due to a decrease in the certainty of 
the GALAXY ZOO classifications for sources which appear fainter as they predominantly lie at greater redshifts and are smaller and less resolved. 
This is underscored by the very low values of contamination achieved for the different combinations. The performance of the cell-based method 
remains easily superior to that of the simple proxies, achieving much greater purity and similar completeness.These results suggest that galaxy 
samples extending faintwards of the SDSS spectroscopic limit can also be classified using the method presented (cf. also Sect.~\ref{zdep}).
\newline
Penultimately, the increased angular resolution and sensitivity of the upcoming surveys with respect to SDSS may allow the method to be 
extended to sources at higher redshifts than the current very local sample. A somewhat similar approach defining subspaces associated with 
early- and late-type galaxies using 
$U-V$ and $V-J$ restframe colours, calibrated using HST ACS imaging, has recently been proposed by 
\citet{PATEL2012} for galaxies at z$\sim 0.9$. Our proposed method may also be helpful in selecting spirals at higher z, especially selections 
using parameters linked to the older stellar populations, as they would be more robust against the increasing occurrence of bursts of star 
formation at higher redshifts. Such a use of structural parameters would, however, demand imaging with spatial resolution similar to that 
attained by SDSS at low redshifts for large samples of galaxies, which may not be available until Euclid.\newline
Finally we note that ,due to the evolution of structural and photometric parameters, it will, in general, be necessary to recalibrate the method at 
higher z, and for new data sets with very different imaging/photometry (angular resolution/filters). In such a case a subset of the new data with 
visual classifications will be required as a training set for the cell-based method.\newline

At this point we emphasize that the use of the parameter space discretizations supplied in Appendix~\ref{appendCell} depends on the 
compatibility of the parameters with those used in this work. When using the discretizations provided, the reader is advised to check for any 
possible systematic offsets between his/her data and the data used in this work.\newline

\begin{table*}
 \centering
	\caption{Purity, completeness, bijective discrimination power, and contamination  for the combinations (log($n$),log($r_e$),log($\mu_*$)), 
	(log($n$),log($r_e$),$M_i$), and (log($n$),log($M_*$),log($\mu_*$)) and the proxies discussed in sect.~\ref{comparison} applied to the 
	faintest 50k galaxies in the \textit{OPTICALsample}, i.e $m_r > 17.24$. The results are presented for calibrations of the cell based method 
	using the brightest 50k galaxies in the \textit{OPTICALsample} ($m_r < 16.48$), as well as for the calibration sample used in 
	sect.~\ref{parameters}. As no calibration is required for the proxies discussed in sect.~\ref{comparison} the results are only listed once.} 
  \begin{tabular}{@{}lcccccccc@{}}
  \hline
  \multicolumn{1}{c}{} &\multicolumn{4}{c}{bright cal}    &     \multicolumn{4}{c}{all cal} \\
 Method & $P_{\mathrm{pure}}$ & $P_{\mathrm{comp}}$ & $P_{\mathrm{bij}}$ & $P_{\mathrm{cont}}$ & $P_{\mathrm{pure}}$ & 
 $P_{\mathrm{comp}}$ & $P_{\mathrm{bij}}$ & $P_{\mathrm{cont}}$ \\
 \hline
 (log($n$), log($r_e$), log($\mu_*$))  & 0.596 & 0.860 & 0.513 & 0.009 & 0.657 & 0.787 & 0.517 & 0.005 \\ 
(log($n$), log($r_e$), $M_i$)  & 0.607 & 0.861 & 0.523 & 0.009 & 0.664 & 0.799 & 0.530 & 0.006 \\
(log($n$), log($M_*$), log($\mu_*$))  & 0.602 & 0.844 & 0.508 & 0.009 & 0.647 & 0.781 & 0.506 & 0.006 \\
Huertas-Company et al., 2011 & 0.477 & 0.934 &  0.446 & 0.078 \\
Baldry et al., 2004 & 0.434 & 0.825 & 0.358 & 0.098 \\
Tempel et al., 2011 & 0.549 & 0.551 & 0.302 & 0.071 \\
$n < 2.5$ & 0.478 & 0.866 & 0.414 &0.066 \\
$C_r < 2.6$ & 0.432 & 0.808 & 0.349 & 0.112\\
\hline                      
\end{tabular}
  \label{tab_faint}
\end{table*}

\subsection{Applicability of the method to the selection of elliptical galaxies}
The cell-based method presented here could, in principle, be adapted to identifying reliable samples of elliptical galaxies in an analogous 
fashion to that described for the identification of spirals. A certain population of the cells, dependent upon the requirements imposed, will not 
be assignable to either the spiral or the elliptical subvolume and will remain undefined. However, it is by no means clear, that the parameter 
combinations which perform best at selecting a pure and complete population of spirals will do the same for ellipticals. As our focus has been to 
identify a method of reliably selecting spirals,
we do not further discuss the selection of ellipticals. We note, however, that it would be straight-forward to implement and optimize 
such a method. We have also supplied the elliptical fractions and relative errors for the three discretizations supplied in 
appendix~\ref{appendCell}.\newline

\subsection{Application to checks of probabilistic parametric methods}
The use of parametric methods, such as linear discriminant analysis for example, in classifying galaxies is attractive, as these methods are 
capable of assigning a probabilistic classification to the morphology of a galaxy, rather than a binary one such as that presented here, which will 
suffer from contamination due to quantization effects. Furthermore, as also discussed in section~\ref{method}, calibrating the cell based 
method requires substantial samples of galaxies with visual classifications, while the training sets for parametric methods can be smaller. 
However, the applicability of such a parametric method depends on the probability distributions of galaxy properties conforming to the assumed 
parameterization, which may not be the case. Obviously, a strength of the non-parametric method presented in this work is that it removes such 
biases arising from assumptions about the correct parameterization.\newline
We suggest, that the non-parametric method presented here can also be used to investigate the performance of parametric methods. If the 
results of both approaches are in reasonable agreement it may be possible to confidently employ the parametric method to selecting samples, 
relaxing the required size of a putative calibration sample. A further investigation into the performance of multi-parameter morphological 
classifications using linear discriminant analysis and the cell-based method presented here as a comparison will be presented in a companion 
paper (Robotham et al., in prep.).\newline

\section{Application to the stellar mass - specific star formation rate relation for spiral galaxies}\label{application}
As an application of the cell-based technique for selecting spiral galaxies we use it to rederive the empirical scaling relation between the specific 
star-formation rate and the stellar mass (the $\psi_* - M_*$ relation) for this class of objects. Previous derivations of the $\psi_* - M_*$ 
relation have used galaxy samples sensitive to star formation properties in their definition, thus potentially biasing the obtained results. A 
further factor influencing the derivation of the $\psi_* - M_*$ relation is the attenuation of stellar emission from the galaxy due to its dust 
content, which introduces a large component of scatter, as well as potentially of bias, into the relation. Here we capitalize on the selection of a 
relatively pure sample of galaxies of known disk-like geometry, by applying a radiation transfer technique to correct for the attenuation of 
stellar emission by dust, utilizing the geometrical information (effective radii \& axis ratio) of each galaxy. To this end, we utilize the method of 
\citet{GROOTES2013}, who have presented a method to obtain highly accurate radiation transfer based attenuation corrections on an object-by-
object basis, using only broadband optical photometric observables not directly linked to star formation, in particular the stellar mass surface 
density. The method of \citet{GROOTES2013}, however, critically relies on the underlying radiation transfer model of \citet{POPESCU2011} being 
applicable to the galaxies considered, and thus requires a clean sample of galaxies with disk geometry not hosting AGN.\newline

\subsection{The intrinsic $\psi_* - M_*$ relation for morphologically-selected spiral galaxies in the local universe}\label{PSIMSTARSPIRALloc} 
Starting from the \textit{OPTICALsample} we define a sample of spirals using the cell-based method and the parameter combination 
(log($n$),log($r_e$),$M_i$) and impose a redshift limit of $z=0.05$. As shown by \citep[e.g.][]{TAYLOR2011}, the SDSS with a limiting depth of 
$r_{petro,0}=17.77$ is $\gtrsim 80\,$\% complete for $M_* \ge 10^{9.5}M_{\odot}$ to this redshift. The sample considered thus represents a 
volume-limited sample for this mass range. The sample is further limited to objects with an NUV detection as well as those for which there is no 
UV counterpart to the SDSS galaxy in the preliminary GCAT MSC (Seibert et al., 2013 in prep.), excluding ambiguous multiple matches which 
would require flux redistribution. For the sources lacking an NUV counterpart, 3-$\sigma$ upper limits have been calculated. Finally, objects 
defined as AGN following the prescription of \citep{KEWLEY2006} using the ratios of [NII] to H$\alpha$ and [OIII] to H$\beta$ have been 
excluded.\newline
This results in a total of 9885 galaxies, 536 of which have no counterpart in the preliminary GCAT MSC. A visual inspection of a random 
selection of these non-detected sources finds that a large fraction ($\sim50\,$\%) of these non-detections lie in the vicinity of bright stars or at 
the very edge of GALEX tiles, so may actually have an NUV counterpart. In the following, we therefore proceed by considering two samples: i) the 
entire selected sample of spiral galaxies, treating all nondetections as real non-detections, and ii) only the subset of spirals with an NUV 
counterpart, implicitly assuming that all non-detections actually possess an NUV counterpart, and can thus be discarded. 
By comparing the $\psi_* - M_*$ relation for the two samples, we will show that the effect of the NUV non-detections is negligible on the 
derivation of the $\psi_* - M_*$ relation.  
\newline
For all spiral galaxies, we have corrected the observed UV photometry (detections and upper limits) for the effects of attenuation by dust using 
the radiation-transfer based method presented in \citet{GROOTES2013}, and have derived values of $\psi_*$ from the de-attenuated UV 
photometry using the conversion factors given in \citet{KENNICUTT1998}, scaled from a \citet{SALPETER1955} IMF to a \citet{CHABRIER2003} 
IMF as in \citet{TREYER2007} and \citet{SALIM2007}. The required stellar masses have been derived as detailed in Sect.~\ref{data}. Inclinations 
(required for the attenuation corrections alongside the effective radii) have been derived from the observed ellipticity  as $i=\mathrm{arccos}(1-
e)$ and subsequently corrected for the effects of finite disk thickness as detailed in Sect.~3 of \citet{DRIVER2007}, using an assumed intrinsic 
ratio of scale-height to semi-major axis of 0.12.\newline

\begin{figure}
\ifhires
\includegraphics[width=0.5\textwidth]{NUVSSFRvSM_test_all_vert_otmoverplot.eps}
\else
\includegraphics[width=0.5\textwidth]{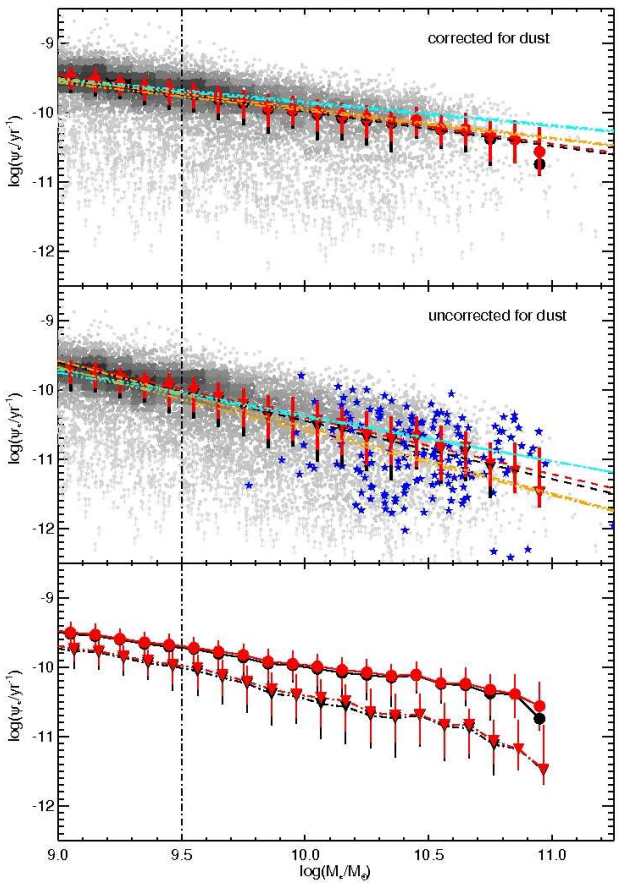}
\fi
\caption{Specific star formation rate ($\psi_*$) versus stellar mass ($M_*$) for a sample of spiral galaxies selected using the cell-based method 
and the parameter combination (log($n$),log($r_e$),$M_i$) and not hosting an AGN following the prescription of \citet{KEWLEY2006}, with 
$z\le 0.05$. Individual sources are plotted as filled circles with the grayscale color indicating the relative source density at their position in the 
$\psi_* - M_*$ plane. Values of $\psi_*$ have been derived from NUV photometry as described in Sect.~\ref{PSIMSTARSPIRALloc}. Galaxies 
without an NUV counterpart in the GCAT MSC (Seibert et al., 2013 in prep.) are show as 3-$\sigma$ upper limits. The limiting stellar mass of 
$M_* \ge 10^{9.5} M_{\odot}$ above which the sample can be considered volume limited is indicated by a vertical dash-dotted line. The median 
value in bins of $0.1\,$dex in $M_*$ is shown as large filled circles, with errorbars depicting the interquartile range in each bin. The medians 
and scatter for the whole sample are shown in black, while those of the sample considering only sources with NUV counterparts are shown in 
red. The top panel shows the distribution and median relations after radiation transfer based attenuation corrections following 
\citet{GROOTES2013} have been applied, while the middle panel shows the uncorrected distribution and median relations. The black and red 
dashed lines in the top and middle panels show power-law fits to the median relation in the mass range $10^{9.5} M_{\odot} \le M_* \le 10^{11} 
M_{\odot}$, corresponding to the volume-limited sample. The bottom panel shows the corrected (circles) and uncorrected (stars) relations to 
facilitate a direct comparison of the slope and scatter before and after correction for dust attenuation. Spirals found to host an AGN following 
the prescription of \citet{KEWLEY2006} are shown by blue stars in the middle panel. The relations found using the prescription of 
\citet{BALDRY2004} and a simple S\'ersic index cut are shown in azure and orange respectively, with the dashed line showing the relation as 
determined from all galaxies considered, and the dash-dotted line indicating the relation as recovered using only the detected sources.}  
\label{fig_SSFRSM}
\end{figure}

Fig.~\ref{fig_SSFRSM} shows the values of $\psi_*$ as a function of $M_*$ before and after correction for dust attenuation (middle and top 
panel, respectively), with the median in bins of $0.1\,$dex in $M_*$ shown as large filled circles with the errorbars indicating the interquartile 
range in logarithmic scatter in each bin. Without attenuation corrections, the $\psi_* - M_*$ relation displays a mean logarithmic 
scatter\footnote{The mean logarithmic scatter is calculated as the difference between the quartiles of the distribution in $\psi_{*}$, averaged 
over 15 equal sized bins in $M_*$ spanning $10^{9.5} M_{\odot} \le M_* \le 10^{11} M_{\odot}$, and weighted by the number of galaxies in 
each bin.} of  $0.70\,$dex ($0.63\,$dex considering only NUV-detected sources) for the volume-limited sample. A pure power-law fit to the 
median distribution of the uncorrected sample finds an index of $\gamma \approx-0.8$, but also shows that a pure power-law is only 
marginally suited to describing the distribution.\newline
After applying attenuation corrections, we find that the mean logarithmic scatter is reduced to $0.48\,$dex ($0.43\,$dex considering only 
NUV-detected sources). In addition to this large reduction in scatter, we find that the median $\psi_* - M_*$ relation for the volume-limited 
corrected sample is well represented by a pure power-law with an index of $\gamma \approx-0.5$ over the entire range in $M_*$, and that 
this power-law also provides a good parameterization of the relation at least down to $M_* = 10^9 M_{\odot}$. The exact value of the power-
law index found using a linear regression analysis of the binwise median of $\psi_*$ as shown in Fig.~\ref{fig_SSFRSM} is $\gamma 
=-0.50\pm0.12$. The quoted error has been derived using the interquartile scatter in each bin and represents a conservative estimate of the 
accuracy. There is no evidence for a break in the power-law over the full range of $M_*$ considered, despite the use of a sample incorporating 
red, quiescent spirals not considered in previous studies.\newline
Both for the corrected and uncorrected samples the median $\psi_* - M_*$ relation is largely invariant between the whole sample, and the 
subsample considering only NUV-detected galaxies, indicating that the true distribution of NUV detections and upper limits would provide 
similar results.\newline

\begin{figure}
\includegraphics[width=0.5\textwidth]{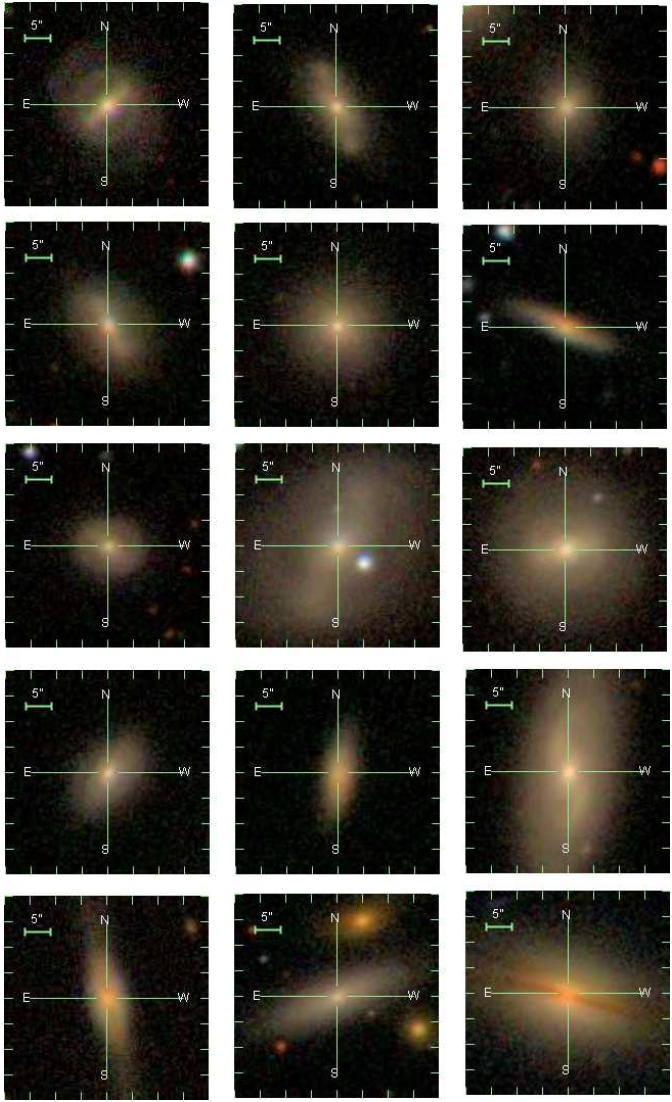}
\caption{SDSS DR7 5 band images of a random selection of 15 spiral galaxies from the sample considered with an NUV counterpart in the GCAT 
MSC, $M_* \ge 10^{9.5} M_{\odot}$, and $\mathrm{log}(\psi_*/ M_{\odot}\,\mathrm{kpc}^{-2}\,) \le -11$ after attenuation corrections have 
been applied. All but two of the sources (top right and second row middle) display a disk-like morphology. The images have been retrieved 
using the SDSS Explore tool.}
\label{fig_thumbnails}
\end{figure}

As the selection of spiral galaxies is purely morphologically based, the sample is capable of including very red and potentially passive spiral 
galaxies and should have a low contamination rate by ellipticals ($\sim2$\%, see Sect.~\ref{parameters}). However, one might expect the small 
number of ellipticals misclassified as spirals to have low values of $\psi_*$, which might
 affect the $\psi_* - M_*$ relation. To investigate to what extent the visible population of passive spirals is in fact a population of misclassified 
 ellipticals, we have visually inspected a random sample of galaxies with NUV detections, $M_* \ge 10^{9.5} M_{\odot}$, and $\mathrm{log}
 (\psi_*/ \mathrm{yr}^{-1}\,) \le -11$ after correction for dust. Fifteen randomly selected such galaxies are shown in Fig.~\ref{fig_thumbnails}. 
 All but two galaxies (top right panel and middle panel of second row) are clearly disk dominated spirals, showing that the large majority of the 
 considered population appear to be disk-like galaxies. This serves as further validation of the cell-based selection technique, and implies that 
 the derived $\psi_* - M_*$ relation is not biased by a large contamination of elliptical galaxies.\newline 

Conversely, even for the combination (log($n$),log($r_e$),$M_i$) a slight bias against early type spirals remains, which could potentially 
affect the $\psi_* - M_*$ relation, in particular if a large fraction of the massive, red spiral population were missed by the cell-based selection 
method. To investigate this potential effect we begin by considering the early-type spirals in the \textit{NAIRsample} (i.e. T-type $> 3$). We find 
$32\,$\% of the early-type spirals recovered from the \textit{NAIRsample} using the cell-based method to be red ($u-r > 2.2$) and massive 
($M_* > 10^{10.5} M_{\odot}$), compared to $38\,$\% red and massive galaxies amongst the early type-spirals NOT recovered by the cell-
based method, implying that the early-type galaxies not recovered are not strongly weighted more towards massive red objects than those 
recovered. To judge the impact of the bias against early-type spirals on the $\psi_* - M_*$ relation, however, it is necessary to consider not 
only the early-type galaxies, but the entire populations of spiral galaxies in the \textit{NAIRsample}  recovered, respectively not recovered by the 
cell-based method. Overall, one finds that for galaxies classified as spirals by \citet{NAIR2010} and recovered by the cell-based method with the 
parameter combination (log($n$), log($r_e$), $M_i$) massive red galaxies constitute $15\,$\% of the sample, while massive red galaxies 
constitute $27\,$\% of the spirals not recovered by the cell-based method. This relatively small shift in weight at the massive red end ($\sim 
12\,$\%) combined with the high completeness fraction ($>65\,$\%) attained by the cell-based selection implies that the results obtained for 
the $\psi_* - M_*$ relation for spiral galaxies in the local universe are robust. Thus, although it is possible that the actual $\psi_* - M_*$ 
relation may still be slightly steeper, this further steepening will be small compared to the steepening to the $\psi_* propto M_*^{-0.5}$ law 
found for the cell-based sample.\newline

Finally, Fig.~\ref{fig_SSFRSM} shows the location of spiral galaxies hosting an AGN on the $\psi_* - M_*$ relation. Although the interpretation of 
the NUV emission of such sources as being indicative of their SFR is by no means secure, since the AGN can also significantly contribute to the 
NUV emission, we find that the ratio of NUV emission to stellar mass of spiral galaxies hosting optically identified AGN is not readily 
distinguishable from that of similar galaxies without an AGN. AGN host galaxies do, however, appear to be more massive than 
$\sim10^{10}M_{\odot}$ as a rule, and display a larger scatter. Fig.~\ref{fig_NUVSSFRvSM_CAL} shows the locations of optically 
identified AGN in a sample of galaxies with the additional requirement of H$\alpha$ and H$\beta$ lines with $S/N > 3$ as described in 
Sect.~\ref{depend}. The similar distribution to that seen in Fig.~\ref{fig_SSFRSM} implies that the predominance of massive galaxies as AGN 
hosts is not a result of selection effects in the spectroscopy used. \newline

\subsubsection{The effects of sample construction: Comparison with the $\psi_* - M_*$ relations for colour-selected and S\'ersic-index 
selected samples}  
\begin{figure*}
\ifhires
\includegraphics[width=0.9\textwidth]{NUVSSFRvSM_test_all_vert_otm2.eps}
\else
\includegraphics[width=0.9\textwidth]{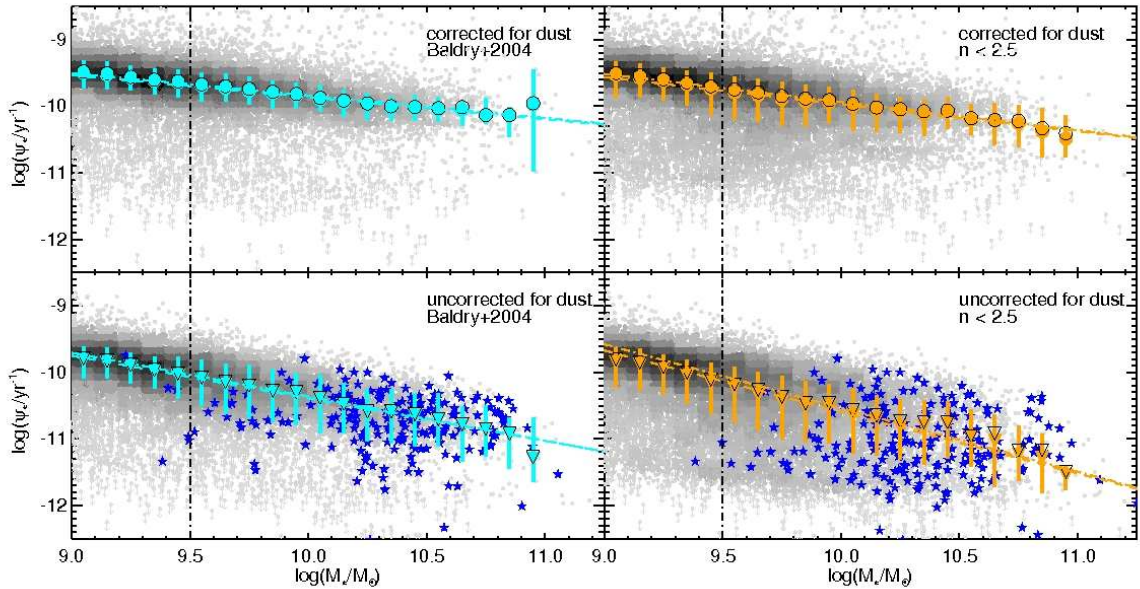}
\fi
\caption{Specific star formation rate ($\psi_*$) versus stellar mass ($M_*$) for a sample of spiral galaxies selected using the method of 
\citep{BALDRY2004} (left top and bottom) and a simple S\'ersic index cut (right top and bottom) and not hosting an AGN following the 
prescription of \citet{KEWLEY2006}, with $z\le 0.05$. Individual sources are plotted as filled circles with the grayscale color indicating the 
relative source density at their position in the $\psi_* - M_*$ plane. Values of $\psi_*$ have been derived as previously detailed. Galaxies 
without an NUV counterpart in the GCAT MSC (Seibert et al., 2013 in prep.) are show as 3-$\sigma$ upper limits. The median values of $\psi_*$ 
in bins of $0.1\,$dex in $M_*$ are shown as large symbols, with errorbars depicting the interquartile range in each bin. The medians and 
scatter for the whole sample are shown in filled symbols and colour, while the medians of the sample considering only sources with NUV 
counterparts are shown as black outlines. The top panels show the distribution and median relations after radiation transfer based attenuation 
corrections following \citet{GROOTES2013} have been applied, while the bottom panels show the uncorrected distribution and median relations. 
The dashed and dash-dotted lines in the top and bottom panels show power-law fits to the median relations in the mass range $10^{9.5} 
M_{\odot} \le M_* \le 10^{11} M_{\odot}$, corresponding to the volume-limited samples, with the dashed line showing the relation derived for 
the entire sample and the dash-dotted line showing the relation as derived only for the detected sources. Spiral galaxies found to host an AGN 
following the prescription of \citet{KEWLEY2006} are shown by blue stars in the bottom panels.} 
\label{fig_otmSSFRSM}
\end{figure*}

We have previously argued and demonstrated, that the cell-based method of selecting pure and complete samples of spiral galaxies is 
capable of including quiescent spirals and is therefore well-suited to investigating the $\psi_* - M_*$ relation for a morphologically defined 
sample of spiral galaxies.
In order to illustrate the effect that the choice of classification method has on the results derived for the $\psi_* - M_*$ relation and 
demonstrate the necessity of an adequate selection method, Fig.~\ref{fig_otmSSFRSM} shows the relation for galaxy samples drawn from the 
\textit{OPTICALsample} and limited to $z\le 0.05$ selected using the prescription of \citet{BALDRY2004} (left) and the S\'ersic index (right). 
Attenuation corrections have been applied using the method of \citet{GROOTES2013} as previously described. The derived relations have also 
been overplotted in Fig.~\ref{fig_SSFRSM} for comparison. For the sample selected following the method of \citet{BALDRY2004} we find a power 
law index of 
$\gamma = -0.64\pm0.15$ before applying attenuation corrections and an index of $\gamma = -0.33\pm0.11$ after applying attenuation 
corrections. Both before and after correction a single power-law appears to be an adequate representation of the $\psi_* - M_*$ relation for this 
sample. Considering the scatter in the $\psi_* - M_*$ relation we find that the relation is tight both before and after applying attenuation 
corrections, with values of $0.52\,$dex interquartile and $0.40\,$dex, respectively.\newline

Using the S\'ersic index to select a sample of spiral galaxies, we find a power-law index of $\gamma = -0.93\pm0.15$ before and 
$\gamma = -0.41 \pm0.14$ after applying attenuation corrections. The $\psi_* - M_*$ relation before correction, however, is not well 
described by a single power-law. For the sample selected in this manner, the $\psi_* - M_*$ relation displays a scatter of $0.89\,$ dex 
interquartile before applying attenuation corrections which is reduced to $0.59\,$dex interquartile by applying attenuation corrections.
\newline
For both these sample selection methods - by S\'ersic-index and by colour - the power-law indices recovered are indicative of a 
shallower relation than for the cell-based selection. Given the similarity of the relations at lower stellar masses ($\sim 10^{9.5} M_{\odot}$) this 
appears to be largely due to a difference in the samples in the high stellar mass range, with the cell-based selection recovering more quiescent 
spirals. This is in line with the finding that the samples selected by these widely used proxies are more strongly biased towards sources with 
large values of H$\alpha$ equivalent width. 
It is particularly note-worthy that the colour based selection of \citet{BALDRY2004} leads to a much shallower slope and a very low scatter, most 
likely due to the exclusion of quiescent galaxies.\newline  

This comparison demonstrates the care necessary in constructing galaxy samples for the purpose of statistical investigations and 
illustrates the suitability of the cell-based method of morphological classification for the investigation of the star formation properties of 
morphologically selected samples of spiral galaxies. A further discussion of the effects of sample construction on the $\psi_* - M_*$ relation is 
given in Sect.~\ref{comparisonwithotherwork}. \newline

\subsection{Dependency on attenuation corrections}\label{depend}
In deriving the intrinsic $\psi_* - M_*$ relation for spiral galaxies in the local universe we have made use of the prescription for obtaining 
attenuation corrections given by \citet{GROOTES2013} and the radiation transfer model of \citet{POPESCU2011}, as empirically 
calibrated on a sample of nearby spirals \citep[see][]{XILOURIS1999,POPESCU2000,POPESCU2004,MISIRIOTIS2001} and incorporating corrections 
for the effects of dust on the perceived effective radii of disks by \citet{PASTRAV2013b}. In order to investigate to what extent the results 
obtained depend on the chosen method of deriving attenuation corrections, we compare the results obtained using the prescription of 
\citet{CALZETTI2000} with those obtained using the method of \citep{GROOTES2013}. These two correction methods, while both being 
empirically based, have a very different basis. Whereas the method of \citet{GROOTES2013} is calibrated on a sample of local universe spirals 
with FIR-UV detections, the method of \citet{CALZETTI2000} is calibrated on a sample of distant starburst galaxies, utilizing measurements of 
emission line fluxes. Furthermore, whereas, by virtue of its radiation transfer treatment, the method of \citet{GROOTES2013} does not assume a 
fixed attenuation law in the UV/optical, this is the case for the method of \citet{CALZETTI2000}. This is potentially a critical factor when 
correcting for dust attenuation in spiral galaxies which lie on the transition between optically thick and thin systems, for which one expects a 
large range in the shape of the attenuation curve. Because of the requirement of emission line fluxes, the comparison must be based on a 
different sample, this time incorporating galaxies with H$\alpha$ and H$\beta$ line fluxes measured at $>3$-$\sigma$, which effectively 
removes the population of red, quiescent galaxies. Thus, we select a sample of spiral galaxies with NUV counterparts, selected using the cell-
based method with the parameter combination (log($n$),log($r_e$),$M_i$), with $z\le 0.05$, not hosting an AGN, and with H$\alpha$ and 
H$\beta$ line fluxes measured at $>3$-$\sigma$ as the basis for the following comparison. We emphasize that the requirements on the 
spectroscopic information serve only to facilitate the comparison with the corrections obtained using the prescription of \citet{CALZETTI2000}.
\newline

\begin{figure*}
\ifhires
\includegraphics[width=0.9\textwidth]{NUVSSFRvSM_test_alt.eps}
\else
\includegraphics[width=0.9\textwidth]{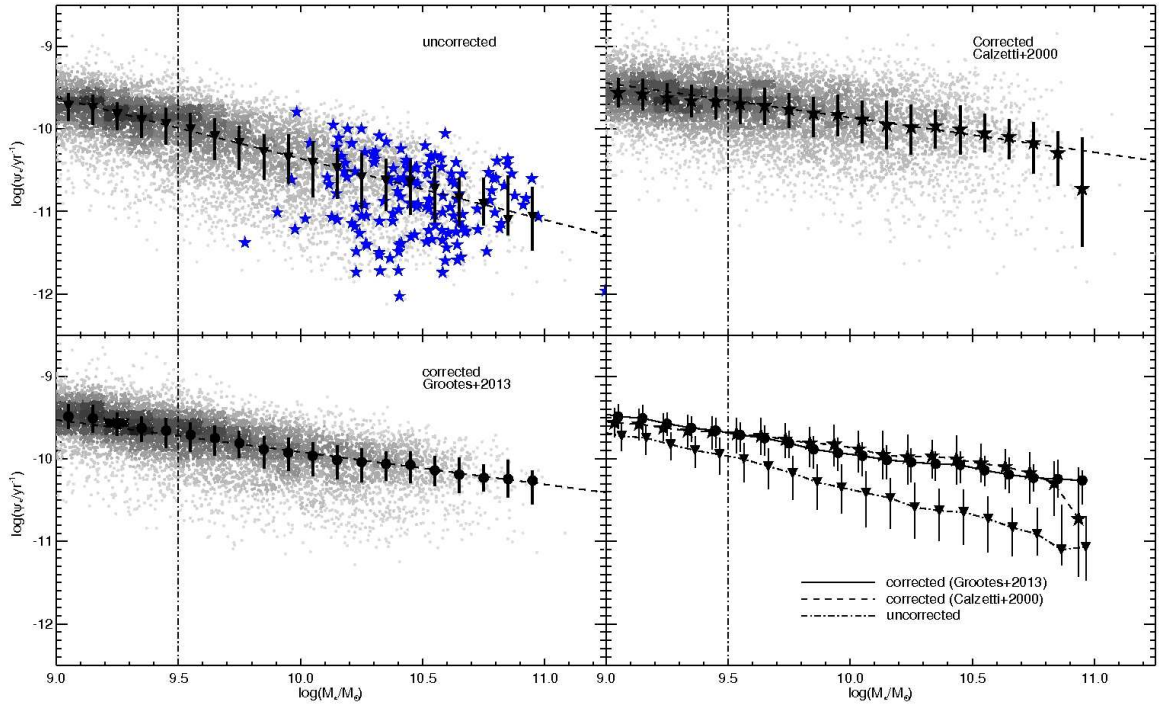}
\fi
\caption{Specific star formation rate $\psi_{*}$ versus stellar mass $M_{*}$ for a subsample of spirals galaxies drawn from the 
\textit{OPTICALsample} using the cell-based method and the parameter combination (log($n$),log($r_e$),$M_i$) with $z\le 0.05$, NUV 
detections and H$\alpha$ and H$\beta$ fluxes at $>3\sigma$, not hosting an AGN. The linear grayscale indicates the relative galaxy density in 
the $\psi_* - M_*$ plane at the position of the galaxy. The same scale has been applied to all panels. The vertical dashed-dotted line indicates 
the stellar mass limit above which the sample can be considered complete. The sources are binned in bins of equal size in $M_{*}$, with the bars 
showing the interquartile range and the filled symbols (stars, inverted triangles and circles) showing the median value of $\psi_*$ in each bin. 
The dashed line in the top panels and the bottom left panel shows a single power-law fit to the binwise median values in the mass range 
$10^{9.5} M_{\odot} \le M_* \le 10^{11} M_{\odot}$. The bottom right panel shows the median relations to facilitate comparison. The 
uncorrected relation is shown as inverted triangles and a dash-dotted line. The relation corrected for dust attenuation following 
\citet{GROOTES2013} is shown as circles and a solid line, while the relation corrected for dust attenuation following \citet{CALZETTI2000} is 
shown as stars and a dashed line. The bin centers have been offset by 0.01 in $\mathrm{log}(M_*)$ for improved legibility. The scatter in the 
relation due to the scatter in the NUV is significantly reduced for the corrections based on the radiation transfer model, while the Balmer 
decrement based corrections have no discernible effect on the scatter. In both cases the intrinsic values of $\psi_{*}$ are shifted upwards w.r.t. 
the uncorrected values. Spiral galaxies fulfilling the criteria of the sample but hosting an AGN have been overplotted as blue stars in the top left 
panel.} 
\label{fig_NUVSSFRvSM_CAL}
\end{figure*}

Fig.~\ref{fig_NUVSSFRvSM_CAL} shows the distributions of $\psi_*$ as a function of $M_*$ without corrections for dust (top left) and with 
corrections obtained using the radiation-transfer based method of \citet{GROOTES2013} (bottom left). The $\psi_* - M_*$ relation obtained 
using the method detailed in \citet{CALZETTI2000} for correcting dust attenuation is shown in the top right panel. As in the case for the full 
sample incorporating red disks, the radiation transfer based corrections lead to a significant tightening of the relation, in this case reducing the 
mean logarithmic scatter from $0.58\,$dex to $0.37\,$dex. This lends confidence that the radiation transfer method also has the ability to 
predict the correct overall shift in the relation (see also discussion in \citealt{GROOTES2013}, Sects.~5 \& 6). By contrast, under the application 
of the corrections based on the Balmer decrement the scatter remains at $0.49\,$dex.\newline
Nevertheless, the overall shift in the relation towards larger values of $\psi_*$ by $0.3\,$dex on average is similar in both cases. This is a 
remarkable result bearing in mind the very different derivations of these methods and shows that the method of \citet{CALZETTI2000} is indeed 
a very robust technique applicable to star-forming galaxies over a wide range of morphology and redshift.\newline 
Both correction methods lead to a similar, shallower dependence of $\psi_*$ on $M_*$ than found for the uncorrected relation, with the slope 
of the relation obtained using the prescription of \citet{CALZETTI2000} being slightly shallower than that of the relation obtained by applying 
the method of \citet{GROOTES2013}. The power-law index found under both corrections is close to $\gamma \approx -0.4$. The flattening 
compared to the power-law index of $\gamma \approx -0.5$ found when applying the corrections of \citet{GROOTES2013} to the full sample 
(as described in Sect.~\ref{PSIMSTARSPIRALloc} and shown in Fig.~\ref{fig_SSFRSM}) may be attributed to the exclusion of red, quiescent 
systems, which tend to be more massive, by the requirement of emission line flux measurements.\newline
The main systematic difference between the two methods for dust corrections is that the relation based on the \citet{CALZETTI2000} attenuation 
corrections shows an indication of a possible break in the power-law at $M_* \approx 10^{10.5} M_{\odot}$,
not found when using the \citet{GROOTES2013} attenuation corrections. The fact that the \citeauthor{GROOTES2013} attenuation 
corrections significantly reduce the overall scatter in the relation, may imply that the break is actually not physical in nature, but rather may be 
an artifact of the application of the \citet{CALZETTI2000} corrections to high mass spiral galaxies.\newline

\subsection{Comparison of the morphologically defined $\psi_* - M_*$ relation with previous determinations }\label{comparisonwithotherwork}
Previous determinations of the $\psi_* - M_*$ relation have generally necessarily been restricted to galaxy samples encompassing the complete 
population of galaxies \citep[e.g.][]{SALIM2007,ELBAZ2007, NOESKE2007}, or to samples selected on the basis of colour or star-formation 
activity \citep[e.g][]{PENG2010,WHITAKER2012}. As such, the $\psi_* - M_*$ relation has been defined in terms of a blue sequence, or more 
generally a sequence of star-forming galaxies, and has been contrasted with a red sequence, or more generally a sequence of non-star-forming 
galaxies (\citealp{PENG2010}, respectively \citealp{NOESKE2007, WHITAKER2012}). However, the more fundamental distinction may be the 
morphology of the galaxy. This is because, while rotationally supported galaxies can support an extended cold ISM which can support 
distributed star-formation, any extended ISM in a spheroid must be hot and tenuous if it is in virial equilibrium with the total mass distribution 
as traced by the stars, in which case it would be expected to be inefficient in forming stars. To constrain processes driving star-formation in 
galaxies, it is therefore instructive to establish the $\psi_* - M_*$ relation for a pure disk sample.\newline
We have found this relation to be a relatively tight ($0.42\,$dex mean logarithmic interquartile range, corresponding to $0.31\,$dex 
1-$\sigma$ for a normal distribution) power-law with an index of $\gamma = -0.5\pm0.12$, with no indication of a cut-off at high stellar 
mass. This result shows that the phenomenon of down-sizing\footnote{Down-sizing describes the phenomenon that star-formation in the 
current epoch is biased towards low mass structures, in contrast to the sequence of growth in dark matter structures, which progresses from 
low mass to high mass} is also exhibited by a morphologically pure sample of disk galaxies, and is not just due to an increasing fraction of 
spheroids with increasing stellar mass in the general galaxy population.\newline
The lack of an obvious turn-off in the $\psi_* - M_*$ relation for spirals, despite the inclusion of red quiescent spirals, suggests that if a 
mechanism exists to restrict the growth of spiral galaxies beyond the stellar mass range probed, such a mechanism must be accompanied by an 
abrupt transformation of galaxy morphology.\newline

As outlined above, previous works addressing the $\psi_* - M_*$ relation have concentrated on the sequence of star-forming galaxies 
rather than a morphologically defined sample. For example \citet{PENG2010} make use of a $U-B$ color selection (their Eq.~2) akin to that of 
\citet{BALDRY2004} investigated in Sect.~\ref{PSIMSTARSPIRALloc} of this paper, applying it to a sample of SDSS galaxies with star formation 
rates derived from H$\alpha$ line measurements as provided by \citet{BRINCHMANN2004}. These authors find a power law index of $\gamma 
= -0.1$, much shallower than the relation found in this work. Similarly, \citet{WHITAKER2012} find that for local universe star-forming galaxies 
selected using $U-V$ \& $V-J$ restframe colors, selecting a blue subset of these galaxies results in a shallow slope similar to that of 
\citet{PENG2010}. However, considering their full sample of star-forming galaxies \citet{WHITAKER2012} find a steeper slope of $\gamma 
\approx -0.4$. Finally, \citet{NOESKE2007} find a slope of $\gamma = -0.33\pm0.08$ for local universe galaxies with indications of on-going 
star-formation either in form of $24\,\mu$m emission and/or H$\alpha$ emission. \newline
The fact that these previously determined values of $\gamma$ are all shallower than the relation found for a morphologically selected sample of 
spirals presented in this work, can be readily understood. By selecting actively star-forming systems, quiescent galaxies of similar morphology 
are excluded from the samples. As passive spirals tend to be more massive, on average, this leads to a flattening of the $\psi_* - M_*$ with 
respect to a morphologically defined, sample, as similarly argued by \citet{WHITAKER2012} in the context of the result of \citet{PENG2010}. 
Indeed, for the sample of spirals selected using the cell-based method with the combination (log($n$), log($r_e$), $M_i$) and the additional 
requirement of H$\alpha$ and H$\beta$ detections, as used in Sect.~\ref{depend}, we find the $\psi_* - M_*$ relation to be well described by a 
single power-law with an index of $\gamma = -0.39\pm0.09$ and a scatter of $0.37\,$dex interquartile ($0.27\,$dex 1-$\sigma$, assuming 
a normal distribution), very similar to the results for star-forming galaxies as obtained by other authors as previously discussed.\newline
Overall, we thus find that the $\psi_* - M_*$ relation for a morphologically selected sample of spiral galaxies with an index of 
$\gamma=-0.5$ is moderately steeper than that found for star-forming galaxies ($\gamma=-0.3\cdots-0.4$), likely due to the inclusion of 
the population of red passive spirals in our analysis. This only moderate increase in slope is not greatly surprising, as the majority of spiral 
galaxies are found to display star-formation activity.\newline

\section{Summary and Outlook}\label{outlook}
We have presented a non-parametric cell-based method of selecting robust, pure, complete, and largely unbiased samples of spirals using 
combinations of three parameters derived from (UV/)optical photometry. 
We find that the parameters log($r_e$), log($\mu_*$), log($n$), and $M_i$ perform well in selecting simultaneously pure and complete 
samples, while the use of the ellipticity $e$ leads to pure yet incomplete samples. These parameters, which are linked to older stellar 
populations, perform at least as well as selections using the $u-r$ colour or the $NUV-r$ colour after NUV preselection. 
The remarkable success/importance of these seldom utilized parameters is consistent with the expected contrast in the structural properties of 
rotationally supported systems (spirals) and pressure supported systems (ellipticals), in agreement with different evolutionary tracks for spiral 
and elliptical galaxies.\newline 
For a selection of combinations of three parameters, the cell-based method is superior to a range of (widely used) photometric morphological 
proxies, and comparable to the algorithmic classification approach using support vector machines presented by \citet{HUERTAS-COMPANY2011} 
in selecting pure and complete samples of spirals from faint galaxy surveys.\newline
The optimum combinations for use with the method may vary according to the science application for which the sample is being constructed.
For application to optically defined galaxy samples comparable in depth or deeper than SDSS we identify the combinations 
(log($n$),log($r_e$),log($\mu_*$)),  (log($n$),log($r_e$),$M_i$), and (log($n$),log($M_*$),log($\mu_*$)) to be the most efficient in selecting a 
sample of spirals balanced between purity and completeness.\newline
While using NUV data can lead to purer samples, it poses the possibility of a bias against UV faint sources and edge-on systems. Furthermore, 
we caution that making use of UV/optical colours additionally poses stringent requirements on the depths of the samples used in order to 
provide complete and unbiased samples.\newline
In this paper, we have used the cell-based classification scheme with the parameter combination (log($n$),log($r_e$),$M_i$) to investigate the 
specific star-formation rate - stellar mass ($\psi_* - M_*$) relation for a purely morphologically defined sample of spiral galaxies. Using this 
approach which is unbiased in terms of star-formation properties and includes red, quiescent spiral galaxies, we find that the intrinsic, i.e. dust 
corrected, $\psi_* - M_*$ relation for spiral galaxies can be represented as a single continuous power-law with an index of $-0.5$ over the 
mass range $10^{9.5} M_{\odot} \le M_* \le 10^{11} M_{\odot}$, likely even extending to $10^{9} M_{\odot} \le M_*$. Despite the 
inclusion of quiescent galaxies, the relation is also found to be very tight, with a mean interquartile range of $0.4\,$dex. The lack of a turn-
over in the relation over the stellar mass range considered, implies that any mechanism terminating the growth of spiral galaxies beyond this 
mass range must be accompanied by a rapid morphological transformation.\newline
We supply the cell based division of the parameter space for the combination (log($n$),log($r_e$),$M_i$), as used in the investigation of the 
$\psi_* - M_*$ relation, as well as for the combinations (log($n$),log($r_e$),log($\mu_*$)) and (log($n$),log($M_*$),log($\mu_*$)) in 
Appendix~\ref{appendCell}, together with a brief instruction on their use.\newline   
Immediate future work will focus on using the method presented to test the performance of linear discriminant analysis using multiple 
parameters in the morphological classification of galaxies (Robotham et al., in prep.), as well as on defining samples of spirals for use in 
applications of radiation transfer modelling techniques \citep{POPESCU2011}, which critically rely on the existence of the appropriate geometry 
(in this case spiral disk geometry), to derive self-consistent corrections of the attenuation of UV/optical light by dust in these objects.

\section*{Acknowledgements} 
We thank Ted Wyder for his assistance in compiling the sample.
Some of the results in this paper have been derived using the HEALPix\footnote{http://healpix.jpl.nasa.gov} \citep{GORSKI2005} package.
Funding for the SDSS and SDSS-II has been provided by the Alfred P. Sloan Foundation, the Participating Institutions, the National Science 
Foundation, the U.S. Department of Energy, the National Aeronautics and Space Administration, the Japanese Monbukagakusho, the Max Planck 
Society, and the Higher Education Funding Council for England. The SDSS Web Site is http://www.sdss.org/.\newline
The SDSS is managed by the Astrophysical Research Consortium for the Participating Institutions. The Participating Institutions are the American 
Museum of Natural History, Astrophysical Institute Potsdam, University of Basel, University of Cambridge, Case Western Reserve University, 
University of Chicago, Drexel University, Fermilab, the Institute for Advanced Study, the Japan Participation Group, Johns Hopkins University, the 
Joint Institute for Nuclear Astrophysics, the Kavli Institute for Particle Astrophysics and Cosmology, the Korean Scientist Group, the Chinese 
Academy of Sciences (LAMOST), Los Alamos National Laboratory, the Max-Planck-Institute for Astronomy (MPIA), the Max-Planck-Institute for 
Astrophysics (MPA), New Mexico State University, Ohio State University, University of Pittsburgh, University of Portsmouth, Princeton University, 
the United States Naval Observatory, and the University of Washington.\newline
GALEX (Galaxy Evolution Explorer) is a NASA Small Explorer, launched in April 2003. We gratefully acknowledge NASA's support for construction, 
operation, and science analysis for the GALEX mission, developed in cooperation with the Centre National d'Etudes Spatiales (CNES) of France 
and the Korean Ministry of Science and Technology.\newline
MWG acknowledges the support of the International Max-Planck Research School in Astronomy and Cosmic Physics Heidelberg (IMPRS-HD) and 
the Heidelberg Graduate School for Fundamental Physics (HGSFP).
We thank the anonymous referee for his/her comments which have helped us improve the paper.

\appendix

\section{Cell decompositions of parameter space}\label{appendCell}
We have found the parameter combinations (log($n$),log($r_e$),$M_i$), (log($n$),log($r_e$), log($\mu_*$)), and (log($n$), log($M_*$), 
log($\mu_*$)) to be most efficient in retrieving a simultaneously pure and complete, largely unbiased sample of spiral galaxies when applied to 
the optically defined galaxy sample used in this work. In addition to the high values of purity and completeness, these selections require a 
minimal amount of spectral coverage, hence can readily be applied to various samples of galaxies.\newline

Tabs.~\ref{tab_logsidxlogreabsi}, \ref{tab_logsidxlogrelogSMSD}, \& \ref{tab_logsidxlogSMlogSMSD} provide the decompositions of the 
parameter space spanned for the combinations (log($n$),log($r_e$),$M_i$), (log($n$),log($r_e$), log($\mu_*$)), and (log($n$), log($M_*$), 
log($\mu_*$))  respectively. These discretizations have been performed using the entire \textit{OPTICALsample} as a calibration sample to 
maximize the purity and completeness. The full tables are available in the online version of the paper and in machine readable form from 
the VizieR Service at the CDS\footnote{Tabs.~\ref{tab_logsidxlogreabsi}, \ref{tab_logsidxlogrelogSMSD}, \& \ref{tab_logsidxlogSMlogSMSD} are available in machine readable form at the CDS via anonymous ftp to cdsarc.u-strasbg.fr (130.79.128.5) or via http://cdsarc.u-strasbg.fr/}. Rather than supply a binary classification into spiral and non-spiral cells we supply the spiral fraction and its relative error for 
each cell, allowing the reader to adapt the classification to his purposes. We do, however note, that the underlying definition of a reliable spiral 
($P_{\mathrm{CS,DB}} \ge 0.7$) is fixed.\newline
In addition, we have chosen to provide the elliptical fraction for each cell and it relative error, where ellipticals are, analogously to spirals, 
defined as sources with $P_{\mathrm{EL,DB}} \ge 0.7$ \newline

We emphasize, that in using the discretizations supplied, it is essential that the reader ensure that there are not significant systematic 
shifts in the parameters being used between the data used in this work, and the data of the sample to be classified. An initial comparison can be 
made using Figs.~\ref{fig_pardistopt} \& \ref{fig_pardistuv}. Similarly, the random uncertainties on the data should not exceed the highest 
resolution cell dimensions as listed in the tables provided.\newline 

The tables supply the front lower left corner of each cell (axis are oriented in a right-hand system), the lengths of the sides in each dimension, 
the spiral fraction $F_{\mathrm{sp}}$, its relative error $\Delta F_{\mathrm{sp,rel}}$, the elliptical fraction $F_{\mathrm{el}}$, its relative error 
$\Delta F_{\mathrm{el,rel}}$, and the resolution level the cell belongs to (1; 1 division per axis, 2; 4 divisions per axis, 3; 8 divisions per axis, 4; 
16 divisions per axis) . With this information the entire grid can, if desired, be reconstructed. For classifying galaxies the tables can be used as 
follows:\newline
\begin{itemize}
\item{select criteria for being a spiral (or elliptical) cell in terms of $F_{\mathrm{sp}}$ and $\Delta F_{\mathrm{sp,rel}}$ (respectively 
$F_{\mathrm{el}}$ and $\Delta F_{\mathrm{el,rel}}$)}
\item{for each source identify the nearest grid point to its forward lower left}
\item{assign the values of $F_{\mathrm{sp}}$ and $\Delta F_{\mathrm{sp,rel}}$ from the corresponding cell to the source in question}
\item{after completion for all sources select those corresponding to the selection criteria determined}  
\end{itemize}

\begin{table*}
 \centering
  \begin{tabular}{@{}lrrrrrrrrrr@{}}
  \hline
  \multicolumn{1}{c}{resolution} &\multicolumn{3}{c}{corner coordinates}    &     \multicolumn{3}{c}{cell dimensions } & \multicolumn{2}{c}{Spiral 
  fractions}& \multicolumn{2}{c}{Elliptical fractions}\\
 & $\mathrm{log}(n)$ & $\mathrm{log}(r_e)$ & $M_i$ & d$\mathrm{log}(n)$ & d$\mathrm{log}(r_e)$ & d$M_i$ & $F_{\mathrm{sp}}$ &   
 $\Delta F_{\mathrm{sp,rel}}$ & $F_{\mathrm{el}}$ &   $\Delta F_{\mathrm{el,rel}}$\\ 
 \hline 
2 & 0.60750 & 0.00000 &-24.5000 & 0.30250 & 0.50000 & 2.25000 & 0.00000 &1.0000e+06 & 0.61290 &2.9136e-01\\
2 &-0.30000 & 0.50000 &-24.5000 & 0.30250 & 0.50000 & 2.25000 & 0.70213 &1.6059e-01 & 0.02128 &7.1459e-01\\
2 &-0.30000 & 1.00000 &-24.5000 & 0.30250 & 0.50000 & 2.25000 & 0.96875 &2.5201e-01 & 0.00000 &1.0000e+06\\
2 &-0.30000 &-0.50000 &-22.2500 & 0.30250 & 0.50000 & 2.25000 & 1.00000 &1.4142e+00 & 0.00000 &1.0000e+06\\
3 & 0.00250 & 0.50000 &-23.3750 & 0.15125 & 0.25000 & 1.12500 & 0.22222 &4.5134e-01 & 0.03704 &1.0184e+00\\
3 & 0.15375 & 0.50000 &-23.3750 & 0.15125 & 0.25000 & 1.12500 & 0.23333 &2.9681e-01 & 0.08333 &4.6547e-01\\
3 & 0.30500 & 0.50000 &-23.3750 & 0.15125 & 0.25000 & 1.12500 & 0.09091 &3.3029e-01 & 0.33636 &1.9005e-01\\
3 & 0.00250 & 0.75000 &-23.3750 & 0.15125 & 0.25000 & 1.12500 & 0.74375 &1.2105e-01 & 0.00625 &1.0031e+00\\
4 & 0.22938 & 0.87500 &-23.3750 & 0.07563 & 0.12500 & 0.56250 & 0.06000 &5.9442e-01 & 0.74000 &2.1686e-01\\
4 & 0.15375 & 0.75000 &-22.8125 & 0.07563 & 0.12500 & 0.56250 & 0.01639 &7.1288e-01 & 0.86885 &1.3278e-01\\
4 & 0.22938 & 0.75000 &-22.8125 & 0.07563 & 0.12500 & 0.56250 & 0.02564 &5.8471e-01 & 0.86325 &1.3582e-01\\
4 & 0.15375 & 0.87500 &-22.8125 & 0.07563 & 0.12500 & 0.56250 & 0.00000 &1.0000e+06 & 0.86441 &1.9120e-01\\
\hline
\hline                      
\end{tabular}
\caption{Excerpt of cell grid for the combination (log($n$),log($r_e$),$M_i$). For cells with a spiral(elliptical) population of 0 the relative error is 
set to 1e6. }
  \label{tab_logsidxlogreabsi}
\end{table*}

\begin{table*}
 \centering
  \begin{tabular}{@{}lrrrrrrrrrr@{}}
  \hline
  \multicolumn{1}{c}{resolution} &\multicolumn{3}{c}{corner coordinates}    &     \multicolumn{3}{c}{cell dimensions } & \multicolumn{2}{c}{Spiral 
  fractions}& \multicolumn{2}{c}{Elliptical fractions}\\
 & $\mathrm{log}(n)$ & $\mathrm{log}(r_e)$ & $\mathrm{log}(\mu_*)$ & d$\mathrm{log}(n)$ & d$\mathrm{log}(r_e)$ & d$\mathrm{log}
 (\mu_*)$ & $F_{\mathrm{sp}}$ &   $\Delta F_{\mathrm{sp,rel}}$ & $F_{\mathrm{el}}$ &   $\Delta F_{\mathrm{el,rel}}$\\  
 \hline
2 &-0.30000 & 1.00000 & 6.25000 & 0.30250 & 0.50000 & 1.25000 & 0.94805 &1.6336e-01 & 0.00000 &1.0000e+06\\
2 &-0.30000 &-0.50000 & 7.50000 & 0.30250 & 0.50000 & 1.25000 & 0.02703 &1.0134e+00 & 0.13514 &4.7647e-01\\
2 & 0.00250 &-0.50000 & 7.50000 & 0.30250 & 0.50000 & 1.25000 & 0.01905 &7.1381e-01 & 0.16190 &2.6143e-01\\
2 & 0.30500 &-0.50000 & 7.50000 & 0.30250 & 0.50000 & 1.25000 & 0.00000 &1.0000e+06 & 0.11236 &3.3352e-01\\
3 & 0.30500 & 0.50000 & 6.87500 & 0.15125 & 0.25000 & 0.62500 & 0.58442 &1.8764e-01 & 0.00000 &1.0000e+06\\
3 & 0.45625 & 0.50000 & 6.87500 & 0.15125 & 0.25000 & 0.62500 & 0.20000 &5.4772e-01 & 0.05000 &1.0247e+00\\
3 & 0.30500 & 0.75000 & 6.87500 & 0.15125 & 0.25000 & 0.62500 & 0.77778 &1.5936e-01 & 0.01111 &1.0055e+00\\
3 & 0.45625 & 0.75000 & 6.87500 & 0.15125 & 0.25000 & 0.62500 & 0.58333 &2.7458e-01 & 0.00000 &1.0000e+06\\
4 & 0.00250 & 0.50000 & 7.18750 & 0.07563 & 0.12500 & 0.31250 & 0.80000 &2.0226e-01 & 0.00000 &1.0000e+06\\
4 & 0.07813 & 0.50000 & 7.18750 & 0.07563 & 0.12500 & 0.31250 & 0.66176 &1.9217e-01 & 0.01471 &1.0073e+00\\
4 & 0.00250 & 0.62500 & 7.18750 & 0.07563 & 0.12500 & 0.31250 & 0.68750 &2.2613e-01 & 0.00000 &1.0000e+06\\
4 & 0.07813 & 0.62500 & 7.18750 & 0.07563 & 0.12500 & 0.31250 & 0.55556 &3.2203e-01 & 0.00000 &1.0000e+06\\
\hline
\hline                      
\end{tabular}
\caption{Excerpt of cell grid for the combination (log($n$),log($r_e$),log($\mu_*$)). For cells with a spiral(elliptical) population of 0 the relative 
error is set to 1e6. }
  \label{tab_logsidxlogrelogSMSD}
\end{table*}

\begin{table*}
 \centering
  \begin{tabular}{@{}lrrrrrrrrrr@{}}
  \hline
  \multicolumn{1}{c}{resolution} &\multicolumn{3}{c}{corner coordinates}    &     \multicolumn{3}{c}{cell dimensions } & \multicolumn{2}{c}{Spiral 
  fractions} & \multicolumn{2}{c}{Elliptical fractions}\\
 & $\mathrm{log}(n)$ & $\mathrm{log}(M_*)$ & $\mathrm{log}(\mu_*)$ & d($\mathrm{log}(n)$) & d$\mathrm{log}(M_*)$ & d$\mathrm{log}
 (\mu_*)$ & $F_{\mathrm{sp}}$ &   $\Delta F_{\mathrm{sp,rel}}$ & $F_{\mathrm{el}}$ &   $\Delta F_{\mathrm{el,rel}}$\\  
 \hline
2 & 0.30500 & 7.50000 & 8.75000 & 0.30250 & 1.12500 & 1.25000 & 0.00000 &1.0000e+06 & 0.00000 &1.0000e+06\\
2 & 0.60750 & 7.50000 & 8.75000 & 0.30250 & 1.12500 & 1.25000 & 1.00000 &1.4142e+00 & 0.00000 &1.0000e+06\\
2 &-0.30000 & 8.62500 & 8.75000 & 0.30250 & 1.12500 & 1.25000 & 0.00000 &1.0000e+06 & 0.00000 &1.0000e+06\\
2 & 0.00250 & 8.62500 & 8.75000 & 0.30250 & 1.12500 & 1.25000 & 0.00000 &1.0000e+06 & 0.16129 &4.8193e-01\\
3 & 0.45625 & 9.75000 & 6.25000 & 0.15125 & 0.56250 & 0.62500 & 0.80000 &4.7434e-01 & 0.00000 &1.0000e+06\\
3 & 0.60750 & 9.75000 & 6.25000 & 0.15125 & 0.56250 & 0.62500 & 0.87500 &5.1755e-01 & 0.00000 &1.0000e+06\\
3 & 0.75875 & 9.75000 & 6.25000 & 0.15125 & 0.56250 & 0.62500 & 0.81250 &3.7339e-01 & 0.00000 &1.0000e+06\\
3 &-0.30000 &10.31250 & 6.25000 & 0.15125 & 0.56250 & 0.62500 & 0.00000 &1.0000e+06 & 0.00000 &1.0000e+06\\
4 &-0.07313 & 9.18750 & 6.87500 & 0.07563 & 0.28125 & 0.31250 & 0.80000 &4.7434e-01 & 0.00000 &1.0000e+06\\
4 &-0.14875 & 9.46875 & 6.87500 & 0.07563 & 0.28125 & 0.31250 & 0.80000 &2.7386e-01 & 0.00000 &1.0000e+06\\
4 &-0.07313 & 9.46875 & 6.87500 & 0.07563 & 0.28125 & 0.31250 & 0.89286 &2.7516e-01 & 0.00000 &1.0000e+06\\
4 &-0.14875 & 9.18750 & 7.18750 & 0.07563 & 0.28125 & 0.31250 & 0.94444 &3.3820e-01 & 0.00000 &1.0000e+06\\
\hline
\hline                      
\end{tabular}
\caption{Excerpt of cell grid for the combination (log($n$),log($M_*$),log($\mu_*$)). For cells with a spiral(elliptical) population of 0 the relative 
error is set to 1e6. }
  \label{tab_logsidxlogSMlogSMSD}
\end{table*}

\newpage

\bsp

\label{lastpage}

\end{document}